\documentclass[article]{aastex63}

\usepackage{ulem}
\usepackage{enumerate}
\usepackage{lineno}

\providecommand{\e}[1]{\ensuremath{\times 10^{#1}}}
\submitjournal{\apj}

\shorttitle{Updates to CLOUDY's molecular network}
\shortauthors{Shaw, Ferland, \& Chatzikos}

\begin{document}
\correspondingauthor{Gargi Shaw}
\email{gargishaw@gmail.com, gary@g.uky.edu, mchatzikos@gmail.com}
\author{Gargi Shaw}
\affiliation{Department of Astronomy and Astrophysics, Tata Institute of Fundamental Research\\
Homi Bhabha Road, Navy Nagar, Colaba, Mumbai 400005, India}
\author{G. J. Ferland}
\affiliation{Department of Physics and Astronomy, University of Kentucky\\
Lexington, KY 40506, USA}
\author{M. Chatzikos}
\affiliation{Department of Physics and Astronomy, University of Kentucky\\
Lexington, KY 40506, USA}

\title{A recent update of gas-phase chemical reactions and molecular lines in CLOUDY: its effects on millimeter and 
sub-millimeter molecular line predictions}

\begin{abstract}
Here we present our current updates of the gas-phase chemical reaction rates and molecular 
lines in the spectral synthesis code \textsc{cloudy}, and its implications in 
spectroscopic modelling of various astrophysical environments. 
We include energy levels, radiative and collisional rates for HF, CF$^+$, HC$_3$N,  ArH$^+$, HCl, HCN, CN, CH, and CH$_2$. Simultaneously, we expand our molecular 
network involving these molecules. For this purpose, we have added 561 new reactions and 
have updated the existing 165 molecular reaction rates involving these molecules. As a result, \textsc{cloudy} now predicts all 
the lines arising from these nine molecules.
In addition, we also update H$_2$--H$_2$ collisional data up to rotational levels $J$=31 for $v$=0.
We demonstrate spectroscopic simulations of these molecules for a few astrophysical environments. 
Our existing model for globules in the Crab nebula 
successfully predicts the observed column density of ArH$^+$. 
Our model predicts a detectable amount of HeH$^+$, OH$^+$, 
and CH$^+$ for the Crab nebula. We also model the ISM towards HD185418, W31C, NGC 253, and our predictions match with most of the observed 
column densities within the observed error bars.
Very often molecular lines trace various physical conditions. Hence, this update will be 
very supportive for spectroscopic modelling of various astrophysical environments, particularly involving 
sub-millimeter and mid-infrared 
observations using ALMA and JWST, respectively.
\end{abstract}

\section{Introduction}
With the advancement of modern technology and state-of-the-art telescopes, a plethora of molecular 
lines have been observed and more will be observed in the future. Very often these lines are tracers of various physical conditions \citep{{2007A&A...468..627V},{2006agna.book.....O},{1999ApJ...517..209G}}.
Hence, these lines need correct identifications and interpretation. 
The interstellar medium (ISM) is dilute and far from thermodynamic equilibrium.
So concepts like a single temperature is not valid. In NLTE cases, the level populations are not determined only by Boltzmann distribution. One needs to utilize all the populating and depopulating mechanisms from a particular level to derive its level population.
In such environments, detailed atomic processes determine
the physical conditions.
This is possible only when a complete set of
atomic and molecular data are available so
that the microphysics can be computed.

The widely used spectroscopic simulation 
code \textsc{cloudy} \citep{{2017RMxAA..53..385F},{2013RMxAA..49..137F}} simulates the conditions in
a non-equilibrium astrophysical plasma 
and predicts the resulting spectrum. 
It solves the detailed microphysics on an ab initio basis, hence needs fundamental atomic
and molecular data.
For molecules, two types of data are needed.
The spectrum depends 
on the density of a species, which in
turn depends
on the chemistry.
The spectrum of a molecule also depends on its internal
structure and excitation/deexcitation rate coefficients.
These are different problems, and \textsc{cloudy} uses different 
databases to obtain the rate coefficients needed
to solve the coupled system of equations. 

The most recent release of \textsc{cloudy},
C17 described by \citet{2017RMxAA..53..385F},
predicts the 
column densities of 92 and spectral lines of 40 molecules.
The hydrogen molecule is treated as a
separate problem and is described by
\citet{2005ApJ...624..794S}.
The heavy-molecule network was originally
described by
\citet{2005ApJS..161...65A}. Most of the chemical-reaction rate coefficients 
in \textsc{cloudy} were updated following
our participation in the Lorentz Center PDR
workshop \citep{2007A&A...467..187R}. 

\textsc{cloudy} determines the density of various
molecules using chemical reaction rate
coefficients from several sources. 
Many are from UMIST Database for Astrochemistry, UDfA \citep{{1997A&AS..121..139M},{2013A&A...550A..36M}}. 
We update our chemical network regularly to predict line intensities
and column densities more accurately.
The current release of \textsc{cloudy} 
\citep{2017RMxAA..53..385F}
uses UDfA rate coefficients from 1999 (RATE99) and 2006 (RATE06). 

Similarly, \textsc{cloudy} uses various databases for energy levels, 
Einstein's radiation coefficients, and collisional rate coefficients. We mainly use five distinct databases: CHIANTI \citep{2012ApJ...744...99L}, 
LAMDA \citep{2005A&A...432..369S}, Stout \citep{2015ApJ...807..118L}, and two internal \textsc{cloudy} databases for H-like and
He-like iso-electronic sequences \citep{2012MNRAS.425L..28P} and the H$_{2}$ molecule \citep{2005ApJ...624..794S}. These are also updated regularly to 
predict more lines with better 
precision. 

In this work, we incorporate the upgraded LAMDA database \citep{2020Atoms...8...15V}\footnote{\url{https://home.strw.leidenuniv.nl/~moldata/}} molecules that 
were not previously included in \textsc{cloudy} \citep{2017RMxAA..53..385F}.
We add molecules HF, CF$^+$, and HC$_3$N from the upgraded LAMDA. We also update the internal structure and chemical reactions for HF, CF$^+$, HC$_3$N, ArH$^+$, HCl, HCN, CN, CH, and CH$_2$.
We update  the earlier rate coefficients involving these above-mentioned molecules 
to UDfA 2012 (RATE12). At the same time, 
we also add new reaction rates from RATE12. 

\textsc{cloudy} can compute the conditions in an immense range of 
physical states extending from highly ionized to 
fully molecular. 
For a particular cloud, different molecules and different lines from the same molecule 
originate at distinct depths in the cloud.  
For a region dominated by interactions with starlight,
such as a PDR,
the conditions depend on the visual extinction, $A_{\rm V}$. 
For instance, in diffuse clouds, ArH$^+$ forms predominantly in the atomic gas with $A_{\rm V}$ $\leq$10$^{-2}$ mag \citep{2016ApJ...826..183N}, 
while HC$_3$N forms at higher values of $A_{\rm V}$ $>1$ mag \citep{2017A&A...605A..88L}. 
We show examples of astronomical environments 
where these updates affect the spectrum (Section~\ref{sec:implications}).


This paper is organized in the following manner: in Section~\ref{sec:Chemistry},
we describe updates of the molecular data (reaction rate coefficients, internal structure including energy levels, Einstein's A coefficients, and collisional 
rate coefficients). 
These updates  change predictions for observables from
a non-equilibrium gas.
Our models with updated data and results are presented in Section~\ref{sec:implications}.
Discussion and summary are given in Section~\ref{sec:summary}. Furthermore, we discuss our future development plans in Section~\ref{sec:future}. 
We list our added and updated chemical reaction rate coefficients in the Appendices.

\section{Updates to the Chemistry and Internal Structure}\label{sec:Chemistry}
In the ISM, gas-phase reactions proceed through various routes. The reactions added/updated from the UDfA for this work can be categorized as
\begin{enumerate}[i)]
    \item ion-neutral (A$^+$ + B $\rightarrow$ C$^+$ + D);
    \item neutral-neutral (A + B $\rightarrow$ C + D);
    \item photo-dissociation (AB + h$\nu$ $\rightarrow$ A + B);
    \item charge transfer (A$^+$ + B $\rightarrow$ A + B$^+$);
    \item radiative-association (A + B $\rightarrow$ AB + h$\nu$);
    \item dissociative-recombination (A$^+$ + e$^{-}$ $\rightarrow$ C + D); and
    \item associative-detachment (A$^-$ + B $\rightarrow$ AB + e$^-$).
\end{enumerate}

The reaction rate coefficients generally depend on temperature. Most of the neutral-neutral reactions are endothermic with high activation barriers for 
the interaction between two
non-radicals. Hence, these types of reaction rate coefficients have positive temperature dependence. The presence of radicals affects the 
rate coefficients by lowering the activation barrier. On the other hand, ion-neutral reactions involve long-range attraction between collisional partners. 
So classically, these rate coefficients are temperature independent \citep{{2010SSRv..156...13W},{2009ApJS..185..273W}}. However, in  the  presence  of directional forces between an ion and a dipolar molecule, the reaction rate coefficients increase as the temperature is lowered \citep{2011IAUS..280..361S}. 

In a nutshell, for a two-body reaction, the rate coefficient $k$ ($\rm{cm^3 \, s^{-1}}$) is given by the usual Arrhenius-type formula,
\begin {equation}
k=\alpha \left(\frac{T}{300}\right)^\beta \, \exp(-\gamma /T),
\end {equation}
where $T$ is the gas temperature.
On the other hand, photoreaction rate coefficients, $k$ ($\rm{s}^{-1}$), are given by
\begin{equation}
k= \alpha \, \exp(-\gamma A_{\rm V}) \, .
\end{equation}
Here, $A_{\rm V}$ is the extinction at visible wavelengths. Details can be found 
in \citet{2013A&A...550A..36M}. It is to be noted that the parameters $\alpha$ and $\gamma$ used in the two-body reactions and photoreactions are not the same. 
For photoreactions, $\alpha$ represents the rate coefficient in the unshielded interstellar ultraviolet radiation field, and $\gamma$ represents the increased dust extinction at ultraviolet wavelengths. Whereas, for a two-body reaction $\gamma$ depends on the activation energy of the reaction. We use the same symbols for ease of presentation 
of the new reactions in the Tables in the Appendices.
Which case they refer to, however, is unambiguous for each reaction.

\textsc{cloudy} does not use the UDfA cosmic-ray ionization rates. We calculate cosmic-ray ionization rates self-consistently just as it calculates
$A_{\rm V}$ self-consistently \citep{2005ApJS..161...65A, 2017RMxAA..53..385F}.
Our cosmic-ray ionization rates are normalised to a total rate for electron production
from cosmic-ray ionization. We adopt the \citet{{2007ApJ...671.1736I}} mean 
cosmic-ray ionization rate of 2$\times$10$^{-16}$ s$^{-1}$ for atomic hydrogen as the default background value. \citet{2021ApJ...908..138S} 
suggest that for the average Galactic PAH abundance,
the cosmic-ray ionization rate of atomic hydrogen is 3.9$\pm$1.9$\times$10$^{-16}$ s$^{-1}$. 
This rate can be
specified with the command, ``\texttt{cosmic rays background 1.95 linear}''.

Uncertainties in chemical reaction rate coefficients influence the 
predicted abundances/column densities of the species involved.
This effect empirically increases as the number of atoms in the interacting molecule
increases \citep{2004AstL...30..566V}. 
Hence, among all the molecules considered here, we expect results to be most uncertain for HC$_3$N, as it involves the 
greatest number of atoms. It is important to note that the abundances of multi-atom organic molecules depend on reactions on dust grain surfaces also. 
\citet{2007msl..confE.112V} have studied the effects of uncertainties in RATE06 
in protoplanetary disks and showed that the dispersion in the column densities 
is not more than
a factor of 4. 
Furthermore, \citet{2010SSRv..156...13W} have studied the uncertainties related to interstellar molecules in detail using the Kinetic Database for Astrochemistry (KIDA)\footnote{\url{https://kida.astrochem-tools.org/}}.

\textsc{cloudy} has a test suite that contains a large number of models covering diverse astrophysical environments. 
These models have been tested since the beginning of \textsc{cloudy} (1978), and they are publicly available with the \textsc{cloudy} download under the directory \texttt{tsuite}. These models are exercised daily.
We compare their predictions every day and monitor any changes
that occur as a result of changes in the source code and atomic and molecular data.
This framework provides an automatic mechanism to determine
changes to species abundances and molecular line intensities as reactions are updated.
We critically examine the update if it results in a relative change of greater than $\pm$ 30$\%$.
In this work, we notice that some reaction rates produce a relative change of more than $\pm$ 90$\%$ in the predicted column density
for some astrophysical models. We mention those rate coefficients in the respective molecule subsections in the
Appendix (A).
 
\section{Implications on modelled predictions} \label{sec:implications}

Here we show the effects of the above-mentioned updates on predictions for a few 
astrophysical models.
One of the models is a generic model of an HII region,
and a photodissociation region (PDR) extending into the deep molecular region.
The others are a model for the Crab nebula, and the ISM towards HD 185418, W31C, and NGC 253. 
These models cover both the Galactic (HD 185418, W31C) and nearby starburst galactic (NGC 253) sight-lines with different star-formation rates that 
provide an unique opportunity to validate our recent updates.

\subsection{An H II and PDR model in pressure equilibrium}
\label{subsec: model1}

To demonstrate the effects of these above-mentioned updates,
we first consider a generic HII region and 
a PDR calculation
that is in pressure equilibrium.
PDRs and HII regions exist side by side in star-forming
regions, so the assumption of pressure equilibrium is a reasonable physical scenario.
This model is similar to  \citet{2005ApJS..161...65A}, but not exactly the same.
The main difference is that the extent of the cloud (implemented in our simulations as a stopping criterion) is different. 
Here, we go deep in the molecular cloud up to $A_{\rm V}$=1000 mag. Furthermore, \citet{2005ApJS..161...65A} used a simple 3-level model for H$_2$ instead of a 
detailed internal microphysical H$_2$ network as described in \citet{2005ApJ...624..794S}.
\citet{2005ApJS..161...65A} showed variation of the number densities of various species across the cloud but did not report column densities 
of those species.

Below we briefly mention the various input parameters of our model.
In our model, the incident ionization is composed of a
stellar continuum with a temperature of 39600 K and Q(H)=10$^{49}$
(ionizing photons per second).
In addition, there is a bremsstrahlung with a temperature of 10$^6$~K
and 10$^{10}$ cm$^{-2}$s$^{-1}$ surface flux of ionizing photons.
The HII region starts at a distance of 10$^{17.45}$~cm away 
from the ionizing star and extends into the PDR and dense molecular region
up to $A_{\rm V}$ = 1000 mag.
The hydrogen density at the ionizing face is 10$^4$ cm$^{-3}$.
The adopted gas-phase abundances are similar to the average gas-phase abundances in the Orion Nebula. 
The gas-phase abundances of a few important elements in terms of number are,
C/H =3$\times$10$^{-4}$ ,
O/H = 4$\times$10$^{-4}$,
N/H = 7$\times$10$^{-5}$ ,
Cl/H = 3$\times$10$^{-4}$,
F/H = 2$\times$10$^{-8}$,
Ar/H = 3$\times$10$^{-6}$.
Moreover, we include larger size grains with
(a$_{min}$ = 0.03 $\mu$m, a$_{max}$ = 0.25 $\mu$m) an MRN size distribution
\citep{1977ApJ...217..425M}, that produces the Orion extinction curve
($R_{\rm V}$=5.5; \citet{1991ApJ...374..580B}). The input script of this model, ``h2$_{-}$orion$_{-}$hii$_{-}$pdr.in'', is publicly available with the \textsc{cloudy} download 
under the directory \texttt{tsuite}.

\begin {figure}
\center
\includegraphics[scale=0.8]{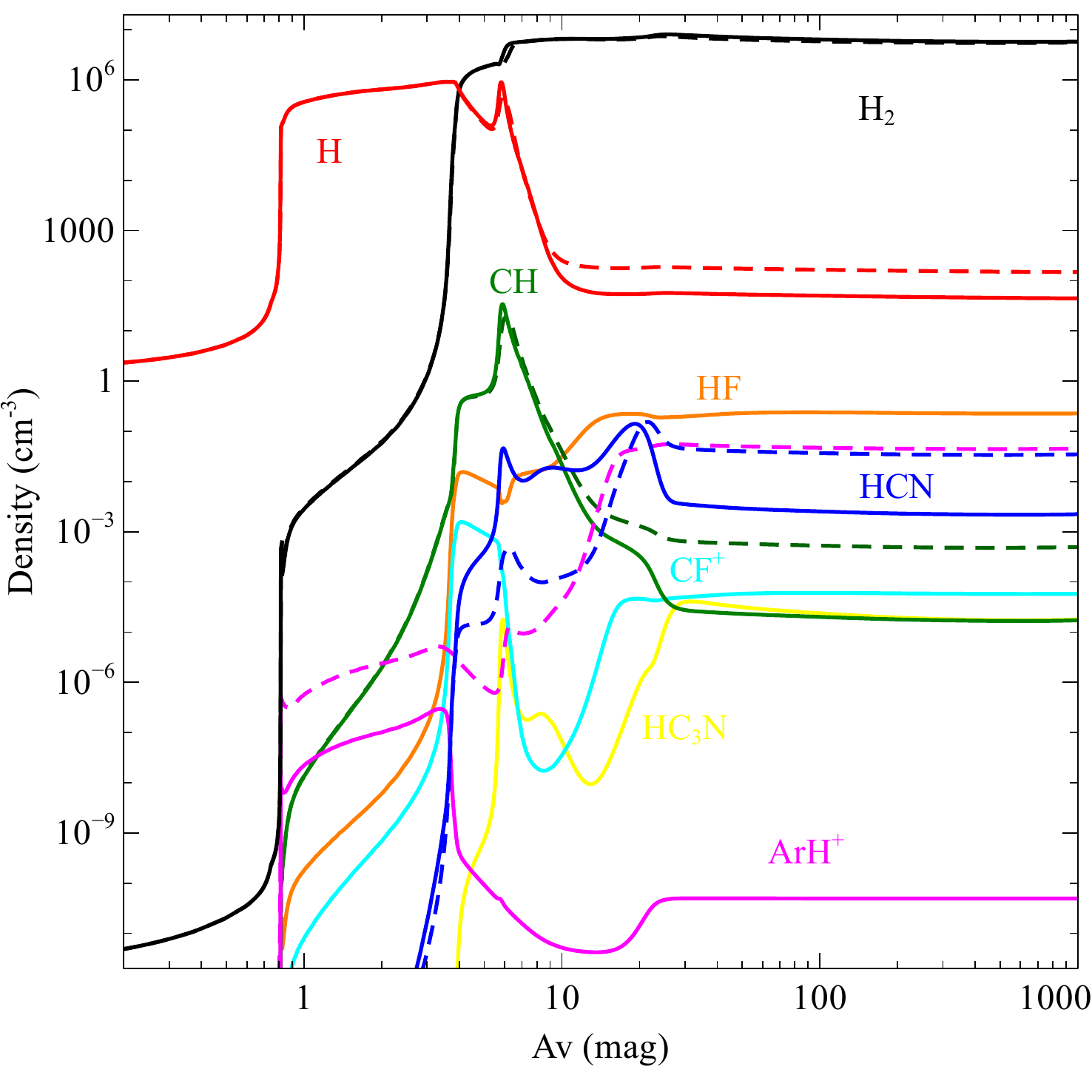}
\caption{Variation of densities of various molecules as a function of $A_{\rm V}$ for an H II and PDR model (``h2$_{-}$orion$_{-}$hii$_{-}$pdr.in'' from the \textsc{cloudy} download under the directory \texttt{tsuite}). 
The name of each molecule and the line representing its density are depicted in the same color. The solid lines represent simulations using this version, 
and the dashed lines represent simulations using an earlier version c17.}
\label{fig:structure_plot}
\end {figure}

\begin {figure}
\center
\includegraphics[scale=0.8]{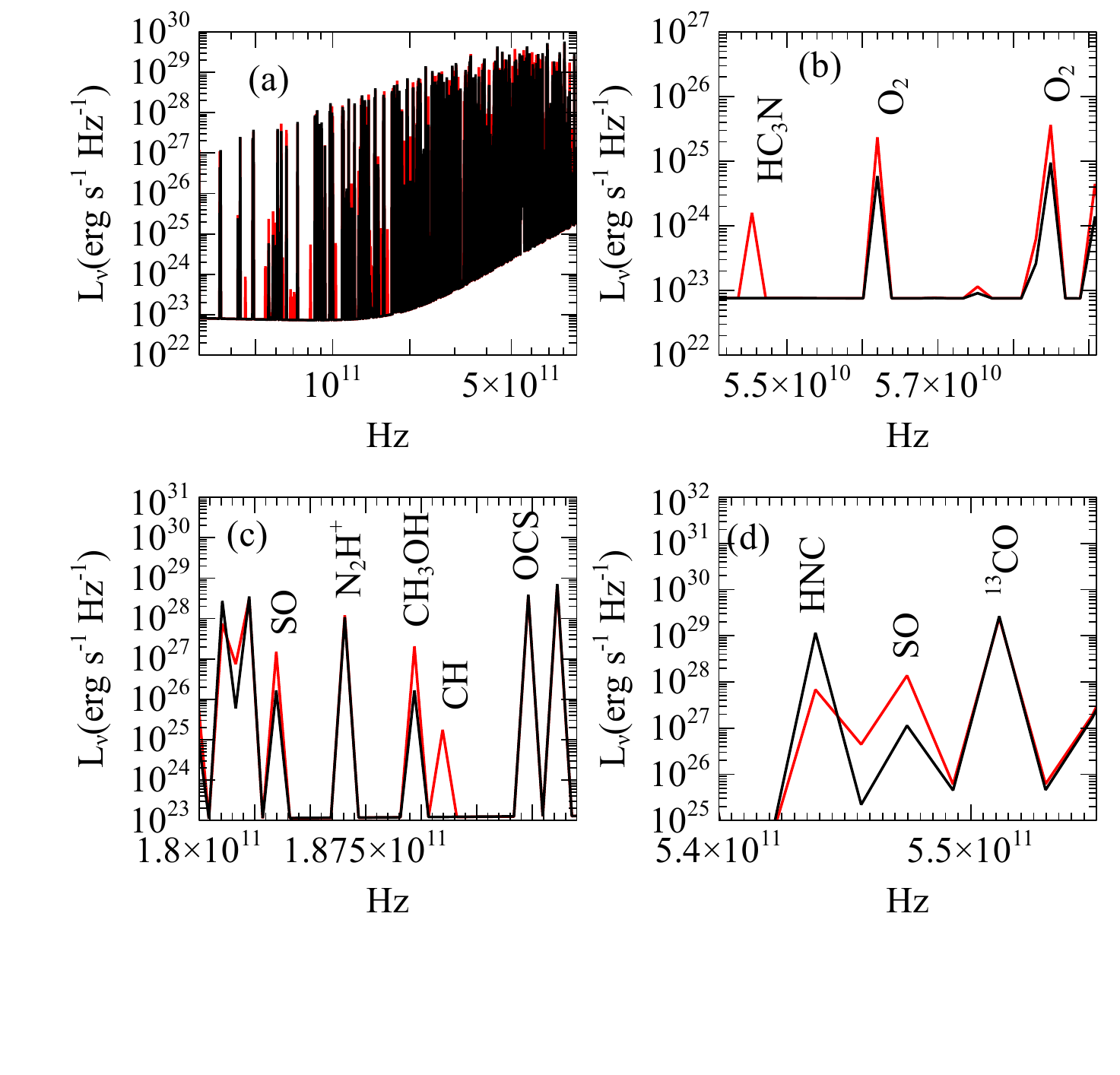}
\caption{The total transmitted continuum L$_\nu$ (erg s$^{-1}$ Hz$^{-1}$) is plotted as a function of frequency (Hz) for an 
H II and PDR model (``h2$_{-}$orion$_{-}$hii$_{-}$pdr.in'' from the \textsc{cloudy} download under the directory \texttt{tsuite}). We can see finer details 
as we zoom out the frequency scale.
To get the spectral flux density (in units of 10$^{23}$ Jy) on the earth, divide by 4$\pi$D$^2$, where D is the distance to earth in cm. The red 
lines are predicted using this update. Whereas, the black lines are predicted using version C17.}
\label{fig:panel_alma_hz}
\end {figure}

\begin {figure}
\center
\includegraphics[scale=0.8]{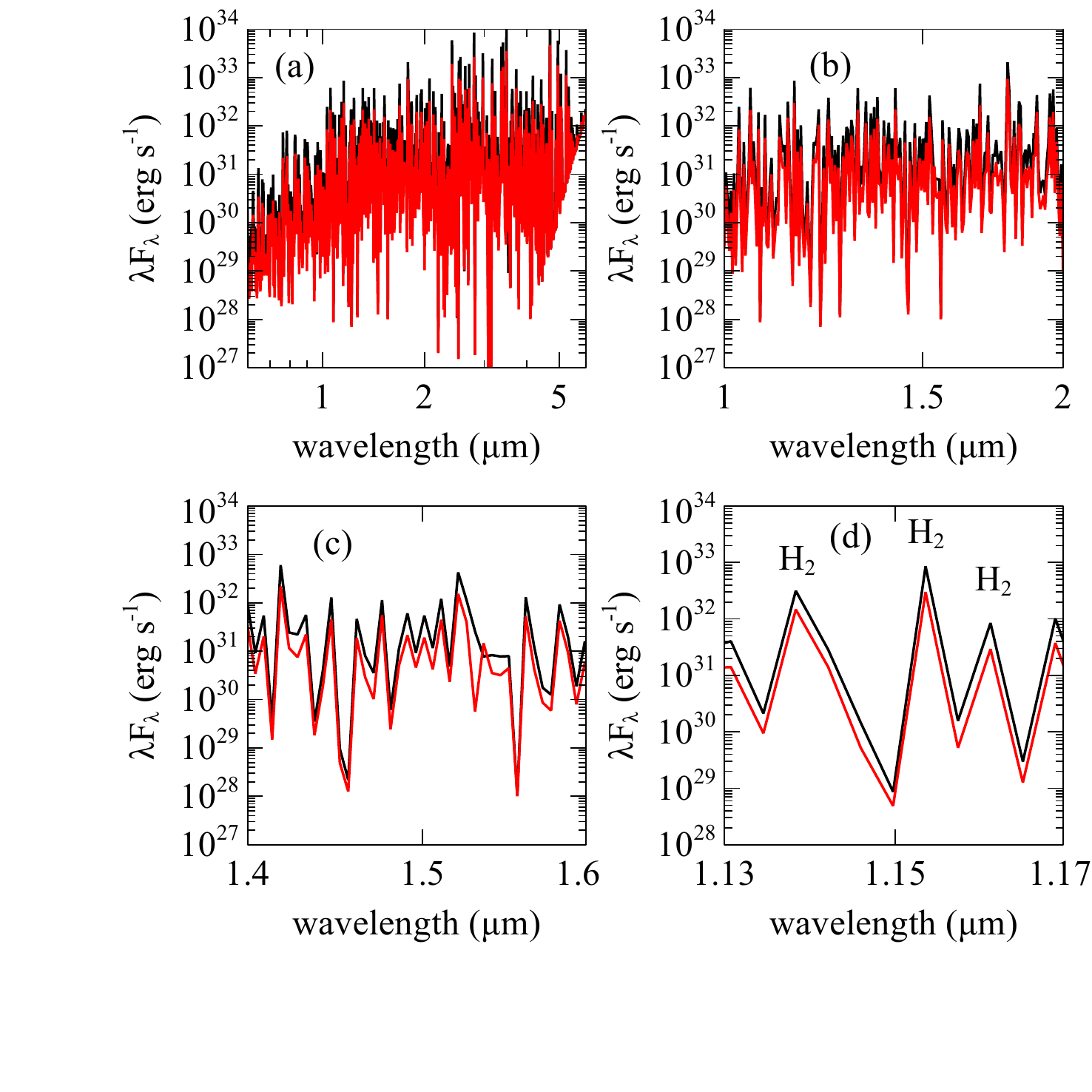}
\caption{Transmitted continuum $\lambda$L$_\lambda$ (erg s$^{-1}$) is plotted as a function of wavelength ($\mu$m) for an H II and PDR model 
(``h2$_{-}$orion$_{-}$hii$_{-}$pdr.in'' from the \textsc{cloudy} download under the directory \texttt{tsuite}). The red lines are predicted using this update.
Whereas, the black lines are predicted using version C17. We can see finer details as we zoom out the wavelength scale.}
\label{fig:panel_JWST_micron}
\end {figure}

\begin{deluxetable*}{lll}
\tablecaption{Comparison of predicted column densities (cm$^{-2}$ in log scale) for HII and PDR in pressure equilibrium with this version and version C17. 
\label{tab:table 1}}
\tablehead{
\colhead{Chemical species} & \colhead{This work} & \colhead{C17}}
\startdata
HF & 16.537 & --\\
CF$^+$ & 13.004 & --\\
HC$_3$N &12.448 & --\\
ArH$^+$ & 8.990 & 15.858\\
HCl & 16.083& 16.464\\
HCN & 14.670 &15.748\\
HNC & 14.566 & 15.988 \\
CN & 13.949& 14.092\\
CH & 15.672& 15.691\\
CH$_2$ & 14.408 & 14.533 \\
H$_2$ & 23.953 & 23.953 \\
NH$_3$ & 17.009 & 17.043\\
H$_2$CO &8.2188 &--\\
HI & 21.771 & 21.771\\
CII  & 18.415 & 18.420\\
CI   & 17.753 & 17.896\\
CO & 20.726 & 20.725\\
\enddata
\end{deluxetable*}

Table \ref{tab:table 1} compares the predicted column densities of the above-mentioned molecules of this 
work with the earlier version, c17. \textsc{cloudy} now predicts column densities of HF,
CF$^+$, HC$_3$N in addition to the other
molecular species. The column densities of H$_2$ and CO remained the same for this model (within a limit of 0.002 dex). The predicted H$_2$ column density 
is 8.97$\times$10$^{23}$ cm$^{-2}$, the same for both cases. Though this is a generic gas-phase model, it is to be noted that a similar H$_2$ column density 
of 5.8$\times$10$^{23}$ \citep{2012A&A...544L..19P} and 3.1$\times$10$^{23}$ \citep{2012ApJ...744...28P} cm$^{-2}$ have been observed towards the Orion ``hot core''. Interestingly, the predicted NH$_3$ column density, (1-4)$\times$10$^{17}$ cm$^{-2}$ \citep{2011ApJ...739L..13G}, is also close to that observed towards the Orion ``hot core''. We notice that the HCN and HNC column densities decreased by more than a factor of 10 from the earlier version. As mentioned before, the HCN/HNC column density ratio is an important diagnostic tool.
This work predicts HCN/HNC = 1.27, while the earlier version produced HCN/HNC $<$ 0.6. 
In general, HCN/HNC= 1 to 20 \citep{2020A&A...635A...4H}. In addition, relative to the previous version of CLOUDY,
the current update predicts a higher CN column density and an ArH$^+$ column density
that is lower by orders of magnitude.
This is because the ArH$^+$ reaction network was incomplete in
previous versions of \textsc{cloudy}.
\citet{2014A&A...566A..29S} reported that ArH$^+$ abundance depends on the molecular fraction. The ArH$^+$ column density predicted by this work is close
to that of \citet{2014A&A...566A..29S} for a fully molecular case. 

Fig.~\ref{fig:structure_plot} shows the variation of densities of several molecules as a function of $A_{\rm V}$.
The name of each molecule and the line representing its density are depicted in the same color. The solid lines represent simulations using this version, and the dashed lines represent simulations using the earlier version c17. 
We show densities of only a few molecules to avoid congestion in the plot. The densities of H, CH, HCN, and ArH$^+$ have decreased deep in the cloud compared to c17. However, the H$_2$ density remains almost the same, and its value is $\sim$ 10$^{6.8}$ cm$^{-3}$ for $A_{\rm V}$ $>$ 7 mag. As mentioned earlier, the previous incompleteness of the ArH$^+$ network resulted in more than 1-dex difference between the two versions. 
We also see that most of the fluorine is in HF. 

Panel (a) of Fig.~\ref{fig:panel_alma_hz} shows the total transmitted continuum L$_\nu$ (erg s$^{-1}$ Hz$^{-1}$) as a 
function of frequency in the range 35 to 950 GHz, covering the operational frequency range of ALMA. This has to be divided by 4$\pi$D$^2$, where D is the distance to earth in cm, to get the spectral flux density on the earth. The resolution of the coarse continuum energy mesh is set at the default value of 300 for this plot. The emission line to continuum contrast is set by ``set save resolving power 30000000''. The resolution of the coarse continuum can be changed using the command ``set continuum resolution''. Here, the red and the black solid lines 
represent data using this work and the earlier version C17, respectively. 
The panel (a) shows the full emission; the purpose of this is to demonstrate our predictability of a large number of spectral lines within the observable 
range of ALMA. However, many lines present for both the cases are superimposed on each other and hence are hard to be distinguished. The spectral density of some lines predicted with c17 changed but with different amounts. The lines that are only present for this version can be identified easily, for example HC$_3$N at 45.55 GHz, as expected. 
However, we can see finer details as we zoom out the frequency range in the subsequent panels. We show some of the 
spectral lines in the frequency ranges 54.1 to 59.1 GHz (panel b), 180 to 197 GHz (panel c), and 540 to 555 GHz (panel d) as demonstrations.
We can clearly see that some new lines are added and some existing line intensities get affected as a result of this current update. For the same model, 
panel (a) of Fig.~\ref{fig:panel_JWST_micron} shows the total transmitted continuum $\lambda$L$_\lambda$ (erg s$^{-1}$) as a function of wavelength in the 
range 0.6 to 6 $\mu$m, in the operational 
wavelength range of JWST. 
The color-coding is the same as the earlier figure.
Although there are far too many lines to discuss
in detail, this figure does show the diagnostic
potential of such a model. We can see finer details as we zoom out the wavelength range in panels (b), (c), and (d). 
There are many H$_2$ lines in the 0.6 to 6 $\mu$m range that are affected. As an example, we show a few H$_2$ lines in the 
wavelength range 1.13 to 1.17 $\mu$m.

\subsection{The Crab nebula}
\label{subsec: model2}

Next, we model Knot 51 (K51) globule in the Crab nebula. The Crab nebula is the nearest and the brightest 
supernova remnant with a rich history and detailed
modeling of H$_2$ emission, e.,g., by \citet{2013MNRAS.430.1257R}.
The argonium molecule, ArH$^+$, was first detected in the Crab nebula by \citet{2013Sci...342.1343B}.
Here we aim to predict the observed column density of ArH$^+$ for K51.
Earlier, \citet{2013MNRAS.430.1257R} modelled H$_2$ and various atomic
and ionic emission lines in K15 region of the Crab nebula using CLOUDY (c10)
but did not concentrate on complex molecules since observations did not exist. 

\citet{2013MNRAS.430.1257R} considered a ``cosmic-ray'' ionized model in some detail.
Several other groups have also modelled the Crab nebula and have suggested
different values for the cosmic-ray ionizing rate towards the Crab nebula.
\citet{2020A&A...643A..91J}  used a cosmic-ray ionization rate
(2.3$\pm$0.3)$\times$10$^{-16}$ s$^{-1}$, and
\citet{2019ApJ...885..109B} also favored a similar
cosmic-ray ionization rate.
On the other hand, \citet{2017MNRAS.472.4444P} suggested a very high cosmic-ray ionization rate ($\geq$ 1.3$\times$10$^{-10}$s$^{-1}$).
\citet{2013MNRAS.430.1257R} proposed several models
for K51, and their ionizing-particle model (CR) and temperature-floor model (TF)
reproduced most of the observed line fluxes.
In the CR model, they included the effects of the high energy
particles that pervade the Crab’s synchrotron plasma and generate heating.
On the contrary, the TF model considered an artificially fixed electron
temperature to the observed H$_2$ temperature of 2800 K.
However, \citet{2013MNRAS.430.1257R} did not report
any of the molecular ions for the Crab nebula, although column densities of some
of the species were available. 

We run both  the models of \citet{2013MNRAS.430.1257R} using this 
version of \textsc{cloudy} to predict column densities of various molecules and molecular ions.
For this modelling purpose, we use the same input parameters as \citet{2013MNRAS.430.1257R}. 
We use the same density law and the same ionizing radiation SED  
with a 10$^{10.56}$ cm$^{-2}$s$^{-1}$ surface flux of H-ionizing photons incident on the illuminated face. 
The Crab nebula has a complex filamentary structure with varying density.
The density varies from 10$^{2.97}$ cm$^{-3}$ to 10$^{5.25}$ cm$^{-3}$ along 
the filament \citep{2013MNRAS.430.1257R}. 
The physical thickness of the cloud considered was 10$^{16.5}$ cm.
Following these authors, we consider dust grains 
with a size distribution altered to produce $R_{\rm V}$=5.5, similar
to the dust found in the Orion nebula. 
\citet{2013MNRAS.430.1257R} have considered a high cosmic-ray ionization rate,
3.99$\times$10$^{-11}$ s$^{-1}$, in their CR model, whereas the TF
model used the standard cosmic-ray ionization rate, 2.$\times$10$^{-16}$ s$^{-1}$.

We compare the predicted results of these two models in 
Table~\ref{tab:table 2} using our new version of \textsc{cloudy}.
We also list the unreported column densities from
\citet{2013MNRAS.430.1257R} (which used version c10) for further comparison.
 The predicted N(H$_2$) for the CR model  decreased by 0.42 dex from that predicted by \citet{2013MNRAS.430.1257R}.
 This is caused by several improvements to the physics.
 Recently, we have modified the mean kinetic energy of the secondary electrons generated by cosmic-rays which leads to
 dissociation of H$_2$ \citep{2020RNAAS...4...78S}. In addition, there have been improvements to the
 physical treatment including self shielding since 
 2013 \citep{2017RMxAA..53..385F}. However, the differences in predicted column densities are less 
 than a factor of 3. As mentioned earlier, \citet{2007msl..confE.112V} have demonstrated that the uncertainties in chemical reaction rates can lead to a 
 factor of $\sim$ 4 difference in the predicted column densities. 
 
 The observed ArH$^+$ column density lies 
 within 10$^{11.6}$ to 10$^{13}$ cm$^{-2}$. The CR model underpredicts
 ArH$^+$ column density by 0.75 dex. However, the TF model predicts ArH$^+$ column density within the observed range. Furthermore, we notice that with a lower cosmic-ray ionization rate compared to \citet{2013MNRAS.430.1257R},
 1$\times$10$^{-11}$ s$^{-1}$, the CR model predicts the observed column density of ArH$^+$. The TF model also predicts a detectable 
amount of HeH$^+$, OH, OH$^+$, CH, CH$^+$, CO, CO$^+$, in addition to ArH$^+$, along this sight-line. 
These results do not by themselves say whether
the temperature floor or cosmic-ray model is to be preferred. Recently \citet{2021arXiv211106033W} have 
observed several molecules in the dense molecular cloud of the Crab supernova remnant using ALMA. 
New models and other future observations using SOFIA, ALMA, and JWST can test
the predicted column densities and determine which one of these scenarios
occurs in the Crab.

\begin{deluxetable*}{lcccc}
\tablecaption{The model predicted column densities for the Crab nebula Knot 51
(cm$^{-2}$ in log scale).
CR and TF represent models with cosmic-ray ionization rates
3.99$\times$10$^{-11}$ s$^{-1}$ and 2.$\times$10$^{-16}$ s$^{-1}$, respectively.
The third and fifth columns show unreported column densities from \citet{2013MNRAS.430.1257R} (obtained through private communication).
}
\label{tab:table 2}
\tablehead{
\colhead{Species} &\colhead{CR}&\colhead{CR}&\colhead{TF}&\colhead{TF}\\
\colhead{} & \colhead{This work}& \colhead{2013}&\colhead{This work}&\colhead{2013}}
\startdata
H$_2$ &16.58 & 17.00&16.85 &16.93\\
H$_2^+$ &12.25& 12.34&11.95&12.02 \\
H$_3^+$ & 9.58& 9.93&9.96&9.90\\
CO & 12.44&  12.71&13.45&13.32\\
ArH$^+$ &10.85&-- & 11.66& --\\
HF & 10.42& --&10.64&--\\
HeH$^+$ & 13.15 & 13.21&12.92&12.98\\
OH & 13.26& 13.64&13.72&13.75\\
OH$^+$&12.64 &13.028&12.99&13.04\\
CH & 11.65& 11.85&11.32&11.41\\
CH$^+$ & 11.24& 11.63&11.70&11.74 \\
NeH$^+$ & 9.48&--&9.39&-- \\
NH$^+$ & 10.65& 11.01&11.15&11.13\\
NH &11.36& 11.72&11.99&11.96\\
CF$^+$ &8.16 &--&8.88&-- \\
CO$^+$ &9.76 &10.13&10.41&10.41\\
SiO$^+$&7.94&8.29&8.87&8.80\\
HS&9.66&9.93 &9.82&9.77\\
HS$^+$ &9.49&9.72&10.26&9.93\\
HCl$^+$&9.00&9.38&9.68&9.65\\
H$_2$O&9.31&9.79&10.33&10.31\\
H$_2$O$^+$&8.94&9.42&9.88&9.86\\
\enddata
\end{deluxetable*}

\subsection{ISM towards HD185418}
\label{subsec: model3}
Here we present a model of a very well studied Galactic ISM toward the star HD 185418. 
HD185418 is a B0.5 V star located at Galactic coordinates (l, b) = (53°, -2.2°) at a distance of 
790 pc from the sun \citep{2003ApJ...596..350S}. This line of sight has many observables \citep{2003ApJ...596..350S}. Earlier \citet{2006ApJ...639..941S} have modelled this line of sight using the C05.08 version of \textsc{cloudy}. 
\citet{2006ApJ...639..941S} considered a plane-parallel geometry with hydrogen density 27 cm$^{-3}$ extending up to a total hydrogen column density of 10$^{21.46}$ cm$^{-2}$. The radiation field is striking from both sides with a standard Galactic strength of 1.1 (in the units of 1.6$\times$10$^{-3}$ erg\,cm$^{-2}$\,s$^{-1}$). Their model could predict most of the ionic, atomic and molecular lines.

Using this update, we run the same model keeping the physical parameters the same as the 
final model of \citet{2006ApJ...639..941S}. We notice a 0.5 dex decrease in CO column density. The H$_2$(0, 2), H$_2$(0, 3), H$_2$(0, 4), and H$_2$(0, 5) column densities decreased by 0.1 to 0.2 dex.

\citet{2006ApJ...639..941S} did not include Polycyclic aromatic hydrocarbons (PAHs) in their final model. However, PAHs are ubiquitous in the ISM. PAHs provide heating and affect the free electron density that further affects column densities of other molecules
\citep{{2006PNAS..10312269D},{2017ApJ...845..163N},{2021ApJ...908..138S}}. \textsc{cloudy} has updated its PAH 
chemistry since C13.03. Hence, we rerun this model with PAHs. In nearby regions where PAH emission is spatially resolved it is found 
to be brightest in atomic regions and faint in molecular regions. Hence, in \textsc{cloudy} we have two options regarding how the PAH abundances vary across different layers of a cloud. We vary PAH abundance as \begin{math} n(PAH)\propto n(H^0)/n_{total} \end{math}  when it is assumed that the PAHs exist only in the atomic region. 
Likewise, we vary PAH abundance as \begin{math}n(PAH)\propto[n(H^0) +2n(H_2))]/n_{total}\end{math} when it is assumed that the PAHs exist both in the atomic and molecular regions.
Here n$_{total}$ is the total density of hydrogen in all forms. We find that the PAH abundance varying as \begin{math}n(PAH)\propto[n(H^0) +2n(H_2))]/n_{total}\end{math} predicts better match with the observed H$_2$(J) column densities. With an abundance of PAHs per hydrogen 10$^{-7.046}$ we reproduce the observed column densities of CO and all the 5 rotational levels of H$_2$(0,J). However, the model with PAHs predicts hydrogen density to be 29 cm$^{-3}$ and radiation field with 1.3 times the standard Galactic field. It is to be noted that \citet{2006ApJ...639..941S} could not reproduce the observed column densities of CH and CH$^+$. We are also unable to reproduce the observed column densities of CH and CH$^+$ using this version. Table \ref{tab:table 3} compares the predicted column densities of some molecules using this updated version and an earlier version C05.08 \citep{2006ApJ...639..941S}.

\begin{deluxetable*}{cccc}

\tablecaption{Comparison of the observed and our model predicted column densities (cm$^{-2}$ in log scale) along the line of sight towards HD 185418. Column 2 represents the observed column densities of various species. Whereas, columns 3 and 4 represent our model predictions using c0.5.08 and this version of \textsc{cloudy}. \label{tab:table 3}}
\tablehead{
\colhead{Chemical species} &\colhead{Column density}&\colhead{Column density}&\colhead{Column density}\\
\colhead{}  & \colhead{ c05.08 \citet{2006ApJ...639..941S}} & \colhead{Observed} &\colhead{Remodelled with this version}}
\startdata
HF &  --&--&12.67\\
CF$^+$ & --&--&10.65\\
HF$^+$  &--&--&6.36\\
ArH$^+$ & --&--&9.88\\
HCl &13.03 &--&12.85\\
HCN &--  &--&7.88\\
HNC  &-- &--&7.92\\
CN  &10.01 &$\leq$11.70&9.31\\
CH  &11.48 &13.11$\pm$0.05&11.03\\
CH$^+$  &9.62  &13.12$\pm$0.02&9.70\\
OH  &15.09&--&15.16\\
H$_3^+$& 13.19&--&14.08\\
HI & 21.24&21.11$\pm$0.15&21.22\\
Cl I & 14.38&14.52$\pm$0.16&14.41\\
Cl II  &13.15&$\leq$ 13.40& 13.31\\
C II  &17.26 &$\leq$17.75& 17.41\\
CI   &15.36 &15.53$\pm$0.09& 15.51\\
CI*   &14.52 &14.45$\pm$0.08& 14.66\\
OI   &18.28 &18.15$\pm$0.08& 18.23\\
CO  &14.82 &14.70$\pm$0.10& 14.66\\
H$_2$(0,0)& 20.40&20.30$\pm$0.10& 20.44\\
H$_2$(0,1)&20.35& 20.50$\pm$0.10& 20.53\\
H$_2$(0,2)&17.96&18.34$\pm$0.10& 18.44\\
H$_2$(0,3)&16.00&16.20$\pm$0.15& 16.27\\
H$_2$(0,4)&14.78&15.00$\pm$0.20& 14.82\\
H$_2$(0,5)&14.42&14.30$\pm$0.80& 14.57\\
\enddata
 
\end{deluxetable*}

\subsection{ISM towards W31C}
\label{subsec: model4}
Here we model a diffuse line of sight towards W31C (G10.6–0.4). It is, one of three bright H\,II regions within the W31 complex, 
located at a distance of 4.8$^{+0.4}_{-0.8}$ kpc \citep{2003ApJ...587..701F}. This line of sight has been extensively observed by many groups using $\it {Herschel}$. Many molecules, such as H$_2$, CH, HF, NH, NH$_2$, NH$_3$, HCl, and HCN have been detected. 
The observed column 
densities of N(H$_2$), N(CH), N(HF), N(NH), N(NH$_2$), N(HCl), N(CO), and N(HCN) are 1.7$\times$10$^{22}$,  5.89$\times$10$^{14 \pm {0.2}}$ \citep{2010A&A...521L..16G}, $\ge$ 1.6$\times$10$^{14}$ \citep{2010A&A...518L.108N}, 1.5$\times$10$^{14}$, 8$\times$10$^{13}$, (2.8 $ \pm $ 0.5)$\times$10$^{13}$ \citep{2013ApJ...767...81M}, and 1.5$\times$10$^{17}$ \citep{2016A&A...585A..80L}, (5.4 $ \pm $ 0.45)$\times$10$^{12}$ cm$^{-2}$ \citep{2010A&A...520A..20G}, respectively. 

To model this line of sight, we assume a plane-parallel geometry with radiation impinging from both sides. The considered gas-phase chemical abundances are an average
for the warm and cold phases of the ISM \citep{{1986ARA&A..24..499C}, {1996ARA&A..34..279S}}. In addition to this, we consider standard size-resolved ISM graphite and silicate grains (Rv = 3.1) with a MRN size distribution {\citep{1977ApJ...217..425M}. Furthermore, we consider a cosmic-ray ionization rate of 2$\times$10$^{-16}$ s$^{-1}$. First, we tried to model this line of sight with a constant density. However, we could not reproduce all the observed column densities with any particular value of density. Hence, we run a grid of models by varying the radiation field $\chi$ (in terms of the standard Galactic field, Habing field) \citep{2006ApJ...639..941S} and the total hydrogen density n(H) cm$^{-3}$. The hydrogen density is varied from 10$^1$ to 10$^3$ cm$^{-3}$ with a step size of 10$^{0.2}$. Whereas, $\chi$ is varied from 10$^0$ to 10$^2$ in a step-size of 10$^{0.5}$. All the models considered here extends up to the observed H$_2$ column density. 

In Fig.~\ref{fig:W31C} we have shown the simulated contour plots of column densities for the molecules HF, CH, NH, NH$_2$, HCN, HCl, 
and CO as a function of n(H) and $\chi$. We present contour plots of different
molecules in different panels of Fig.~\ref{fig:W31C} with different colors. Contour plots of NH, NH$_2$, HF, CH, HCl, HCN, and CO are represented by red, blue, 
green, pink, grey, brown, and black lines, respectively. Most of the contour lines are vertical which means that these column densities are not dependent of the incident radiation field in the given range. 
The colored-filled areas represent respective observed 
column density $\pm$ observed error bars. All the contour plots do not intersect at a single point which 
confirms that the observed species are coming from regions with different densities.
Our simulations reproduce the observed column densities for a range of densities, 10$^{1.6}$ to 10$^{3}$, for $\chi$ 1 to 10. 
HF and CH are produced in the low-density region (n(H) $<$ 10$^{1.7}$ cm$^{-3}$). Whereas, HCl is produced in the high-density region (n(H) $>$ 10$^{2.7}$ cm$^{-3}$). This might suggest that either density gradient or clumps exist along this sight-line. We also predict the unobserved column densities of 
N(ArH$^+$)$\leq$10$^{10}$, and 3$\times$10$^{10}$ $\leq$N(CF$^+$)$\leq$ 1.5$\times$10$^{11}$ cm$^{-2}$ over this range of density and radiation field.

\begin {figure}
\center
\includegraphics[scale=0.8]{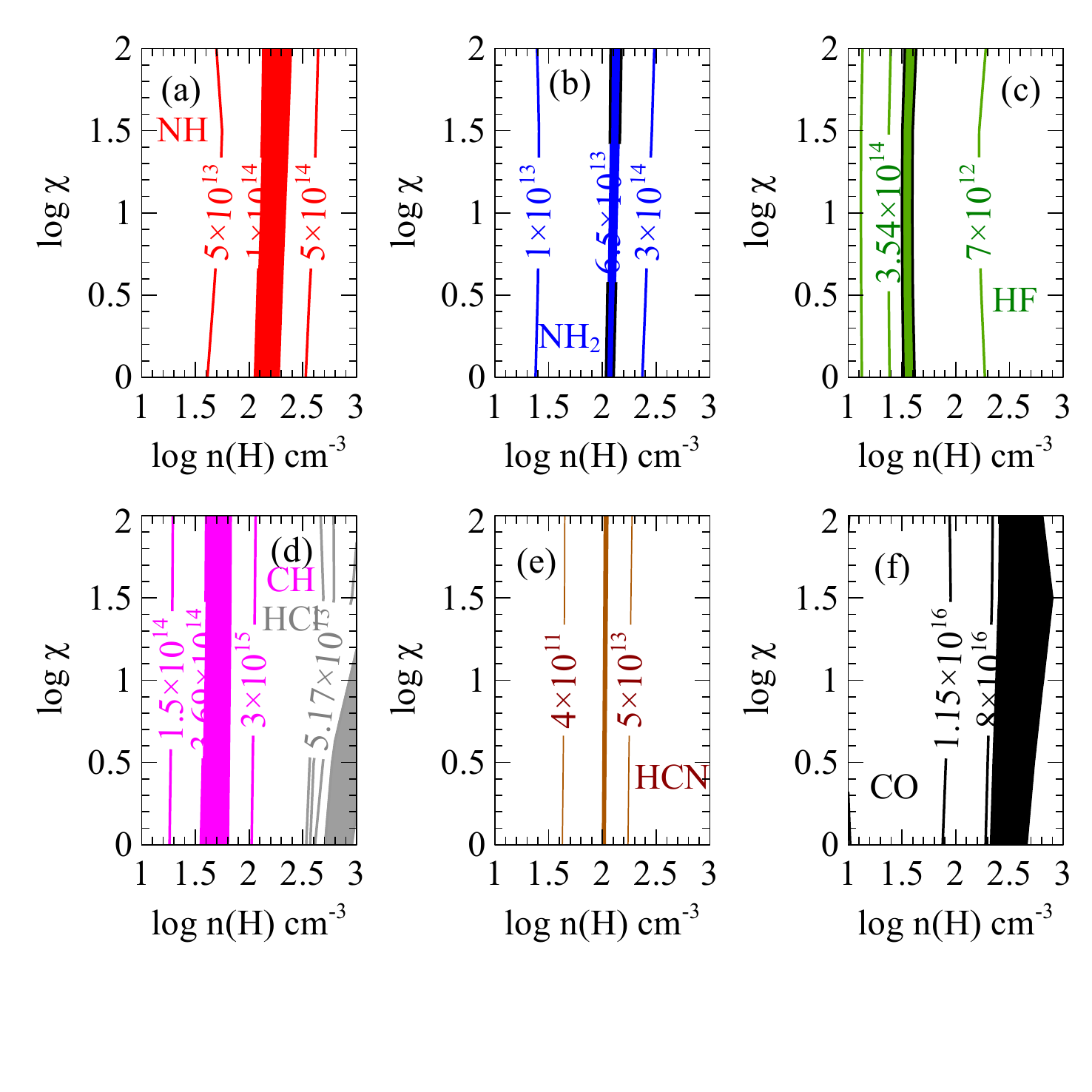}
\caption{Contour plots of NH, NH$_2$, HF, CH, HCl, HCN, and CO column densities are represented by red, blue, 
green, pink, grey, brown, and black lines, respectively. The colored-filled areas represent respective observed 
column density $\pm$ observed error bars. }
\label{fig:W31C}
\end {figure}

\subsection{ISM towards NGC 253}
\label{subsec: model5}
Recently, \citet{2021arXiv211215546J} have observed ArH$^+$, HF, OH$^+$, H I and, o--H$_2$O$^+$, p--H$_2$O$^+$ towards the 
nearby luminous starburst galaxy NGC 253 at a distance of 3.94 Mpc \citep{2003A&A...404...93K}. Here we try to model their observed column densities within the starburst galaxy NGC 253. We consider a plane-parallel slab with total hydrogen density n(H) cm$^{-3}$ illuminated by radiation field $\chi$ (in terms of the standard Galactic field, Habing field) \citep{2006ApJ...639..941S}. 
The extent of the cloud is fixed by the observed N(H I) column density. We vary the cosmic-ray ionization rate of hydrogen, 
$\chi_{CR}$ (s$^{-1}$), together with n(H) and $\chi$. The gas-phase abundances of a few important elements in terms of number are 
Ar/H = 10$^{-4.6}$, F/H = 10$^{-6.8}$, C/H = 10$^{-3.6}$, O/H = 10$^{-3.4}$. Abundances of the other elements are fixed at 
their standard ISM value. We consider standard size-resolved ISM graphite and silicate grains (Rv = 3.1) with a MRN size distribution \citep{1977ApJ...217..425M}. Moreover, we include PAHs (10$^{-6.125}$ number per hydrogen). }

Table \ref{tab:table 4} compares the predicted column densities, using this update, to the observed data \citep{2021arXiv211215546J}. Besides ArH$^+$, HF, OH$^+$, and H I, our model predicts observable amount of HeH$^+$ and CF$^+$. It is to be noted that \textsc{cloudy}
does not treat o--H$_2$O$^+$ and p--H$_2$O$^+$ separately. Hence we do not report the column densities of o--H$_2$O$^+$ and p--H$_2$O$^+$. Our final model predicts n(H) = 42.7 cm$^{-3}$, $\chi$ = 0.01, $\chi_{CR}$ = 1.$\times$10$^{-13}$ (s$^{-1}$) and molecular fraction $\it f_{H2}$= 0.013. Our model predicts Av = 0.526 {\textcolor{blue}{mag}}. \citet{2021arXiv211215546J} have derived n(H) = 35 cm$^{-3}$, $\chi_{CR}$ = 2.2$\times$ 10$^{-16}$ (s$^{-1}$). But the other assumptions like electron fraction in their model is not well constrained. However, \citet{2021ApJ...923...24H} have derived $\chi_{CR}$ $\geq$ 10$^{-14}$ (s$^{-1}$).

\begin{deluxetable*}{lll}

\tablecaption{Comparison of predicted column densities (cm$^{-2}$) along the line of sight towards NGC 253 
\label{tab:table 4}
}
\tablehead{
\colhead{Chemical species} &\colhead{Column density}&\colhead{Column density}\\
\colhead{} & \colhead{Observed} &\colhead{This work}}
\startdata
HF & (1.40$\pm$0.60)$\times$ 10$^{14}$ & 0.96$\times$ 10$^{13}$\\
ArH$^+$ &(3.35$\pm$0.31)$\times$ 10$^{12}$ & 2.34$\times$ 10$^{12}$\\
OH$^+$ & $>$1.57$\times$ 10$^{14}$ & 2.24$\times$ 10$^{14}$\\
HI & (4.57$\pm$3.80)$\times$ 10$^{20}$ & 8.32$\times$ 10$^{20}$ \\
H$_2$ &-- & 5.5$\times$ 10$^{18}$\\
HeH$^+$& --& 1.48$\times$ 10$^{13}$\\
CF$^+$& --& 1.02$\times$ 10$^{11}$\\
\enddata

\end{deluxetable*}

\section {Discussion and Summary} \label{sec:summary} 
In this work, we have updated gas-phase chemical reaction rate {\linenumberfont{}} and molecular data (including energy levels, Einstien's coefficients, and collisional rate coefficients) for HF, CF$^+$, HC$_3$N,  ArH$^+$, HCl, HCN, CN, CH and CH$_2$ in the spectral synthesis code \textsc{cloudy}. We have added 561 new reactions and have updated 165 molecular reaction rates involving these nine molecules. We have updated the H$_2$--H$_2$ collisional rate coefficients as well.
Our main conclusions from this work are listed below:

\begin{enumerate}
\item The earlier version, c17 \citep{2017RMxAA..53..385F}, was used to predict column densities of 92 molecules. \textsc{cloudy} now predicts column densities of 100 molecules. As a result of this update, roughly 650 new molecular lines are added to the predicted spectrum.

\item
This update produces $\geq$ 90$\%$ relative change to some molecular column densities and line intensities. We have documented these changes. 

\item
Our model for Knot 51 of the Crab nebula predicts ArH$^+$ within the range observed by \citet{2013MNRAS.430.1257R}. We show how its column density is effected by changing the cosmic-ray ionization rate. We also predict the detectable amount of HeH$^+$, OH$^+$, CH$^+$, NH$^+$, and CO$^+$ along this sight-line. These molecular ions provide new diagnostics of the importance of cosmic-rays or energetic particles entering a molecular environment.

\item
We reproduce observed column densities \citep{2003ApJ...596..350S} of various species including the rotational levels of H$_2$ 
for ISM towards HD185418. Our model suggests the presence of PAHs with an abundance of PAHs per hydrogen 10$^{-7.046}$ along this line of sight. It also favours PAHs 
abundance varying as \begin{math}n(PAH)\propto[n(H^0) +2n(H_2))]/n_{total}\end{math}.

\item
We model many molecules, such as H$_2$, CH, HF, NH, NH$_2$, HCl, CO, and HCN for a diffuse 
line of sight towards W31C (G10.6–0.4). Our models suggest that either a density gradient or clumps exist along this sight-line. The density varies across a range, 10$^{1.6}$ to 10$^{3}$ cm$^{-3}$.

\item
We reproduce the observed column densities \citep{2021arXiv211215546J} of HF, OH$^+$, H I for the ISM towards the starburst 
galaxy NGC 253. Our predicted ArH$^+$ column density is within the same order of magnitude. We find $\it n$(H) = 42.7 cm$^{-3}$, $\chi$ = 0.01, $\chi_{CR}$ = 1.$\times$ 10$^{-13}$ (s$^{-1}$) and molecular fraction $\it f_{H2}$= 0.013 towards this sight-line. 
\end{enumerate} 

\section {Future development} \label{sec:future}
In future, we plan to predict more molecular lines. For that purpose, we will first update existing gas-phase reaction rates for other chemical reactions and include available energy levels, radiative and collisional rates. 

Next, we will expand our chemical reaction network further by incorporating chemical reactions mediated on dust grains. This is important as complex molecules  interact on grain surfaces
and form more complex biological molecules in a dense medium \citep{2009ARA&A..47..427H}. 

Predicting more spectral lines is very challenging. Many molecules lack collisional data. Hence, we will add a g-bar approximated collisional rate coefficients to molecular data provided by CDMS \footnote{\url{https://cdms.astro.uni-koeln.de/}}and JPL\footnote{\url{https://spec.jpl.nasa.gov/}}. In appendix A.7, we have demonstrated that the g-bar approximation worked very well for H$_2$ \citep{{2005ApJ...624..794S}, {2018ApJ...862..132W}}, and the predictions do not differ much. This motivates us to apply a g-bar approximation that will make it possible to model a large number of spectral lines. We will replace the g-bar approximated data with accurate theoretical or experimental data, when they become available.

\acknowledgments
We thank the reviewers for their constructive comments and suggestions. We thank P.C. Stancil and Ziwei E. Zhang for providing the H$_2$--H$_2$ collisional data.
We also thank \citet{2013MNRAS.430.1257R} for providing their unreported molecular column densities.
GS acknowledges WOS-A grant from Department of Science and Technology (SR/WOS-A/PM-2/2021). 
GJF acknowledges support by NSF (1816537, 1910687), NASA (ATP 17-ATP17-0141, 19-ATP19-0188), and STScI (HST-AR- 15018).
MC acknowledges support by STScI (HST-AR14556.001-A), NSF (1910687), and NASA (19-ATP19-0188).

\appendix
\section{Appendix (A)}
Here we briefly describe the added and updated molecules for this work. We also mention reactions that produce a relative change in the column density of more than  $\pm$ 90$\%$ for some models compared to the previous 
version. For this work, the reference models are the bench mark models from the Lorentz Center PDR workshop \citep{2007A&A...467..187R} and a few other PDR models. Some of the models are constant temperature model and in the other models the temperature is derived self-consistently from heating ad cooling balance. To avoid repetition, we only show a few plots comparing the old and updated rates. 
\subsection{CN}

CN is an important ubiquitous molecule in the ISM \citep{1989ApJ...340..273V}.  
The internal structure including energy levels, Einstein's A coefficient, and
collisional rate coefficients from the LAMDA database have been part of the \textsc{cloudy} chemical network 
since version c96. In the old LAMDA data,  
energy levels are not hyperfine split (total number of energy levels = 41) and collisional rate coefficients with electron and He are present. 
However, the newly updated LAMDA data for CN (10sep2018) has hyperfine energy levels (total number of energy levels = 73) with collisional rate coefficients with H$_2$, but it lacks electron-collisional rate coefficients. 

To quantify the effects of the hyperfine levels and different collisional partners, we ran the Orion bar model. Orion bar is an interface between ionized and molecular gas which is excited by the radiation from the Trapezium cluster. It is one of the nearest and best-studied 
photodissociation or photon-dominated regions(PDRs). This is viewed nearly edge-on, and hence provides a testing ground for the transition from
H$^+$ to H$^0$ to H$_2$ regions. Earlier \citet{2009ApJ...693..285P} have modelled this line of sight in detail using \textsc{cloudy} and reproduced several multiwavelength observations \citep{{2009ApJ...693..285P},{2009ApJ...701..677S}}. We use the same input parameters from \citet{2009ApJ...693..285P}, and run the Orion bar PDR model (the input script of this model, ``h2$_{-}$orion$_{-}$bar.in'', is publicly available with the \textsc{cloudy} download under the directory \texttt{tsuite}) with these two data sets, separately. As expected, we notice more CN lines 
with the updated LAMDA data as it has more energy levels.
Although the total CN column density does not change within a limit of 0.001 dex, individual CN 
line intensities do change. As an example, with the old data the luminosity of 880.858 $\mu$m line 
((transition from quantum numbers (N=3, J=2.5) to (N=2, J=1.5)) is 10$^{32.907}$ erg~s$^{-1}$, while with the updated data, with hyperfine structure splitting, there are six closely spaced lines at 880.817 $\mu$m, 
880.825 $\mu$m, 880.856 $\mu$m, 880.864 $\mu$m, 880.864 $\mu$m, 881.407 $\mu$m  with 
luminosities 10$^{31.982}$, 10$^{31.856}$, 10$^{32.104}$, 10$^{32.459}$, 10$^{32.296}$, 10$^{32.203}$ erg s$^{-1}$, respectively. 
The total intensity of the resolved lines (10$^{32.973}$ erg~s$^{-1}$) is close to that of the unresolved line. This implies that the H$_2$ collisions are just as important
as the e$^-$ and He collisions.
\textsc{cloudy} use the CN data with no hyperfine energy levels with
electron collisions as default. However, the option to use hyperfine levels and H$_2$ collisional data is still available. 
We advocate for using hyperfine level data with H$_2$, 
as well as e$^-$ and He collisions. For that purpose we urge theoreticians/experimentalists to provide hyperfine level data with e$^-$ and He collisions.

\begin{figure}
\center
\includegraphics[scale=0.8]{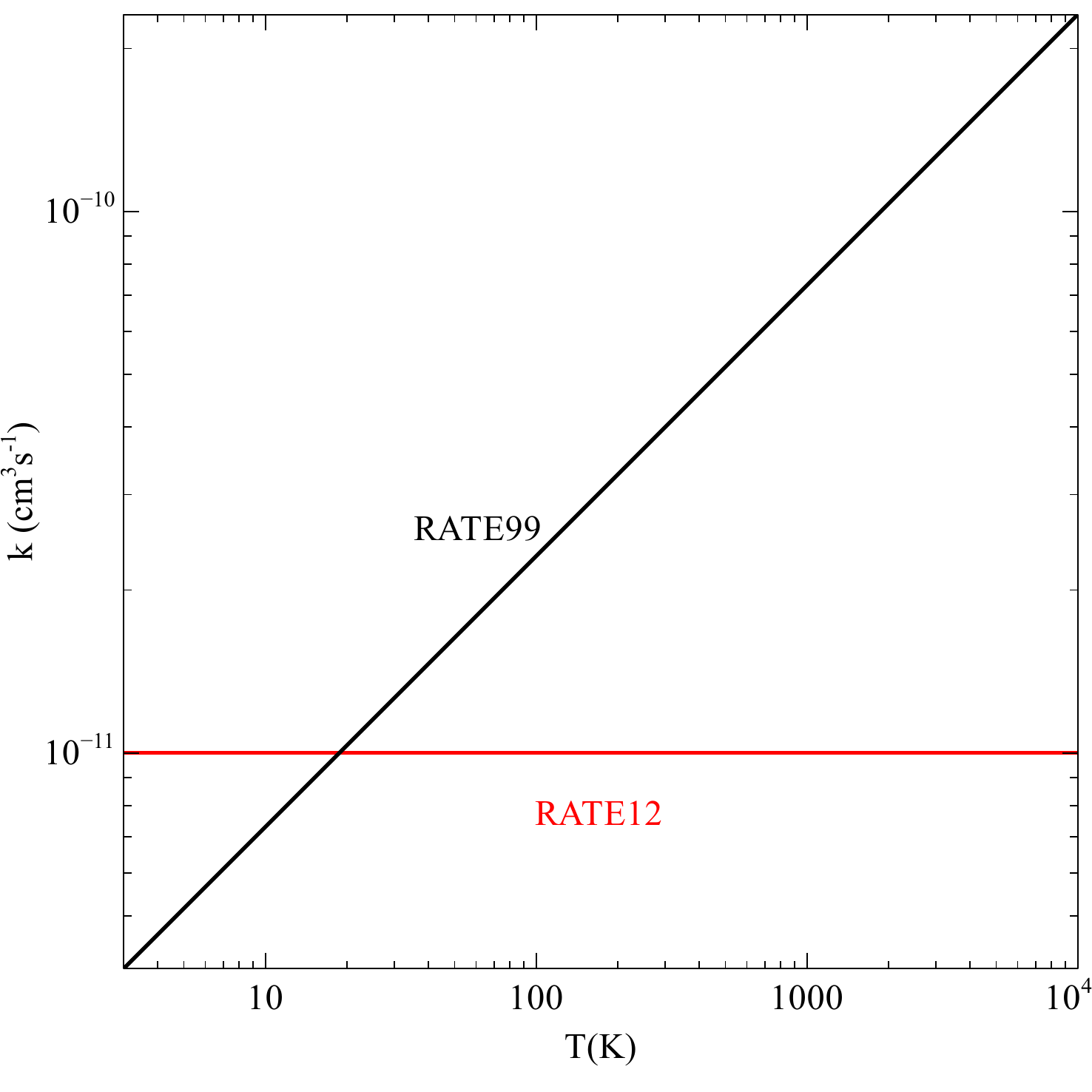}
\caption{Rate coefficients (RATE99 and RATE12) as a function of temperature for the reaction C + OCN $\rightarrow$ CO + CN.}
\label{fig:OCN_8}
\end {figure}

In addition, we have included new reactions for better prediction of CN abundances, listed in the Appendix B.
Some of these reactions produce relative changes $\geq$ 90$\%$ compared to earlier versions of \textsc{cloudy}. 
Below we discuss them one by one.
\begin{equation}
    \rm C + OCN \rightarrow CO + CN    
\end{equation}
The earlier rate coefficient, 4$\times$10$^{-11} \, \left(T /300\right)^{0.5}$ ($\rm{cm^3 \, s^{-1}}$),
was from RATE99.
We now use the updated rate coefficient from RATE12, which is a constant,
1$\times$10$^{-11}$ $\rm{cm^3 \, s^{-1}}$.
Fig.~\ref{fig:OCN_8} compares these rates as a function of temperature.
It is clear that RATE99 is higher than RATE12 for temperatures $\gtrsim 20$~K.

\begin {figure}
\center
\includegraphics[scale=0.8]{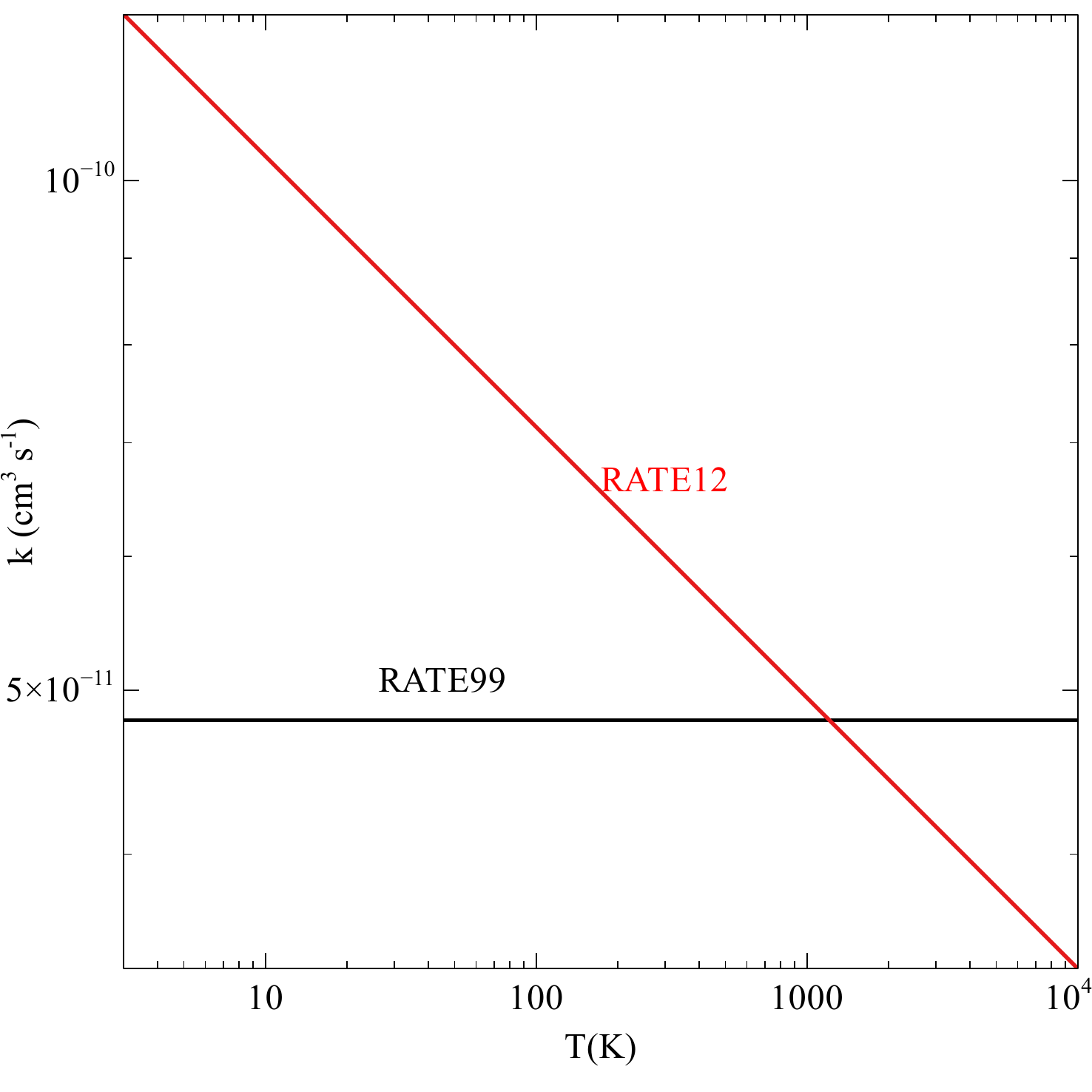}
\caption{Rate coefficient as a function of temperature for the reaction C + NO $\rightarrow$ CN + O.}
\label{fig:CN_O}
\end {figure}

\begin{equation}
    \rm H + HCN \rightarrow CN + H_2   
\end{equation}
This reaction was not included in the earlier versions of \textsc{cloudy}, but it is now.
The rate coefficient for this reaction is the same for all of RATE99, RATE06, and
RATE12, namely, 6.2$\times$$10^{-10} \, \exp(-12500/T)$$\rm{cm^3 \, s^{-1}}$.

\begin{equation}
    \rm C + NO \rightarrow CN + O    
\end{equation}
The earlier rate coefficient was from RATE99, a constant 4.8$\times$10$^{-11}$~$\rm{cm^3 \, s^{-1}}$.
However, in RATE12 it is
6$\times$$10^{-11} \, \left(T / 300\right)^{-0.16}$~$\rm{cm^3 \, s^{-1}}$.
\textsc{cloudy} now employs the RATE12 data.
Fig.~\ref{fig:CN_O} compares the two rate coefficients as a function of temperature.
With the updated rate coefficient, the reaction is more important than before for temperatures $\lesssim 1000$~K.

\begin{equation}
    \rm C + NS \rightarrow S + CN
\end{equation}

The earlier rate coefficient was from RATE99, 2$\times$$10^{-11} \, \left( T / 300 \right)^{0.5}$ ($\rm{cm^3 \, s^{-1}}$). 
We update this to RATE12, 1.5$\times$$10^{-10} \, \left( T / 300 \right)^{-0.16}$ ($\rm{cm^3 \, s^{-1}}$). 
This leads to a relative change of around 95$\%$ in NS column density.

\subsection{CH}

CH traces low-density gas and is also a good tracer of shocks \citep{1998MNRAS.297.1182F}. Strong CH lines can be found in diffuse clouds.
CH chemistry has been a part of \textsc{cloudy} since version c96, although we did not include a model for the internal structure of this molecule.
We now include its energy levels, Einstein's coefficients,
and collisional rate coefficients with different partners (H and H$_2$) \citep{{2018MNRAS.475.5480D},{2015PCCP...1721583M}} from the
LAMDA database (18jul2019, total number of energy levels = 32). 
As a result, \textsc{cloudy} now predicts CH spectral lines.
At the same time, we update the old chemical rate coefficients to the more recent RATE12 data
and add some new reaction rate coefficients.
Here, we mention reactions that produce relative change in the column density by more than 90$\%$ in some models compared to the previous version.

\begin{equation}
    \rm CH + SO \rightarrow OCS + H    
\end{equation}
The reaction rate coefficient is newly added in RATE12 and inclusion of this constant rate
(1.1$\times$$10^{-10}$~$\rm{cm^3 \, s^{-1}}$) leads to a relative change of nearly 99$\%$ to the column density of OCS.
We adopt this rate.

\begin{equation}
    \rm CH + NO \rightarrow HCN + O   
\end{equation}
The earlier rate coefficient was from RATE99, 5.59$\times$$10^{-9} \, \exp(-10814/T)$~$\rm{cm^3 \, s^{-1}}$.
This neutral-neutral reaction involves an active radical CH.
Hence, one can expect the activation energy to be close to zero.
The recent RATE12 is temperature-dependent,
1.2$\times$$10^{-10} \left(T/300\right)^{-0.13}$~$\rm{cm^3 \, s^{-1}}$,
without any activation barrier.
We adopt the RATE12 data, even though it leads to
a relative change of nearly 150$\%$ in the column density of HCN.

\begin{equation}
    \rm CH + NO \rightarrow OCN + H   
\end{equation}
Following the same logic as the above reaction we adopt the current RATE12,
3.49$\times$$10^{-11} \, \left(T/300\right)^{-0.13}$~$\rm{cm^3 \, s^{-1}}$.

\begin{equation}
    \rm CH + S \rightarrow CS + H   
\end{equation}
The earlier rate coefficient was adopted from RATE99, 1.1$\times$10$^{-12}$ ($\rm{cm^3 \, s^{-1}}$). The recent RATE12 is nearly 50 times higher, 
5$\times$10$^{-11}$ ($\rm{cm^3 \, s^{-1}}$). This is a rapid neutral-neutral reaction \citep{2004MNRAS.350..323S} and produces a relative change 
of 100$\%$ in CS column density. We adopt the RATE12 rate.

\begin{equation}
    \rm C^+ + CH \rightarrow C_2^+ + H    
\end{equation}
The old rate coefficient was a constant 3.8$\times$10$^{-10}$ ($\rm{cm^3 \, s^{-1}}$) from RATE06.
The newly adopted rate is 3.89$\times$10$^{-10} \, \left( T/300 \right)^{-0.5}$ ($\rm{cm^3 \, s^{-1}}$) from RATE12.

\begin{equation}
    \rm CH + S^+ \rightarrow CS^+ + H    
\end{equation}
The old rate coefficient was a constant 6.2$\times$10$^{-10}$ ($\rm{cm^3 \, s^{-1}}$) from RATE06.
The new adopted rate coefficient is 6.2$\times$10$^{-10} \, \left( T/300 \right)^{-0.5}$ ($\rm{cm^3 \, s^{-1}}$) from RATE12.

\subsection{\texorpdfstring{CH$_2$}{CH2}}

CH$_2$ is an important chemical species which is a precursor to many other molecules. However, it is difficult to detect in the ISM \citep{1996A&A...307..645V}. CH$_2$ chemistry has been a part of \textsc{cloudy} since version c96.
We now include energy levels, Einstein's coefficients,
and collisional rate coefficients with different partners (ortho-H$_2$ and para-h$_2$) from the LAMDA database
(18oct2019, total number of energy levels = 24).
As a result, numerous lines of CH$_2$ are now predicted.
At the same time, we update the old chemical rate coefficients and add some new reaction rate coefficients from the recent version of RATE12. 
Here, we mention reactions that produce more than 90$\%$ relative change in the column density in some models compared to the previous version.
\begin{equation}
    \rm CH_2 + S \rightarrow CS + H_2    
\end{equation}   
This reaction was not included in \textsc{cloudy} by c17.
The RATE12 provides a constant rate coefficient, 1.$\times$10$^{-10}$ ($\rm{cm^3 \, s^{-1}}$),
which produces a relative change of nearly 94$\%$ to the CS column density.

\begin{equation}
    \rm N + CH_2 \rightarrow HCN + H    
\end{equation} 
This reaction was not a part of \textsc{cloudy} by c17.
We incorporate the rate coefficient from RATE12,
3.95$\times$10$^{-11} \, \left( T/300 \right)^{-0.17}$ ($\rm{cm^3 \, s^{-1}}$).
This produces a relative change of nearly 103$\%$ to the HCN column density.

\subsection{Fluorine chemistry and \texorpdfstring{CF$^+$}{CF+} }

Fluorine is the most reactive neutral species observed in the diffuse ISM \citep{2007ApJ...655..285S}.
It exothermally reacts with H$_2$ to form HF \citep{2019ApJ...872..203D}. 
HF further reacts with C$^+$ to produce CF$^+$.
CF$^+$ is the second most abundant fluorine molecule after HF.
CF$^+$ has been observed both in extragalactic \citep{2016A&A...589L...5M},
and Galactic environments \citep{{2006A&A...454L..37N},{2015A&A...579A..12L}}.
\textsc{cloudy} \citep{2017RMxAA..53..385F} has been predicting
all the ionization states of atomic F since version c90, but Fluoride molecules
have not been included.
Here, we extend our chemical network of F by incorporating chemical reactions from the
UDfA (RATE12) for the formation and destruction of HF and CF$^+$.
The detailed energy levels and corresponding transition rate coefficients 
of HF (19may2014, total energy levels = 16),
and CF$^+$ (08jul2016, total energy levels = 11)
are from the LAMDA database. 
As a result of these updates, \textsc{cloudy} now predicts column densities of HF$^+$, HF, H$_2$F$^+$, and CF$^+$. 
It also predicts numerous lines of HF and CF$^+$.

\subsection{\texorpdfstring{HC${_3}$N}{HC3N}}

HC$_3$N plays an important role as a density tracer in dense clouds.
It is observed in Galactic star-forming regions \citep{2016ApJ...830..106T},
proto-planetary disks \citep{2012ApJ...756...58C},
and extragalactic environments \citep{2011A&A...527A.150L}.
Because of its importance, we incorporate the HC$_3$N molecule in \textsc{cloudy}.
The energy levels, Einstein's A coefficient, and collisional rate coefficients for HC$_3$N are from 
the LAMDA database (21dec2018, total energy levels = 61).
The energy levels include hyperfine levels.
The formation and destruction rates of HC$_3$N are from UDfA (RATE12).
As a result, CLOUDY now predicts column density and line intensities of HC$_3$N.

\subsection{\texorpdfstring{ArH$^+$}{ArH+}}

ArH$^+$ is ubiquitous in the diffuse ISM.
It is a noble gas compound, first detected in the Crab nebula
\citep{2013Sci...342.1343B}. UDfA (RATE12) does not have any reaction rate coefficients involving ArH$^+$.
However, \textsc{cloudy} already contains some chemical reactions involving
ArH$^+$ \citep{1993JPCRD..22.1469A} and predicts column densities of ArH$^+$ since version c17. 
With the inclusion of detailed energy levels, Einstein's A coefficients, and collisional rate coefficients
from the LAMDA database (29sep2020, total number of energy levels = 12), CLOUDY also predicts 
ArH$^+$ line intensities.
We have added a few new reactions and updated some existing rate coefficients
\citep{{2014A&A...566A..30R}, {2014A&A...566A..29S}, {2017MNRAS.472.4444P}}.
 
\subsection{\texorpdfstring{H$_2$}{H2}}
 The detailed H$_2$ microphysics implemented in \textsc{cloudy} is described in \citet{2005ApJ...624..794S}, \citet{2012ApJ...746...78G}, and \citet{2020RNAAS...4...78S}. 
Prior to this work, we supplemented missing H$_2$  collisional data by the g-bar approximation following \citet{1962ApJ...136..906V}. The g-bar approximation is a semi-empirical formula that is widely used when theoretical or experimental data is unavailable. This applies to dipole-allowed transitions.
In this work, we update our H$_2$--H$_2$ collisional rate coefficients to \citet{2021ApJ...912..116Z} and \citet{2018ApJ...862..132W}.
The earlier version of \textsc{cloudy} had de-excitation
rate coefficients from $v$=0, $J$=1-8 \citep{2008ApJ...689.1105L}.
Recently, \citet{2021ApJ...912..116Z} have provided updated collisional
rate coefficients with ortho-H$_2$ and para-H$_2$ for $v$=0, $J$=0-31.
So hereafter, the g-bar approximated data for H$_2$--H$_2$ collisions are replaced
with these novel data.
To check the effects of this change, we ran the Orion bar model 
\citep{{2009ApJ...693..285P},{2009ApJ...701..677S}} using 
\citet{2008ApJ...689.1105L} and \citet{2018ApJ...862..132W} collisional data, separately. 
The changes in H$_2$ line intensities are
less than 0.3$\%$.
This shows that, in this case, the g-bar approximation was surprisingly accurate.

\subsection{HCl} 
HCl can be found in both diffuse and dense ISM regions.
It is believed to be the dominant reservoir of chlorine in the dense clouds \citep{1974ApJ...192L..37D}. 
HCl has been a part of \textsc{cloudy} since version c96.
We upgrade the chemical network of HCl by adding a few new rate coefficients
and updating some of the existing ones.
The updated LAMDA data set has more energy levels
(07jun2021, total number of energy levels = 40)
than the existing 
(till c17) energy levels (total number of energy levels = 28) in \textsc{cloudy}.
The new data use ortho-H$_2$ and para-H$_2$ collisional rate coefficients
separately. However, the older data set uses 
just H$_2$ collisional data, scaled from HCl-He collisional data.
Like the previous case, we ran the Orion bar model \citep{{2009ApJ...693..285P},{2009ApJ...701..677S}} 
with the two HCl data sets.
We noticed that many HCl line luminosities changed by a factor of 10,
even though the total HCl column density remained the same.
From now on, we use the updated HCl data set.

\subsection{HCN and HNC} 
HCN is a good tracer of density in molecular clouds and galaxies. 
The updated LAMDA database uses HCN data following 
\citet{2017MNRAS.468.1084H} (20dec2018). However, the new  model has fewer energy levels 
(a total of 25) than the LAMDA model used in older versions of \textsc{cloudy} 
(total energy levels = 32, with energies obtained from CDMS, JPL).  
The updated LAMDA data includes electron collisions in addition to H$_2$ collisions. 
We continue to use the old LAMDA model as our default since it has more energy levels. 
However, we also
provide the new LAMDA model with electron collisions and updated energy levels. 

HNC is a structural isomer, a tautomer, of HCN. The HNC/HCN line ratio is used to 
distinguish between photodissociation 
regions and X-ray-dissociation regions \citep{{2005A&A...436..397M}, {2010MNRAS.404..518S}}. Furthermore, 
\citet{2020A&A...635A...4H} have suggested the intensity ratio (I(HCN)/I(HNC)) of their J = 1–0 lines as a gas kinetic temperature tracer for the molecular ISM. It is valid for 15 K$\leq$ T$_{K}\leq$40 K, and the intensity ratio
increases for higher temperatures. \citet{2020A&A...635A...4H} found 
that the observed intensities are reproduced with an abundance ratio of HCN/HNC = 1-20. So, even though there is no update for energy levels 
and transition rates in the LAMDA for HNC after 2015, we update the 
chemical reaction rate coefficients from RATE12 to be consistent and predict HNC/HCN ratio accurately. 
Here, we discuss the reaction rates whose inclusion results in
more than 90$\%$ relative change in predicted column density and line intensities compared to the
previous versions of \textsc{cloudy}.
\begin{equation}
    \rm Na + HCNH^+ \rightarrow HCN + Na^+ + H.  
\end{equation}
The RATE12 provides a constant rate coefficient
1.35$\times$10$^{-9}$ ($\rm{cm^3 \, s^{-1}}$) over a temperature range 10 to 41000 K.
Inclusion of this rate coefficient increases the Na$^+$ column density by 1.39 dex for a model at constant 
temperate 1000K (\texttt{coll\_t3.in}).
In general, for this type of ion-neutral reaction, the rate coefficients are 
temperature-independent.
We incorporate this rate coefficient even though it produces a relative change of nearly 100$\%$ in HCN column density at high temperature.

\begin{equation}
    \rm N + CH_2 \rightarrow HCN + H
\end{equation}
This is a neutral-neutral reaction involving a radical atom and an unsaturated molecule.
Hence, one expects the activation energy to be close to zero.
The form (RATE12), 3.95$\times$10$^{-11} \, \left( T/300 \right)^{0.17}$ ($\rm{cm^3 \, s^{-1}}$),
is well behaved over a wide range of temperature.
So we do include it in \textsc{cloudy} even though the inclusion of this rate
changes HCN column density of nearly 100 $\%$ compared to the previous versions of \textsc{cloudy}.

\begin{equation}
    \rm CH + NO \rightarrow HCN + O   
\end{equation}
The old rate coefficient was adopted from UDfA RATE99,
5.59$\times$10$^{-9} \, \exp(-10814./T)$ ($\rm{cm^3 \, s^{-1}}$), 
which is very close to zero at low temperature because of the exponential term.
The updated rate, RATE12, is 2.$\times$10$^{-10} \left( T /300 \right)^{-0.13}$ ($\rm{cm^3 \, s^{-1}}$).
This reaction is a neutral-neutral reaction between a radical and a molecule with an unpaired electron.
The rate coefficient is expected to increase as temperature is lowered \citep{2011IAUS..280..361S}.
Even though the RATE12 rate coefficient produces a relative change of nearly 103 $\%$ for HCN column density,
we update it as it is a well behaved rate.

\begin{equation}
    \rm H + HNC \rightarrow HCN + H    
\end{equation}
Inclusion of this rate coefficient introduces instabilities in \textsc{cloudy}'s chemistry solver.
The rate coefficient, 1.14$\times$10$^{-13} \left( T/300 \right)^{4.23} \, \exp(114.6/T)$ ($\rm{cm^3 \, s^{-1}}$),
looks unrealistic due to the huge positive exponential term. 
We do not include this and suggest a further investigation of this rate coefficient.
We searched the literature for this rate coefficient and found that
\citet{2012A&A...541A..21H} estimated the rate coefficient to be
4.$\times$10$^{-11} \, \exp(-1200/T)$ ($\rm{cm^3 \, s^{-1}}$).
We update to the rate of \citet{2012A&A...541A..21H}.
\begin{equation}
    \rm HCN + HCO^+ \rightarrow HCNH^+ + CO
    \label{eqn:HNC_HCO+}
\end{equation} 
The old rate coefficient was adopted from RATE06,
3.1$\times$10$^{-9}$ ($\rm{cm^3 \, s^{-1}}$),  which is constant over temperature.
The updated rate coefficient in RATE12 is 
3.1$\times$10$^{-9} \left( T/300 \right)^{-0.5}$ ($\rm{cm^3 \, s^{-1}}$).
The RATE12 produces a relative change of nearly 90$\%$ for HCNH$^+$ column density for HII-like
environments as the rate increases below 300K and decreases above 300K.
Generally, ion-neutral reactions are temperature-independent as there is
no activation barrier involved.
However, if the neutral possesses a dipole moment the rate coefficient
becomes temperature-dependent and higher at a lower temperature \citep{2011IAUS..280..361S}.
Hence, we update our rate coefficient from RATE06 to RATE12.

\begin{equation}
    \rm HNC + HCO^+ \rightarrow HCNH^+ + CO   
\end{equation} 
The same argument as for the earlier reaction (\ref{eqn:HNC_HCO+}) applies here.
Hence, we update our rate coefficient from RATE06 to RATE12.

\begin{equation}
    \rm N + CH_2 \rightarrow HNC + H    
\end{equation} 
The old reaction rate coefficient was taken from UDfA RATE99.
The recent UDfA rate coefficient, RATE12, is 0.5 times the old rate coefficient and
causes a relative change of 98 $\%$ to the HNC column density.
We adopt the RATE12 rate coefficient.

\section{Appendix (B)}
Here we list our added and modified rate coefficients for this work.
Some reactions involve more than one species discussed above. 
To avoid double-counting, we list them only once. The rate coefficients for the two-body reactions are derived using equation 1, 
and the photoreaction rate coefficients are derived using equation 2. 
It is to be noted that we do not report the dead-end molecules that are produced by sink reactions.

\begin{deluxetable*}{lccc}
\tablecaption{List of added chemical reactions involving CF$^+$ \label{tab:5}}
\tablehead{
\colhead{Reactions}&\colhead{$\alpha$}&\colhead{$\beta$}&\colhead{$\gamma$}
}
\startdata
H$_2$ + F$^+$ $\rightarrow$ HF$^+$ + H & 3.8e-10 & 0 & 0 \\
H$_2$ + HF$^+$ $\rightarrow$ H$_2$F$^+$ + H &1.3e-9 & 0 &0 \\
H$_2$ + HF$^+$ $\rightarrow$ H$_2$F$^+$ + H &1.3e-9 & 0 & 0 \\
H$_3^+$ + HF  $\rightarrow$ H$_2$F$^+$ + H$_2$ &1.2e-8 &-0.5 & 0 \\
H$_2$F$^+$ + e$^-$ $\rightarrow$ H$_2$ + F & 3.5e-7& -0.5 & 0 \\
H$_2$F$^+$ + e$^-$ $\rightarrow$ HF + H & 3.5e-7 &-0.5 & 0 \\
H$_2$ + F $\rightarrow$ HF + H &1.e-10 & 0 & 400. \\
H$_2$O + F $\rightarrow$ HF + OH & 1.4e-11 & 0 & 0\\
OH + F $\rightarrow$ HF + O &1.6e-10 & 0 & 0 \\
C$^+$ + HF $\rightarrow$ CF$^+$ + H &7.2e-9 &-0.5 & 0 \\
CF$^+$ + e$^-$ $\rightarrow$ C + F & 5.2e-8 & -0.80 & 0 \\
HF + PHOTON $\rightarrow$ H + F & 1.17e-10 & 0 & 2. \\
\enddata
\tablecomments{rate coefficients are from RATE12 }
\end{deluxetable*}
\vspace*{0.2in}

\startlongtable
\begin{deluxetable*}{lccc}
\tablecaption{List of added chemical reactions involving HC$_3$N \label{tab:6}}
\tablehead{
\colhead{Reactions}&\colhead{$\alpha$}&\colhead{$\beta$}&\colhead{$\gamma$}}
\startdata
C$_2$H + CN  $\rightarrow$ HC$_3$N + PHOTON &1.e-16&0&0 \\
N + CH$_2$CCH  $\rightarrow$ HC$_3$N + H$_2$ &8.e-11&0&0\\
N + C$_3$H$_2$ $\rightarrow$ HC$_3$N + H & 1.e-13&0&0  \\
HNC$_3$ + H $\rightarrow$ HC$_3$N + H &1.e-11&0&0  \\
CN + C$_2$H$_2$ $\rightarrow$ HC$_3$N + H &2.7e-10&-0.52&19. \\
C + CH$_2$CN $\rightarrow$ HC$_3$N + H  &1.e-10&0&0 \\
C$_2$H + NCCN $\rightarrow$ HC$_3$N + CN &2.e-13&0&0  \\
C$_2$H + HNC $\rightarrow$ HC$_3$N + H &2.5e-10&-0.2&0 \\
C$_2$H + HCN $\rightarrow$ HC$_3$N + H &5.3e-12&0&770.  \\
NH$_3$ + HC$_3$NH$^+$ $\rightarrow$ NH$_4^+$ + HC$_3$N &2.e-9&-0.5&0 \\
HNC$_3$ + H$^+$ $\rightarrow$ HC$_3$N + H$^+$ &3.23e-9&-0.5&0 \\
H$_2$O + C$_4N^+$ $\rightarrow$ HC$_3$N + HCO$^+$ &7.5e-10&-0.5&0 \\
H$^+$ + CH$_3$C$_3$N $\rightarrow$ HC$_3$N + CH$_3^+$ &8.e-9&0&0 \\
CH$_4$ + C$_4$N$^+$ $\rightarrow$ C$_3$H$_3^+$ + HC$_3$N &1.43e-10 &0&0 \\
CH$_3$OH + HC$_3$NH$^+$ $\rightarrow$ CH$_3$OH$_2^+$ + HC$_3$N &1.9e-9&-0.5&0 \\
CH$_3$CN + HC$_3$NH$^+$ $\rightarrow$ CH$_3$CNH$^+$ + HC$_3$N & 3.6e-9&-0.5&0\\
HC$_5$NH$^+$ + e$^-$ $\rightarrow$ HC$_3$N + C$_2$H &1.2e-7&-0.7&0 \\
HC$_3$NH$^+$ + e$^-$ $\rightarrow$ HC$_3$N + H &7.31e-7&-0.58&0 \\
H$_2$C$_4$N$^+$ + e$^-$ $\rightarrow$ HC$_3$N + CH & 3.e-7&-0.5&0 \\
CH$_3$C$_3$NH$^+$ + e$^-$ $\rightarrow$ HC$_3$N + CH$_3$ &1.e-6&-0.3&0 \\
CH$_2$CHCN$^+$ + e$^-$ $\rightarrow$ HC$_3$N + H$_2$& 1.5e-7&-0.5&0\\
NH$_3$ + HC$_3$N$^+$ $\rightarrow$ HC$_3$N + NH$_3^+$ &1.7e-9&-0.5&0 \\
HC$_4$H + HC$_3$N$^+$ $\rightarrow$ C$_4$H$_2^+$ + HC$_3$N &8.9e-10&0&0 \\
C$_2$H$_4$ + HC$_3$N$^+$ $\rightarrow$ C$_2$H$_4^+$ + HC$_3$N &5.36e-10&0&0\\
C$_2$H$_2$ + HC$_3$N$^+$ $\rightarrow$ C$_2$H$_2^+$ + HC$_3$N &1.28e-10&0&0 \\
N + C$_3$H$^-$ $\rightarrow$ HC$_3$N + e$^-$ &5.e-12&0&0 \\
H + C$_3$N$^-$ $\rightarrow$ HC$_3$N + e$^-$ &5.4e-10&0&0 \\
H$^+$ + HC$_3$N $\rightarrow$ HC$_3$N$^+$ + H &4.e-9&-0.5&0 \\
C$^+$ + HC$_3$N $\rightarrow$ C$_2$N$^+$ + C$_2$H & 1.e-10&-0.5&0 \\
C$^+$ + HC$_3$N $\rightarrow$ C$_3^+$ + HCN &2.5e-10&-0.5&0 \\
C$^+$ + HC$_3$N $\rightarrow$ C$_3$H$^+$ + CN &3.25e-9&-0.5&0\\
C$^+$ + HC$_3$N $\rightarrow$ C$_4$N$^+$ + H &1.4e-9&-0.5&0 \\
C$_2$H$^+$ + HC$_3$N $\rightarrow$ C$_4$H$^+$ + HCN &7.6e-10&-0.5&0 \\
C$_2$H$^+$ + HC$_3$N $\rightarrow$ C$_4$H$_2^+$ + CN &4.56e-10&-0.5&0 \\
C$_2$H$^+$ + HC$_3$N $\rightarrow$ HC$_3$NH$^+$ + C$_2$ & 1.416e-9&-0.5&0  \\
C$_2$H$^+$ + HC$_3$N $\rightarrow$ HC$_5$N$^+$ + H &1.186e-9&-0.5&0 \\
C$_2$H$_3^+$ + HC$_3$N $\rightarrow$ HC$_3$NH$^+$ + C$_2$H$_2$ &3.8e-9&-0.5&0 \\
C$_2$H$_4^+$ + HC$_3$N $\rightarrow$ HC$_3$NH$^+$ + C$_2$H$_3$ &1.1e-9&-0.5&0 \\
C$_2$H$_5^+$ + HC$_3$N $\rightarrow$ HC$_3$NH$^+$ + C$_2$H$_4$ &3.3e-9&-0.5&0 \\
C$_3^+$ + HC$_3$N $\rightarrow$ C$_5$H$^+$ + CN & 3.2e-9&-0.5&0\\
C$_3$H$_5^+$ + HC$_3$N $\rightarrow$ HC$_3$NH$^+$ + CH$_3$CCH &1.e-10&-0.5&0 \\
CH$_3^+$ + HC$_3$N $\rightarrow$ C$_3$H$_3^+$ + HCN  &1.e-9&-0.5&0\\
H$_3^+$ + HC$_3$N $\rightarrow$ HC$_3$NH$^+$ + H$_2$ &9.1e-9&-0.5&0 \\
H$_3$O$^+$ + HC$_3$N $\rightarrow$ HC$_3$NH$^+$ + H$_2$O &4.e-9&-0.5&0 \\
HCNH$^+$ + HC$_3$N $\rightarrow$ HC$_3$NH$^+$ + HCN  &1.7e-9& -0.5&0 \\
HCNH$^+$ + HC$_3$N $\rightarrow$ HC$_3$NH$^+$ + HNC  & 1.7e-9& -0.5& 0 \\
HCO$^+$ + HC$_3$N $\rightarrow$ HC$_3$NH$^+$ + CO & 4.e-9&-0.5&0 \\
He$^+$ + HC$_3$N $\rightarrow$ C$_2$N$^+$ + CH + He & 2.8e-9&-0.5&0 \\
He$^+$ + HC$_3$N $\rightarrow$ C$_3$H$^+$ + N + He &4.e-10&-0.5&0  \\
He$^+ $+ HC$_3$N $\rightarrow$ CN + C$_2$H$^+$ + He &2.2e-9&-0.5&0 \\
He$^+$ + HC$_3$N $\rightarrow$ C$_3$N$^+$ + He + H  &2.45e-9&-0.5&0 \\
HN$_2$+ + HC$_3$N $\rightarrow$ HC$_3$NH$^+$ + N$_2$ &4.2e-9&-0.5&0  \\
HC$_3$N + PHOTON $\rightarrow$ CN + C$_2$H & 5.6e-9&0&2.2\\
C$_2$H$_2^+$ + HC$_3$N $\rightarrow$ H$_3$C$_5$N$^+$ + PHOTON &2.e-12&-2.5.&0\\  
C$_4$H$_2^+$ + HC$_3$N $\rightarrow$ H$_3$C$_7$N$^+$ + PHOTON &2.e-12&2.5.&0\\  
CH$_3^+$ + HC$_3$N $\rightarrow$ CH$_3$C$_3$NH$^+$ + PHOTON & 8.6e-11&-1.4.&0 \\
\enddata
\tablecomments{Rate coefficients are from RATE12  }
\end{deluxetable*}
\clearpage
\begin{deluxetable*}{lccc}
\tablecaption{List of added chemical reactions involving HCl label{tab:7}}
\tablehead{
\colhead{Reactions}&\colhead{$\alpha$}&\colhead{$\beta$}&\colhead{$\gamma$}}
\startdata
NH$_3$ + CCl$^+$ $\rightarrow$ HCNH$^+$ + HCl &1.3e-9&-0.5&0 \\   
CH$_5^+$ + HCl $\rightarrow$ H$_2$Cl$^+$ + CH$_4$  &2.06e-9&-0.5&0\\  
H$_2$ + Cl $\rightarrow$ HCl + H &5.27e-12&1.4&1760\\  
\enddata
\tablecomments{Rate coefficients are from RATE12 }
\end{deluxetable*}
\begin{deluxetable*}{lccc}
\tablecaption{List of chemical reactions with updated reaction rate coefficients involving HCl \label{tab:8}}
\tablehead{
\colhead{Reactions}&\colhead{$\alpha$}&\colhead{$\beta$}&\colhead{$\gamma$}}
\startdata
H$_2$Cl$^+$ + e$-$ $\rightarrow$ HCl + H &1e-8&-0.85&0\\    
H$_3^+$ + HCl $\rightarrow$ H$_2$Cl$^+$ + H$_2$ & 3.8e-9&-0.5&0\\    
He$^+$ + HCl $\rightarrow$ Cl$^+$ + He + H & 3.3e-9 &0&0 \\   
C$^+$ + HCl $\rightarrow$ CCl$^+$ + H & 1.1e-9&-0.5&0\\     
CH$_3^+$ + HCl $\rightarrow$ H$_2$CCl$^+$ + H$_2$ & 5.72e-11&-2.25&47.5\\  
H$_2$O + H$_2$Cl$^+$ $\rightarrow$ HCl + H$_3$O$^+$ & 2e-9&-0.5&0\\   
\enddata
\tablecomments{Rate coefficients are from RATE12  }
\end{deluxetable*}
\startlongtable
\begin{deluxetable*}{lccc}
\tablecaption{List of added chemical reactions involving HCN and HNC \label{tab:9}. Rate coefficients are from RATE12. }
\tablehead{
\colhead{Reactions}&\colhead{$\alpha$}&\colhead{$\beta$}&\colhead{$\gamma$}}
\startdata
C$^+$ + HCN $\rightarrow$ CNC$^+$ + H &3.e-9&-0.5&0\\
C$_2^+$ + HCN $\rightarrow$ C$_3$N$^+$ + H &2.6e-9&-0.5&0\\
C$_2$H$^+$ + HCN $\rightarrow$ HC$_3$N$^+$ + H &1.35e-9&-0.5&0\\
C$_2$H$_2^+$ + HCN $\rightarrow$ HC$_3$NH$^+$ + H &1.3e-10&-0.5&0\\
CH$^+$ + HCN $\rightarrow$ C$_2$N$^+$ + H$_2$ &3.6e-10&-0.5&0\\
CH$^+$ + HCN $\rightarrow$ C$_2$NH$^+$ + H &2.4e-10&-0.5&0\\
CH$_2^+$ + HCN $\rightarrow$ CH$_2$CN$^+$ + H & 1.8e-9&-0.5&0\\
CN$^+$ + HCN $\rightarrow$ C$_2$N$_2^+$ + H &3.15e-10&-0.5&0\\
H$^-$ + HCN $\rightarrow$ CN$^-$ + H$_2$ &3.8e-9&0&0\\
HCN + C$_2$H$_5^+$ $\rightarrow$ C$_2$H$_4$ + HCNH$^+$ & 2.7e-9&-0.5&0 \\
HCN + C$_2$H$_7^+$ $\rightarrow$ CH$_3$CNH$^+$ + CH$_4$ &2.2e-10&-0.5&0\\
HCN + C$_2$H$_7^+$ $\rightarrow$ HCNH$^+$ + CH$_3$CH$_3$ & 1.98e-9&-0.5&0\\
HCN + C$_2$N$_2^+$ $\rightarrow$ HCN$^+$ + NCCN &5.4e-10&-0.5&0\\
HCN + C$_2$N$_2^+$ $\rightarrow$ NCCNH$^+$ + CN &2.03e-9&-0.5&0 \\
HCN + C$_3^+$ $\rightarrow$ C$_3$H$^+$ + CN &2.6e-10&-0.5&0\\
HCN + C$_3^+$ $\rightarrow$ C$_4$N$^+$ + H &1.04e-9&-0.5&0 \\
HCN + C$_4$H$^+$ $\rightarrow$ C$_4$H$_2^+$ + CN &9.45e-11&-0.5&0\\
HCN + C$_4$H$^+$ $\rightarrow$ HC5N$^+$ + H &1.23e-9&-0.5&0\\
HCN + CH$_3$CH$_3^+$ $\rightarrow$ HCNH$^+$ + C$_2$H$_5$ & 1.20e-9&-0.5&0\\
HCN + CH$_3$OH$_2^+$ $\rightarrow$ CH$_3$CNH$^+$ + H$_2$O & 2.30e-11&-0.5&0\\
HCN + H$_2$CO$^+$ $\rightarrow$ HCO + HCNH$^+$ &1.4e-9&-0.5&0\\
HCN + H$_3$S$^+$ $\rightarrow$ H$_2$S + HCNH$^+$ &1.5e-9&-0.5&0\\
HCN + HSiS$^+$ $\rightarrow$ HCNH$^+$ + SiS &6.1e-10&-0.5&0\\
HCN + NCCNH$^+$ $\rightarrow$ HCNH$^+$ + NCCN & 2.e-9&-0.5&0\\
HCN + O$_2$H$^+$ $\rightarrow$ O$_2$ + HCNH$^+$ &9.7e-10&-0.5&0\\
PH$^+$ + HCN $\rightarrow$ HCNH$^+$ + P & 3.06e-10&-0.5&0\\
HCN + Si$^+$ $\rightarrow$ SiNC$^+$ + H & 1.40e-12&-0.5&0\\
O$^-$ + HCN $\rightarrow$ CN$^-$ + OH & 1.20e-9&-0.5&0\\
OH$^-$ + HCN $\rightarrow$ CN$^-$ + H$_2$O & 1.2e-9&0&0\\
C$_2$ + HCN $\rightarrow$ C$_3$N + H &1.11e-10&-0.82&9.7\\
CH$_3^+$ + HCN $\rightarrow$ CH$_3$CNH$^+$ + PHOTON &9.e-9&-0.50&0\\
HCN + Si$^+$ $\rightarrow$ SiNCH$^+$ + PHOTON &6.0e-15&-1.5&0\\
H + CN$^-$ $\rightarrow$ HCN + e$^-$ &6.3e-10&0&0 \\
C$_2$H$_2$ + HCN$^+$ $\rightarrow$ C$_2$H$_2^+$ + HCN & 1.5e-9&0&0\\
CH$_2$CHCNH$^+$ + e$^-$ $\rightarrow$ C$_2$H$_2$ + HCN + H &4.45e-7&-0.8&0\\
CH$_2$CN$^+$ + e$^-$ $\rightarrow$ HCN + CH &1.5e-7&-0.5&0\\
CH$_3$CN$^+$ + e$^-$ $\rightarrow$ HCN + CH$_2$ & 1.5e-7&-0.5&0\\
CH$_4$N$^+$ + e$^-$ $\rightarrow$ HCN + H$_2$ + H &3.e-7&-0.5&0 \\
HC$_5$NH$^+$ + e$^-$ $\rightarrow$ HCN + C$_4$H &4.4e-7&-0.7&0\\
HCNOH$^+$ + e$^-$ $\rightarrow$ HCN + OH &1.e-7&-0.5&0 \\
NH$_2$CNH$^+$ + e$^-$ $\rightarrow$ NH$_2$ + HCN &1.e-7&-0.5&0 \\
C$^+$ + CH$_2$CHCN $\rightarrow$ C$_3$H$_2^+$ + HCN & 9.84e-10&-0.5&0\\
C$_2$H$_2^+$ + CH$_3$CN $\rightarrow$ C$_3$H$_4^+$ + HCN & 1.06e-9&-0.5&0\\
C$_2$H$_2$ + C$_2$N$^+$ $\rightarrow$ C$_3$H$^+$ + HCN & 8.e-10&0&0\\
C$_2$H$_2$ + C$_4$N$^+$ $\rightarrow$ C$_5$H$^+$ + HCN &8.e-10&0&0 \\
C$_2$H$_2$ + HC$_3$N$^+$ $\rightarrow$ C$_4$H$_2^+$ + HCN & 5.12e-10&0&0\\
CH$_3^+$ + CH$_3$CN $\rightarrow$ C$_2$H$_5^+$ + HCN & 6.66e-10&-0.5&0\\
CH$_4$ + C$_4$N$^+$ $\rightarrow$ C$_4$H$_3^+$ + HCN &1.71e-10&0&0 \\
CH$_4$ + HC$_3$N$^+$ $\rightarrow$ C$_3$H$_4^+$ + HCN &8.3e-11&0&0\\
CN$^+$ + H$_2$CO $\rightarrow$ HCO$^+$ + HCN &5.2e-10&-0.5&0 \\
H$^+$ + CH$_3$CN $\rightarrow$ CH$_3^+$ + HCN &3.e-9&-0.5&0\\
H$_2$ + C$_4$N$^+$ $\rightarrow$ C$_3$H$^+$ + HCN &2.2e-11&0&0\\
H$_2$O + C$_2$N$^+$ $\rightarrow$ HCO$^+$ + HCN & 2.3e-10&-0.5&0\\
CNC$^+$ + H$_2$O $\rightarrow$ HCO$^+$ + HCN &1.63e-9&-0.5&0 \\
H$_2$S + C$_2$N$^+$ $\rightarrow$ HCS$^+$ + HCN &1.2e-9&-0.5&0\\
HCNH$^+$ + CH$_2$CHCN $\rightarrow$ CH$_2$CHCNH$^+$ + HCN & 2.25e-9&-0.5&0\\
HCNH$^+$ + CH$_3$CN $\rightarrow$ CH$_3$CNH$^+$ + HCN & 1.9e-9&-0.5&0\\
HCNH$^+$ + H$_2$CO $\rightarrow$ H$_3$CO$^+$ + HCN &1.05e-9&-0.5&0\\
HCNH$^+$ + H$_2$S $\rightarrow$ H$_3$S$^+$ HCN &1.7e-10&-0.5&0\\
He$^+$ + CH$_2$CHCN $\rightarrow$ HCN + C$_2$H$_2^+$ + He &3.5e-9&-0.5&0 \\
N + C$_3$H$_5^+$ $\rightarrow$ C$_2$H$_4^+$ + HCN &1.1e-10&0&0 \\
N + C$_4$H$_2^+$ $\rightarrow$ C$_3$H$^+$ + HCN & 1.71e-10&0&0\\
N + C$_6$H$_2^+$ $\rightarrow$ C$_5$H$^+$ + HCN & 1.9e-10&0&0\\
N + C$_6$H$_5^+$ $\rightarrow$ CH$_3$C$_4$H$^+$ + HCN &3.7e-11&0&0 \\
N + HC$_3$N$^+$ $\rightarrow$ C$_2$N$^+$ + HCN & 1.44e-10&0&0\\
N + SiCH$_2^+$ $\rightarrow$ Si$^+$ + HCN + H & 7.6e-10&0&0\\
NH$_3$ + C$_2$N$^+$ $\rightarrow$ HCNH$^+$ + HCN & 1.8e-9&-0.5&0\\
NH$_3$ + HCNH$^+$ $\rightarrow$ HCN + NH$_4^+$ &1.1e-9&-0.5&0 \\
Na + HCNH$^+$ $\rightarrow$ HCN + Na$^+$ + H & 1.35e-9&0&0\\
C$^-$ + HCNH$^+$ $\rightarrow$ C + HCN + H & 3.76e-8&-0.5&0\\
C$_{10}^-$ + HCNH$^+$ $\rightarrow$ C$_{10}$ + HCN + H & 3.76e-8&-0.5&0\\
C$_{10}$H$^-$ + HCNH$^+$ $\rightarrow$ C$_{10}$H + HCN + H & 3.76e-8&-0.5&0\\
C$_2^-$ + HCNH$^+$ $\rightarrow$ C$_2$ + HCC + H & 3.76e-8&-0.5&0\\
C$_2$H$^-$ + HCNH$^+$ $\rightarrow$ C$_2$H + HCN + H & 3.76e-8&-0.5&0\\
C$_3^-$ + HCNH$^+$ $\rightarrow$ C$_3$ + HCN + H & 3.76e-8&-0.5&0\\
C$_3$H$^-$ + HCNH$^+$ $\rightarrow$ C$_3$H + HCN + H &3.76e-8&-0.5&0 \\
C$_3$N$^-$ + HCNH$^+$ $\rightarrow$ C$_3$N + HCN + H & 3.76e-8&-0.5&0\\
C$_4^-$ + HCNH$^+$ $\rightarrow$ C$_4$ + HCN + H &3.76e-8&-0.5&0 \\
C$_4$H$^-$ + HCNH$^+$ $\rightarrow$ C$_4$H + HCN + H &3.76e-8&-0.5&0 \\
C$_5^-$ + HCNH$^+$ $\rightarrow$ C$_5$ + HCN + H & 3.76e-8&-0.5&0\\
C$_5$H$^-$ + HCNH$^+$ $\rightarrow$ C$_5$H + HCN + H & 3.76e-8&-0.5&0\\
C$_5$N$^-$ + HCNH$^+$ $\rightarrow$ C$_5$N + HCN + H & 3.76e-8&-0.5&0\\
C$_6^-$ + HCNH$^+$ $\rightarrow$ C$_6$ + HCN + H & 3.76e-8&-0.5&0\\
C$_6$H$^-$ + HCNH$^+$ $\rightarrow$ C$_6$H + HCN + H & 3.76e-8&-0.5&0\\
C$_7^-$ + HCNH$^+$ $\rightarrow$ C$_7$ + HCN + H & 3.76e-8&-0.5&0\\
C$_7$H$^-$ + HCNH$^+$ $\rightarrow$ C$_7$H + HCN + H & 3.76e-8&-0.5&0\\
C$_8^-$ + NCNH$^+$ $\rightarrow$ C$_8$ + HCN + H & 3.76e-8&-0.5&0\\
C$_8$H$^-$ + HCNH$^+$ $\rightarrow$ C$_8$H + HCN + H & 3.76e-8&-0.5&0 \\
C$_9^-$ + HCNH$^+$ $\rightarrow$ C$_9$ + HCN + H & 3.76e-8&-0.5&0 \\
C$_9$H$^-$ + HCNH$^+$ $\rightarrow$ C$_9$H + HCN + H & 3.76e-8&-0.5&0\\
CH$^-$ + HCNH$^+$ $\rightarrow$ CH + HCN + H &3.76e-8&-0.5&0\\
H$^-$ + HCNH$^+$ $\rightarrow$ H + HCN + H & 3.76e-8&-0.5&0\\
O$^-$ + HCNH$^+$ $\rightarrow$ O + HCN + H & 3.76e-8&-0.5&0\\
O$_2^-$ + HCNH$^+$ $\rightarrow$ O$_2$ + HCN + H & 3.76e-8&-0.5&0\\
OH$^-$ + HCNH$^+$ $\rightarrow$ OH + HCN + H & 3.76e-8&-0.5&0\\
S$^-$ + HCNH$^+$ $\rightarrow$ S + HCN + H & 3.76e-8&-0.5&0\\
C$_2$H$_2$ + NO $\rightarrow$ HCN + CO + H & 8.97e-12&0&18973\\
C + NH$_2$ $\rightarrow$ HCN + H & 3.26e-11&-0.1&-9.0\\
CH$_2$ + N$_2$ $\rightarrow$ HCN + NH & 8.e-12&0&18000 \\
CH$_3$ + NO $\rightarrow$ HCN + H$_2$O & 4.e-12&0&7900\\
CN + C$_2$H$_4$ $\rightarrow$ C$_2$H$_3$ + HCN & 1.25e-10&0.7&30.\\
CN + H$_2$CO $\rightarrow$ HCO + HCN &2.6e-10&-0.47.&826\\
H + H$_2$CN $\rightarrow$ HCN + H$_2$ & 1.e-10&0&0\\
H + NCCN $\rightarrow$ HCN + CN & 1.48e-10&0&3588.\\
H + OCN $\rightarrow$ HCN + O & 1.87e-11&0.9&2924.\\
HOCN + C $\rightarrow$ CO + HCN &3.33e-11&0&0 \\
N + C$_2$H$_4$ $\rightarrow$ HCN + CH$_3$ &3.69e-14&0&161.\\
N + CH$_2$ $\rightarrow$ HCN + H & 3.95e-11&0.17&0\\
N + H$_2$CN $\rightarrow$ HCN + NH & 1.e-10&0&200.\\
N + HCO $\rightarrow$ HCN + O & 1.7e-10&0&0\\
N + HCS $\rightarrow$ S + HCN &1.e-10&0&0\\
OH + NCCN $\rightarrow$ HCN + OCN &3.11e-13&0&1450 \\
H$_2$CN + PHOTON $\rightarrow$ HCN + H &5.48e-10&0&2\\
\enddata
\end{deluxetable*}
\startlongtable
\begin{deluxetable*}{lccc}
\tablecaption{List of updated chemical reactions involving HCN $\&$ HNC \label{tab:10}}
\tablehead{
\colhead{Reactions}&\colhead{$\alpha$}&\colhead{$\beta$}&\colhead{$\gamma$}
}
\startdata
HNC + HCO$^+$ $\rightarrow$ HCNH$^+$ + CO &3.1e-9&-0.5&0 \\
HCN + HCO$^+$ $\rightarrow$ HCNH$^+$ + CO &3.1e-9&-0.5&0\\
O$^+$ + HCN $\rightarrow$ NO$^+$ + CH &1.2e-9&-0.5&0\\
NH + HCN$^+$ $\rightarrow$ CN + NH$_2^+$&6.5e-10&-0.5&0\\
CN + HNO $\rightarrow$ NO + HCN & .0e-11&0&0 \\
NH$_3$ + CN $\rightarrow$ HCN + NH$_2$ &2.75e-11&-1.14&0 \\
CH$_2$ + NO $\rightarrow$ HCN + OH &3.65e-12&0&0 \\
CH + NO $\rightarrow$ HCN + O &1.2e-10&-0.13&0 \\
HCN + C$_2$H$_3^+$ $\rightarrow$ HCNH$^+$ + C$_2$H$_2$ &2.9e-9&-0.5&0 \\
C$_2$H$^+$ + HCN $\rightarrow$ C$_2$H$_2^+$ + CN &1.4e-9&-0.5&0\\
C$_2$H$_2^+$ + HCN $\rightarrow$ HCNH$^+$ + C$_2$H &2.3e-10&-0.5&0\\
C$_2$H$^+$ + HNC $\rightarrow$ HCNH$^+$ + C$_2$ &1.4e-9&-0.5&0\\
C$_2$H$^+$ + HCN $\rightarrow$ HCNH$^+$ + C$_2$ &1.4e-9&-0.5&0 \\
HCN + HS$^+$ $\rightarrow$ S + HCNH$^+$ &8.6e-10&-0.5&0\\
HCN + HNO$^+$ $\rightarrow$ NO + HCNH$^+$ &9.9e-10&-0.5&0 \\
H$_3$O$^+$ + HCN $\rightarrow$ HCNH$^+$ + H$_2$O &3.8e-9&-0.5&0\\
H$_2$O$^+$ + HCN $\rightarrow$ HCNH$^+$ + OH &2.1e-9&-0.5&0\\
CH$_5^+$ + HCN $\rightarrow$ HCNH$^+$ + CH$_4$ &1.2e-9&-0.5&0\\
OH$^+$ + HCN $\rightarrow$ HCNH$^+$ + O &1.2e-9&-0.5&0 \\
NH$_2$ + HCNH$^+$ $\rightarrow$ HCN + NH$_3^+$ &4.45e-10&-0.5&0 \\
NH$_2^+$ + HCN $\rightarrow$ HCNH$^+$ + NH &1.2e-9&-0.5&0 \\
NH$^+$ + HCN $\rightarrow$ HCNH$^+$ + N &1.8e-9&-0.5&0\\
CH + HCNH$^+$ $\rightarrow$ HCN + CH$_2^+$ &3.15e-10&-0.5&0\\
HNC + HS$^+$ $\rightarrow$ S + HCNH$^+$ &8.6e-10&-0.5&0 \\
HNC + HNO$^+$ $\rightarrow$ NO + HCNH$^+$ &9.9e-10&-0.5&0\\
H$_2$O$^+$ + HNC $\rightarrow$ HCNH$^+$ + OH &1.1e-9&-0.5&0 \\
CH$_5^+$ + HNC $\rightarrow$ HCNH$^+$ + CH$_4$ &1.2e-9&-0.5&0 \\
NH$_3$ + HCNH$^+$ $\rightarrow$ HNC + NH$_4^+$ &1.1e-9&-0.5&0 \\
OH$^+$ + HNC $\rightarrow$ HCNH$^+$ + O &1.2e-9&-0.5&0\\
NH$_2$ + HCNH$^+$ $\rightarrow$ HNC + NH$_3^+$ &4.45e-10&-0.5&0\\
NH$_2^+$ + HNC $\rightarrow$ HCNH$^+$ + NH &1.2e-9&-0.5&0\\
NH$^+$ + HNC $\rightarrow$ HCNH$^+$ + N &1.8e-9&-0.5&0\\
CH + HCNH$^+$ $\rightarrow$ HNC + CH$_2^+$ &3.15e-10&-0.5&0\\
CH$^+$ + HNC $\rightarrow$ HCNH$^+$ + C&1.8e-9&0&0\\
HCN + N$_2^+$ $\rightarrow$ N$_2$ + HCN$^+$ &3.9e-10&-0.5&0\\
CN$^+$ + HCN $\rightarrow$ HCN$^+$ + CN &1.79e-9&-0.5&0\\
H$_2$O + HCN$^+$ $\rightarrow$ HCN + H$_2$O$^+$ &1.8e-9&-0.5&0 \\
NH$_3$ + HCN$^+$ $\rightarrow$ HCN + NH$_3^+$ &1.68e-9&-0.5&0 \\
N$^+$ + HCN $\rightarrow$ HCN$^+$ + N &3.7e-9&-0.5&0\\
H$_3^+$ + HCN $\rightarrow$ HCNH$^+$ + H$_2$ &8.1e-9&-0.5&0 \\
H + HNC $\rightarrow$ HCN + H &4.e-11&0&1200\\
H$_3^+$ + HNC $\rightarrow$ HCNH$^+$ + H$_2$ &8.1e-9&-0.5&0\\
H$_2^+$ + HCN $\rightarrow$ HCN$^+$ + H$_2$ &2.7e-9&-0.5&0\\
He$^+$ + HCN $\rightarrow$ CN$^+$ + He + H&1.46e-9&-0.5&0\\
He$^+$ + HCN $\rightarrow$ N + C$^+$ + He + H&7.75e-10&-0.5&0\\
He$^+$ + HCN $\rightarrow$ N$^+$ + CH + He&2.17e-10&-0.5&0 \\
He$^+$ + HCN $\rightarrow$ N + CH$^+$ + He & 6.51e-10&-0.5&0 \\
H$_3^+$ + CN $\rightarrow$ HCN$^+$ + H$_2$ &2.0e-9&-0.5&0\\
H$_2^+$ + CN $\rightarrow$ HCN$^+$ + H &1.2e-9&-0.5&0\\
\enddata
\tablecomments{Rate coefficients are from RATE12}
\end{deluxetable*}
\startlongtable
\begin{deluxetable*}{lccc}
\tablecaption{List of added chemical reactions involving CH \label{tab:11}. Rate coefficients are from RATE12.}
\tablehead{
\colhead{Reactions}&\colhead{$\alpha$}&\colhead{$\beta$}&\colhead{$\gamma$}
}
\startdata
H + C$^-$ $\rightarrow$ CH + e$^-$ & 5.0e-10&0&0 \\
CH$^+$ + HCO $\rightarrow$ HCO$^+$ + CH & 4.6e-10&-0.5&0\\ 
CH$^+$ + Mg $\rightarrow$ Mg$^+$ + CH&3.6e-10&0&0 \\
CH$^+$ + Na $\rightarrow$ Na$^+$ + CH &3.5e-10&0&0\\
C$_2$H$_3^+$ + e$^-$ $\rightarrow$ CH$_2$ + CH &1.5e-8&-0.84&0 \\
C$_2$H$_4^+$ + e$^-$ $\rightarrow$ CH$_3$ + CH &1.12e-8&-0.76&0 \\
C$_3$H$_4^+$ + e$^-$ $\rightarrow$ C$_2$H$_3$ + CH &4.e-8&-0.67&0 \\
C$_3$H$_7^+$ + e$^-$ $\rightarrow$ CH$_3$CH$_3$ + CH &9.2e-9&-0.73&0 \\
C$_4$H$_5^+$ + e$^-$ $\rightarrow$ CH$_3$CCH + CH &1.5e-7&-0.5&0 \\
CH$_2$CCH$^+$ + e$^-$ $\rightarrow$ C$_2$H$_2^+$ + CH&5.e-8&-0.5&0 \\
CH$_3^+$ + e$^-$ $\rightarrow$ CH + H$_2$ &1.95e-7&-0.5&0 \\
CH$_3^+$ + e$^-$ $\rightarrow$ CH + H + H&2.5e-7&-0.4&0 \\
CH$_3$OH$^+$ + e$^-$ $\rightarrow$ CH + H$_2$O + H&6.49e-7&-0.66&0 \\
CH$_5^+$ + e$^-$ $\rightarrow$ CH + H$_2$ + H$_2$ &8.4e-9&-0.52&0 \\
H$_2$C$_4$N$^+$ + e$^-$ $\rightarrow$ HC$_3$N + CH &3.e-7&-0.5&0 \\
H$_3$CO$^+$ + e$^-$ $\rightarrow$ CH + H$_2$O &1.4e-8&-0.78&0 \\
HC$_2$O$^+$ + e$^-$ $\rightarrow$ CO + CH &1.e-7&-0.5&0 \\
HC$_2$P$^+$ + e$^-$ $\rightarrow$ CP + CH &1.5e-7&-0.5&0 \\
HC$_2$S$^+$ + e$^-$ $\rightarrow$ CS + CH &1.5e-7&-0.5&0 \\
HC$_3$S$^+$ + e$^-$ $\rightarrow$ C$_2$S + CH &1.5e-7&0.5&0 \\
HC$_4$S$^+$ + e$^-$ $\rightarrow$ C$_3$S + CH &1.e-7&-0.5&0 \\
HCNO$^+$ + e$^-$ $\rightarrow$ CH + NO &1.5e-7&-0.5&0 \\
HCP$^+$ + e$^-$ $\rightarrow$ P + CH &1.5e-7&-0.5&0 \\
HCSi$^+$ + e$^-$ $\rightarrow$ Si + CH &1.5e-7&-0.5&0 \\
PC$_3$H$^+$ + e$^-$ $\rightarrow$ CCP + CH &1.e-7&-0.5&0 \\
PC$_4$H$^+$ + e$^-$ $\rightarrow$ C$_3$P + CH &7.5e-8&0&0 \\
SiC$_4$H$^+$ + e$^-$ $\rightarrow$ SiC$_3$ + CH &1.5e-7&-0.5&0 \\
SiNCH$^+$ + e$^-$ $\rightarrow$ SiN + CH &1.5e-7&-0.5&0 \\
C$^+$ + C$_2$H$_4$ $\rightarrow$ C$_2$H$_3^+$ + CH &8.5e-11&0&0 \\
C$^+$ + CH$_2$CCH$_2$ $\rightarrow$ C$_3$H$_3^+$ + CH &3.5e-10&0&0 \\
C$^+$ + CH$_3$C$_4$H $\rightarrow$ C$_5$H$_3^+$ + CH &7.5e-10&0&0 \\
C$^+$ + C$_7$H$_4$ $\rightarrow$ C$_7$H$_3^+$ + CH &7.5e-10&0&0 \\
C$^+$ + CH$_3$CCH $\rightarrow$ CH$_2$CCH$^+$ + CH &3.8e-10&0&0 \\
C$^+$ + H$_2$CO $\rightarrow$ HCO$^+$ + CH & 7.8e-10&-0.5&0 \\
C$^+$ + HC$_2$P $\rightarrow$ CCP$^+$ + CH &5.e-10&0&0 \\
CH$_2^+$ + H$_2$S $\rightarrow$ H$_3$S$^+$ + CH &2.3e-10&-0.5&0 \\
HNC$_3$ + C$^+$ $\rightarrow$ C$_3$N$^+$ + CH &1.55e-9&-0.5&0 \\
He$^+$ + C$_4$H$_3$ $\rightarrow$ C$_3$H$_2^+$ + CH + He &6.7e-10&0&0 \\
He$^+$ + C$_6$H$_6$ $\rightarrow$ C$_3$H$_2^+$ + CH + He &6.7e-10&0&0 \\
He$^+$ + HC$_2$P $\rightarrow$ CP$^+$ + CH + He &5.e-10&0&0 \\
He$^+$ + HCP $\rightarrow$ P$^+$ + CH + He &5.e-10&0&0 \\
He$^+$ + HCSi $\rightarrow$ Si$^+$ + CH + He &1.e-9&0&0 \\
He$^+$ + SiCH$_2$ $\rightarrow$ SiH$^+$ + CH + He &1.e-9&0&0\\
O + C$_4$H$_2^+$ $\rightarrow$ HC$_3$O$^+$ + CH &1.35e-11&0&0 \\
Si$^+$ + CH$_3$CCH $\rightarrow$ SiC$_2$H$_3^+$ + CH &7.2e-10&0&0 \\
CH$^-$ + C$^+$ $\rightarrow$ C + CH &7.51e-8&-0.5&0\\
CH$^-$ + C$_2$H$_2^+$ $\rightarrow$ CH + C$_2$H$_2$ &7.51e-8&-0.5&0 \\
CH$^-$ + C$_2$H$_3^+$ $\rightarrow$ CH + C$_2$H$_3$ &7.51e-8&-0.5&0 \\
CH$^-$ + C$_4$H$_2^+$ $\rightarrow$ CH + HC$_4$H &7.51e-8&-0.5&0 \\
CH$^-$ + C$_4$H$_3^+$ $\rightarrow$ CH + C$_4$H$_3$&7.51e-8&-0.5&0 \\
CH$^-$ + C$_4$S$^+$ $\rightarrow$ CH + C$_4$S &7.51e-8&-0.5&0 \\
CH$^-$ + CH$_2$CCH$^+$ $\rightarrow$ CH + CH$_2$CCH &7.51e-8&-0.5&0 \\
CH$^-$ + CH$_3^+$ $\rightarrow$ CH + CH$_3$ &7.51e-8&-0.5&0 \\
CH$^-$ + CNC$^+$ $\rightarrow$ CH + C$_2$N &7.51e-8&-0.5&0 \\
CH$^-$ + Fe$^+$ $\rightarrow$ CH + Fe &7.51e-8&-0.5&0 \\
CH$^-$ + H$^+$ $\rightarrow$ CH + H &7.51e-8&-0.5&0 \\
CH$^-$ + H$_2$CO$^+$ $\rightarrow$ CH + H$_2$CO & 7.51e-8&-0.5&0 \\
CH$^-$ + H$_2$S$^+$ $\rightarrow$ CH + H$_2$ S&7.51e-8&-0.5&0 \\
CH$^-$ + H$_3^+$ $\rightarrow$ CH + H$_2$ + H &7.51e-8&-0.5&0 \\
CH$^-$ + H$_3$O$^+$ $\rightarrow$ CH + H + H$_2$O &7.51e-8&-0.5&0 \\
CH$^-$ + HC$_2$S$^+$ $\rightarrow$ CH + C$_2$S + H&7.51e-8&-0.5&0 \\
CH$^-$ + HCNH$^+$ $\rightarrow$ CH + HNC + H &3.76e-8&-0.5&0 \\
CH$^-$ + HCO$^+$ $\rightarrow$ CH + H + CO &3.76e-8&-0.5&0 \\
CH$^-$ + HCO$^+$ $\rightarrow$ CH + HCO &3.76e-8&-0.5&0 \\
CH$^-$ + He$^+$ $\rightarrow$ CH + He &7.51e-8&-&0 \\
CH$^-$ + N$^+$ $\rightarrow$ CH + N &7.51e-8&-0.5&0 \\
CH$^-$ + N$_2$H$^+$ $\rightarrow$ CH + N$_2$ + H &7.51e-8&-0.5&0 \\
CH$^-$ + NH$_3^+$ $\rightarrow$ CH + NH$_3$ &7.51e-8&-0.5&0 \\
CH$^-$ + NH$_4^+$ $\rightarrow$ CH + NH$_3$ + H &7.51e-8&-0.5&0 \\
CH$^-$ + NO$^+$ $\rightarrow$ CH + NO &7.51e-8&-0.5&0 \\
CH$^-$ + Na$^+$ $\rightarrow$ CH + Na &7.51e-8&-0.5&0 \\
CH$^-$ + O$^+$ $\rightarrow$ CH + O &7.51e-8&-0.5&0 \\
CH$^-$ + S$^+$ $\rightarrow$ CH + S &7.51e-8&-0.5&0 \\
CH$^-$ + SO$^+$ $\rightarrow$ CH + SO &7.51e-8&-0.5&0 \\
CH$^-$ + Si$^+$ $\rightarrow$ CH + Si &7.51e-8&-0.5&0 \\
CH$^-$ + SiOH$^+$ $\rightarrow$ CH + SiO + H &7.51e-8&-0.5&0 \\
CH$^-$ + SiS$^+$ $\rightarrow$ CH + SiS &7.51e-8&-0.5&0 \\
C + HCO $\rightarrow$ CO + CH &1.e-10&0&0 \\
N + CH$_2$ $\rightarrow$ NH + CH &9.96e-13&0&20380. \\
O + HCSi $\rightarrow$ SiO + CH &2.e-11&0&0 \\
CH$^-$ + PHOTON $\rightarrow$ CH + e$^-$ &1.36e-8&0&1.50 \\
CH$_2^+$ + PHOTON $\rightarrow$ CH + H$^+$ &4.67e-11&0&2.2 \\
HCNO + PHOTON $\rightarrow$ CH + NO &1.e-9&0&1.7 \\
C$^-$ + CH $\rightarrow$ C$_2$H + e$^-$ &5.e-10&0&0\\
CH + O$^-$ $\rightarrow$ HCO + e$^-$ &5.e-10&0&0 \\
CH + OH$^-$ $\rightarrow$ H$_2$CO + e$^-$ &5.e-10&0&0 \\
CH + H$_2$CO$^+$ $\rightarrow$ H$_2$CO + CH$^+$ &3.1e-10&-0.5&0 \\
CH + C$_2$H$_2^+$ $\rightarrow$ C$_3$H$_2^+$ + H &6.4e-10&-0.5&0 \\
CH + H$_2$CO$^+$ $\rightarrow$ HCO + CH$_2^+$ &3.1e-10&-0.5&0 \\
CH + H$_3$CO$^+$ $\rightarrow$ H$_2$CO + CH$_2^+$ &6.2e-10&-0.5&0 \\
CH + HN$_2^+$ $\rightarrow$ N$_2$ + CH$_2^+$ &6.3e-10&-0.5&0 \\
CH + O$_2$H$^+$ $\rightarrow$ O$_2$ + CH$_2^+$ &6.2e-10&-0.5&0 \\
CH + Si$^+$ $\rightarrow$ SiC$^+$ + H &6.3e-10&-0.5&0 \\
CH + SiH$^+$ $\rightarrow$ Si + CH$_2^+$ &6.0e-10&-0.5&0 \\
CH + C$_2$H$_2$ $\rightarrow$ C$_3$H$_2$ + H&3.59e-10&0&0\\
CH + C$_2$H$_2$ $\rightarrow$ H$_2$CCC + H &3.92e-10&-0.45&30.40 \\
CH + C$_2$H$_4$ $\rightarrow$ CH$_2$CCH$_2$ + H &1.11e-10&-0.62&32.90 \\
CH + C$_2$H$_4$ $\rightarrow$  CH$_3$CCH + H &2.21e-10&-0.62&32.90 \\
CH + CH$_2$CCH$_2$ $\rightarrow$ CH$_2$CHCCH + H &3.17e-10&-4.03&397.80 \\
CH + CH$_3$CCH $\rightarrow$ CH$_2$CHCCH + H &4.2e-10&-.23&16.00 \\
CH + CH$_3$CH$_3$ $\rightarrow$ C$_2$H$_4$ + CH$_3$&2.47e-11&-0.52&29.20 \\
CH + CH$_3$CH$_3$ $\rightarrow$ CH$_3$CHCH$_2$ + H&6.17e-11&-0.52&29.20 \\
CH + CH$_3$CHCH$_2$ $\rightarrow$ CH$_2$CCH$_2$ + CH$_3$&9.0e-11&0&0 \\
CH + CH$_3$CHCH$_2$ $\rightarrow$ CH$_2$CHCHCH$_2$ + H&3.1e-10&0&0 \\
CH + CH$_3$OH $\rightarrow$ CH$_3$CHO + H&2.49e-10&-1.93&0 \\
CH + CH$_4$ $\rightarrow$ C$_2$H$_4$ + H&1.05e-10&-1.04&36.0 \\
CH + CO$_2$ $\rightarrow$ HCO + CO &2.94e-13&0.5&3000. \\
CH + H$_2$CO $\rightarrow$ HCO + CH$_2$ &9.21e-12&0.7&2000. \\
CH + HCO $\rightarrow$ CO + CH$_2$&2.87e-12&0.7&500. \\
CH + NH$_3$ $\rightarrow$ CH$_2$NH + H &1.69e-10&-0.41&19.0 \\
CH + NO $\rightarrow$ HCO + N&1.16e-11&-0.13&0 \\
CH + O$_2$ $\rightarrow$ CO$_2$ + H&1.14e-11&-0.48&0 \\
CH + O$_2$ $\rightarrow$ CO + O + H&1.14e-11&-0.48&0 \\
CH + O$_2$ $\rightarrow$ HCO + O&7.6e-12&-0.48&0 \\
CH + OCS $\rightarrow$ CO + CS + H&4.e-10&0.&0 \\     
CH + OH $\rightarrow$ HCO + H&1.44e-11&0&0 \\
\enddata
\end{deluxetable*}
\startlongtable
\begin{deluxetable*}{lccc}
\tablecaption{List of updated chemical reactions involving CH \label{tab:12}}
\tablehead{
\colhead{Reactions}&\colhead{$\alpha$}&\colhead{$\beta$}&\colhead{$\gamma$}}
\startdata
CH + CO$^+$ $\rightarrow$ CO + CH$^+$&3.2e-10&-0.5&0\\
CH + HCO$^+$ $\rightarrow$ CO + CH$_2^+$&6.3e-10&-0.5&0  \\
CH + O$_2$ $\rightarrow$ CO + OH &7.6e-12&-0.48&0\\
O + C$_2$H $\rightarrow$ CO + CH &1.e-10&0&0 \\
C$_2$H$^+$ + e$^-$ $\rightarrow$ CH + C &1.53e-7&-0.76&0 \\
CH + PHOTON $\rightarrow$ CH$^+$ + e$^-$ &7.6e-10&0&3.3 \\
CH + PHOTON $\rightarrow$ C + H &9.2e-10&0&1.7 \\
PHOTON + CH$_2$ $\rightarrow$ CH + H &5.8e-10&0&2. \\
PHOTON + CH$_3$ $\rightarrow$ CH + H$_2$ &1.35e-10&0&2.3 \\
PHOTON + CH$_4$ $\rightarrow$ CH + H$_2$ + H &2.2e-10&0&2.6 \\
O + CH $\rightarrow$ CO + H &6.02e-11&0.1.&-4.5\\
C$^+$ + CH $\rightarrow$ C$_2^+$ + H &3.8e-10&-0.50&0\\
C$^+$ + CH $\rightarrow$ CH$^+$ + C&3.8e-10&-0.5.&0 \\
H$^+$ + CH $\rightarrow$ CH$^+$ + H&1.9e-9&-0.5&0 \\
H$_2^+$ + CH $\rightarrow$ CH$_2^+$ + H &7.1e-10&-0.5&0 \\
H$_3^+$ + CH $\rightarrow$ CH$_2^+$ + H$_2$&1.2e-9&-0.5&0\\
He$^+$ + CH $\rightarrow$ CH$^+$ + He &5.0e-10&-0.5&0  \\
He$^+$ + CH $\rightarrow$ C$^+$ + H + He&1.1e-9&-0.5&0\\
He$^+$ + HCN $\rightarrow$ N$^+$ + CH + He &2.17e-10&-0.5&0 \\
He$^+$ + C$_2$H $\rightarrow$ CH + C$^+$ + He &5.1e-10&0&0 \\
He$^+$ + C$_2$H$_2$ $\rightarrow$ CH$^+$ + CH + He &7.7e-10&0&0 \\
CH + NH$_2^+$ $\rightarrow$ NH$_2$ + CH$^+$ &3.5e-10&-0.5&0 \\
CH$^+$ + NH$_3$ $\rightarrow$ NH$_3^+$ + CH &4.59e-10&-0.5&0\\
CH + CN$^+$ $\rightarrow$ CN + CH$^+$ &6.4e-10&-0.5&0\\
CH + N$_2^+$ $\rightarrow$ N$_2$ + CH$^+$ &6.3e-10&-0.5&0 \\
CH + N$^+$ $\rightarrow$ N + CH$^+$ &3.6e-10&-0.5&0\\
CH + HCNH$^+$ $\rightarrow$ HNC + CH$_2^+$ &3.15e-10&-0.5&0 \\
CH + HCNH$^+$ $\rightarrow$ HCN + CH$_2^+$&3.15e-10&-0.5&0 \\
CH + C$_2$H$^+$ $\rightarrow$ C$_2$ + CH$_2^+$ &3.2e-10&-0.5&0 \\
CH + C$_2^+$ $\rightarrow$ C$_3^+$ + H &3.2e-10&0&0\\
CH + C$_2$H$^+$ $\rightarrow$ C$_3$H$^+$ + H &3.2e-10&-0.5&0\\
CH + CH$_3^+$ $\rightarrow$ C$_2$H$_2^+$ + H$_2$ &7.1e-10&-0.5&0\\
CH + C$_2^+$ $\rightarrow$ C$_2$ + CH$^+$ &3.2e-10&-0.5&0\\
CH$^+$ + CH $\rightarrow$ C$_2^+$ + H$_2$ &7.4e-10&-0.5&0\\
O + CH $\rightarrow$ HCO$^+$ + e$^-$ &1.09e-11&-2.19&165.1\\
O$^+$ + CH $\rightarrow$ O + CH$^+$ &3.5e-10&-0.5&0\\
O$^+$ + CH $\rightarrow$ CO$^+$ + H &3.5e-10&-0.5&0 \\
CH + CO$^+$ $\rightarrow$ HCO$^+$ + C &3.2e-10&-0.5&0\\
CH + H$_2$O$^+$ $\rightarrow$ H$_2$O + CH$^+$ &3.4e-10&-0.5&0	\\
CH + H$_2$O$^+$ $\rightarrow$ OH + CH$_2^+$ & 3.4e-10&-0.5&0\\
CH + H$_3$O$^+$ $\rightarrow$ H$_2$O + CH$_2^+$ &6.8e-10&-0.5&0\\
CH + O$_2^+$ $\rightarrow$ O$_2$ + CH$^+$ &3.1e-10&-0.5&0\\
CH + O$_2^+$ $\rightarrow$ HCO$^+$ + O &3.1e-10&-0.5&0\\
CH + OH$^+$ $\rightarrow$ OH + CH$^+$ &3.5e-10&-0.5&0 \\
CH + OH$^+$ $\rightarrow$ O + CH$_2^+$ &3.5e-10&-0.5&0\\
CH + SiO$^+$ $\rightarrow$ HCO$^+$ + Si &5.9e-10&-0.5&0\\
CH$_5^+$ + CH $\rightarrow$ CH$_4$ + CH$_2^+$ &6.9e-10&-0.5&0\\
CH + NO $\rightarrow$ HCN + O &1.2e-10&-0.13&0\\
CH + NO $\rightarrow$ OCN + H &3.49e-11&-0.13&0\\
CH + HNO $\rightarrow$ NO + CH$_2$ &1.73e-11&0&0\\
CH + S $\rightarrow$ CS + H &5e-11&0&0 \\
CH + N$^+$ $\rightarrow$ CN$^+$ + H &3.6e-10&-0.5&0\\
CH + NH$^+$ $\rightarrow$ CH$_2^+$ + N &9.9e-10&-0.5&0\\
CH + NH$_2^+$ $\rightarrow$ NH + CH$_2^+$ &3.5e-10&-0.5&0 \\
CH + NH$_3^+$  $\rightarrow$ NH$_4^+$ + C &6.9e-10&-0.5&0 \\
CH + HCN$^+$ $\rightarrow$ CN + CH$_2^+$ &6.3e-10&-0.5&0 \\
CH + HNO$^+$ $\rightarrow$ NO + CH$_2^+$ &6.2e-10&0&0\\
CH + S$^+$ $\rightarrow$ CS$^+$ + H &6.2e-10&-0.50 &0\\
CH + HS$^+$ $\rightarrow$ S + CH$_2^+$&5.8e-10&-0.5&0\\
CH$_2^+$ + NH$_3$ $\rightarrow$ NH$_4^+$ + CH &1.26e-9&-0.5&0\\
O$^+$ + HCN $\rightarrow$ NO$^+$ + CH &1.2e-9&-0.5&0\\
\enddata
\tablecomments{ Rate coefficients are from RATE12  }
\end{deluxetable*}
\startlongtable
\begin{deluxetable*}{lccc}
\tablecaption{List of added chemical reactions involving CH$_2$ \label{tab:13}}
\tablehead{
\colhead{Reactions}&\colhead{$\alpha$}&\colhead{$\beta$}&\colhead{$\gamma$}}
\startdata
H$_2$ + C$^-$ $\rightarrow$ CH$_2$ + e$^-$&1.e-13&0&0 \\
H + CH$^-$ $\rightarrow$ CH$_2$ + e$^-$ &1.e-10&0&0 \\
C$_2$H$_5^+$ + e$^-$ $\rightarrow$ CH$_3$ + CH$_2$ &4.76e-8&-0.79&0 \\
C$_3$H$_2^+$ + e$^-$ $\rightarrow$ C$_2$ + CH$_2$&3.e-8&-0.5&0 \\
C$_3$H$_7^+$ + e$^-$ $\rightarrow$ C$_2$H$_4$ + CH$_2$ + H&9.2e-9&-0.73&0 \\
CH$_2$CO$^+$ + e$^-$ $\rightarrow$ CO + CH$_2$ &2.e-7&-0.5&0 \\
CH$_2$NH$_2^+$ + e$^-$ $\rightarrow$ NH$_2$ + CH$_2$ &1.5e-7&-0.5&0 \\
CH$_3$CHOH$^+$ + e$^-$ $\rightarrow$ CH$_2$ + H$_2$CO + H &8.47e-7&-0.74&0 \\
CH$_3$COCH$_3^+$ + e$^-$ $\rightarrow$ CH$_3$CHO + CH$_2$ &1.5e-7&-0.5&0 \\
CH$_3$OCH$_3^+$ + e$^-$ $\rightarrow$ CH$_3$OH + CH$_2$&1.5e-7&-0.5&0 \\
CH$_3$OH$_2^+$ + e$^-$ $\rightarrow$ CH$_2$ + H$_2$O + H &.87e-7&-0.59&0 \\
CH$_5^+$ + e$^-$ $\rightarrow$ CH$_2$ + H$_2$ + H &4.76e-8&-0.52&0 \\
H$_2$CNO$^+$ + e$^-$ $\rightarrow$ CH$_2$ + NO &1.5e-7&-0.50&0 \\
H$_2$CO$^+$ + e$^-$ $\rightarrow$ CH$_2$ + O &2.5e-8&-0.70&0 \\
H$_3$CO$^+$ + e$^-$ $\rightarrow$ CH$_2$ + OH &4.2e-8&-0.78&0 \\
SiCH$_2^+$ + e$^-$ $\rightarrow$ Si + CH$_2$ &2.e-7&-0.5&0 \\
C$^+$ + CH$_3$CH$_3$ $\rightarrow$ C$_2$H$_4^+$ + CH$_2$ &1.16e-10&0&0 \\
C$_2$H$_2^+$ + C$_2$H$_3$ $\rightarrow$ C$_2$H$_4^+$ + CH$_2$ &3.3e-10&-0.5&0 \\
C$_2$H + C$_2$H$_4^+$ $\rightarrow$ CH$_2$CCH$^+$ + CH$_2$ &5.e-10&0&0 \\
CH$^+$ + CH$_3$OH $\rightarrow$ H$_3$CO$^+$ + CH$_2$ &2.9e-10&-0.5&0 \\
CH$^+$ + H$_2$CO $\rightarrow$ HCO$^+$ + CH$_2$ &9.6e-10&-0.5&0 \\
CH$_3^+$ + CH$_3$COCH$_3$ $\rightarrow$ CH$_3$COCH$_4^+$ + CH$_2$ &2.4e-10&-0.5&0 \\
H$^+$ + CH$_2$NH $\rightarrow$ NH$_2^+$ + CH$_2$ &1.e-9&-0.5&0 \\
He$^+$ + C$_2$H$_4$ $\rightarrow$ NH$_2^+$ + CH$_2$ &4.8e-10&0&0 \\
He$^+$ + H$_2$CCO $\rightarrow$ CO$^+$ + CH$_2$ + He &1.e-9&-0.5&0 \\
He$^+$ + CH$_2$PH $\rightarrow$ PH$^+$ + CH$_2$ + He &5.e-10&0&0\\
He$^+$ + CH$_2$PH $\rightarrow$ S$^+$ + CH$_2$ + He &5.e-10&0&0\\
He$^+$ + H$_2$CS $\rightarrow$ PH$^+$ + CH$_2$ + He &8.1e-10&-0.5&0 \\
He$^+$ + SiCH$_3$ $\rightarrow$ SiH$^+$ + CH$_2$ + He &1.e-9&-0.5&0 \\
N$^+$ + H$_2$CO $\rightarrow$ NO$^+$ + CH$_2$ &2.9e-10&-0.5&0 \\
CH$_3^+$ + O$_2$ $\rightarrow$ O$_2$H + CH$_2$ &5.3e-12&0&34975.0 \\
H + HCO $\rightarrow$ O + CH$_2$ &6.61e-11&0&51598.0 \\
O + C$_2$H$_4$ $\rightarrow$ H$_2$CO + CH$_2$ &4.2e-11&0&2520.0 \\
O + CH$_2$CCH$_2$ $\rightarrow$ CH$_2$ + CH$_2$CO &1.27e-11&-0.09&798.7 \\
H$_2$CCO + PHOTON $\rightarrow$ CO + CH$_2$ &1.4e-9&0&2.7 \\
CH$_2$PH + PHOTON $\rightarrow$ CH$_2$ + PH &9.54e-10&0&1.8 \\
CH$_2$CHCH$_2$ + PHOTON $\rightarrow$ C$_2$H$_4$ + CH$_2$ &1.13e-9&0&1.6 \\
C$^-$ + CH$_2$ $\rightarrow$ C$_2$H$_2$ + e$^-$ &5.e-10&0&0 \\
CH$_2$ + O$^-$ $\rightarrow$ H$_2$CO + e$^-$ &5.e-10&0&0 \\
CH$_2$ + H$_2$CO$^+$ $\rightarrow$ H$_2$CO + CH$_2^+$ &4.3e-10&0&0 \\
CH$_2$ + C$_2$H$^+$ $\rightarrow$ C$_3$H$_2^+$ + H &4.4e-10&0&0 \\
CH$_2$ + C$_2$H$_2^+$ $\rightarrow$ CH$_2$CCH$^+$ + H &8.8e-10&0&0 \\
CH$_2$ + H$_2$CO$^+$ $\rightarrow$ HCO + CH$_3^+$ &4.3e-10&0&0 \\
CH$_2$ + HN$_2^+$ $\rightarrow$ N$_2$ + CH$_3^+$ &8.6e-10&0&0 \\
CH$_2$ + O$_2^+$ $\rightarrow$ H$_2$CO$^+$ + O &4.3e-10&0&0 \\
CH$_2$ + O$_2$H$^+$ $\rightarrow$ O$_2$ + CH$_3^+$ &8.5e-10&0&0 \\
CH$_2$ + Si$^+$ $\rightarrow$ HCSi$^+$ + H&8.7e-10&0&0 \\
CH$_2$ + SiO$^+$ $\rightarrow$ H$_2$CO + Si$^+$&8.2e-10&0&0 \\
CH$_2$ + C$_2$H$_2$ $\rightarrow$ CH$_2$CCH + H&2.64e-10&-0.9&0 \\
CH$_2$ + C$_2$H$_2$ $\rightarrow$ CH$_3$CHCH$_2$&8.e-10&-2.&350 \\
CH$_2$ + CH$_2$ $\rightarrow$ C$_2$H$_3$ + H&3.32e-11&0&0 \\
CH$_2$ + H$_2$CO $\rightarrow$ HCO + CH$_3$&3.3e-13&0&3270\\
CH$_2$ + HCO $\rightarrow$ CO + CH$_3$&3.e-11&0&0 \\
CH$_2$ + NO$_2$ $\rightarrow$ H$_2$CO + NO&6.91e-11&0&0 \\
CH$_2$ + NO $\rightarrow$ H$_2$CO + N&2.7e-12&0&3500 \\
CH$_2$ + NO $\rightarrow$ HCNO + H&3.65e-11&0&0 \\
CH$_2$ + O$_2$ $\rightarrow$ CO$_2$ + H$_2$&2.92e-11&-3.3&1443.\\
CH$_2$ + O$_2$  $\rightarrow$ CO$_2$ + H + H&3.65e-11&-3.3&1443.\\
CH$_2$ + O$_2$ $\rightarrow$  H$_2$CO + O&3.65e-11&-3.3&1443.\\
CH$_2$ + O$_2$ $\rightarrow$ HCO + OH&4.1e-11&0&750.\\
CH$_2$ + O $\rightarrow$  HCO + H&5.01e-11&0&0\\
CH$_2$ + S $\rightarrow$ CS + H2&1.e-10&0&0 \\
CH$_2$ + S $\rightarrow$ HCS + H&1.e-10&0&0 \\
CH$_2$ + Si $\rightarrow$ HCSi + H&1.e-10&0&0 \\
\enddata
\tablecomments{Rate coefficients are from RATE12  }
\end{deluxetable*}
\startlongtable
\begin{deluxetable*}{lccc}
\tablecaption{List of updated chemical reactions involving CH$_2$\label{tab:14}}
\tablehead{
\colhead{Reactions}&\colhead{$\alpha$}&\colhead{$\beta$}&\colhead{$\gamma$}}
\startdata
O + C$_2$H$_2$ $\rightarrow$ CO + CH$_2$&1.15e-12&1.4&1110 \\
CH$_2$ + PHOTON $\rightarrow$ CH + H&5.8e-10&0&2. \\
CH$_3$ + PHOTON $\rightarrow$ CH$_2$ + H&1.35e-10&0&2.3\\
CH$_4$ + PHOTON $\rightarrow$ CH$_2$ + H2&9.8e-10&0&2.6 \\
e$^-$ + CH$_3^+$ $\rightarrow$ CH$_2$ + H&7.75e-8&-0.5&0\\
N + CH$_2$ $\rightarrow$ HNC + H&3.95e-11&0.17&0 \\
C + CH$_2$ $\rightarrow$ C$_2$H + H&1.e-10&0&0 \\
CH + HNO $\rightarrow$ NO + CH$_2$&1.73e-11&0&0\\
CH$_2$ + NO $\rightarrow$ HCN + OH&3.65e-12&0&0\\
CH$_3^+$ + NH$_3$ $\rightarrow$ NH$_4^+$ + CH$_2$&3.4e-10&-0.5&0\\
\enddata
\tablecomments{Rate coefficients are from RATE12  }
\end{deluxetable*}
[]
\startlongtable
\begin{deluxetable*}{lccc}
\tablecaption{List of added chemical reactions involving ArH$^+$\label{tab:15}}
\tablehead{
\colhead{Reactions}&\colhead{$\alpha$}&\colhead{$\beta$}&\colhead{$\gamma$}}
\startdata
ArH$^+$ + H$_2$ $\rightarrow$ Ar + H$_3^+$ &8.e-10 & 0 & 0\\ 
ArH$^+$ + CO $\rightarrow$ Ar + HCO$^+$ &1.25e-9&0&0\\ 
ArH$^+$ + C $\rightarrow$ Ar + CH$^+$ &8.e-10 & 0 & 0\\ 
ArH$^+$ + O $\rightarrow$ Ar + OH$^+$ &8.e-10&0&0\\ 
ArH$^+$ + PHOTON $\rightarrow$ Ar$^+$ + H &4.2e-12&0&3.27\\
\enddata
\tablecomments{For references, see section 2.6} 
\end{deluxetable*}
\startlongtable
\begin{deluxetable*}{lccc}
\tablecaption{List of updated chemical reactions involving ArH$^+$\label{tab:16}}
\tablehead{
\colhead{Reactions}&\colhead{$\alpha$}&\colhead{$\beta$}&\colhead{$\gamma$}
}
\startdata
H$_2^+$ + Ar $\rightarrow$ ArH$^+$ + H&1.e-9&0&0\\ 
H$_3^+$ + Ar $\rightarrow$ ArH$^+$ + H$_2$&8.5e-10&0&6400\\ 
H$_2$ + Ar$^+$ $\rightarrow$ ArH$^+$ + H & 8.4e-10& 0.16 & 0\\
ArH$^+$ + e$^-$ $\rightarrow$ H + Ar & 1.0e-9& 0& 0\\
\enddata
\tablecomments{For references, see section 2.6} 
\end{deluxetable*}
\startlongtable
\begin{deluxetable*}{lccc}
\tablecaption{List of added chemical reactions involving CN\label{tab:17}}
\tablehead{
\colhead{Reactions}&\colhead{$\alpha$}&\colhead{$\beta$}&\colhead{$\gamma$}}
\startdata
O$^-$ + CN $\rightarrow$ CN$^-$ + O &1.e-9&0&0 \\
OH$^-$ + CN $\rightarrow$ CN$^-$ + OH & 1.e-9&0&0\\
C$_2$H$^+$ + CN $\rightarrow$ C$_3$N$^+$ + H & 9.e-10&-0.5&0\\
CH$^+$ + CN $\rightarrow$ C$_2$N$^+$ + H & 5.5e-10&-0.5&0\\
CH$^+$ + CN $\rightarrow$ CNC$^+$ + H & 5.5e-10&-0.5.&0\\
CH$_3^+$ + CN $\rightarrow$ CH$_2$CN$^+$ + H & 1.1e-9&-0.5&0\\
CN + C$_2$H$_2^+$ $\rightarrow$ HC$_3$N$^+$ + H & 9.e-10&-0.5&0\\
CN + O$_2$H$^+$ $\rightarrow$ O$_2$ + HCN$^+$ & 8.6e-10&-0.5&0\\
NH$_2^+$ + CN $\rightarrow$ H$_2$NC$^+$ + N & 1.e-10&-0.5&0\\
CN + C$_2$H$_4$ $\rightarrow$ C$_2$H$_3$ + HCN & 1.25e-10&0.7&30\\
CN + C$_2$H$_4$ $\rightarrow$ CH$_2$CHCN + H & 1.25e-10&0.7&30\\
CN + C$_6$H$_2$ $\rightarrow$ HC$_7$N + H & 2.72e-10&-0.52&19\\
CN + C$_8$H$_2$ $\rightarrow$ HC$_9$N + H & 2.72e-10&-0.52&19\\
CN + CH$_2$CCH$_2$ $\rightarrow$ CH$_3$C$_3$N + H & 4.1e-10&0&0\\
CN + CH$_3$CCH $\rightarrow$ CH$_3$C$_3$N + H & 4.1e-10&0&0\\
CN + CH$_3$CH$_3$ $\rightarrow$ C$_2$H$_5$ + HCN & 2.34e-11&1.02&-35\\
CN + CH$_3$CHCH$_2$ $\rightarrow$ CH$_2$CHCN + CH$_3$ & 2.23e-10&0&0\\
CN + CN $\rightarrow$ N$_2$ + C$_2$ & 2.66e-9&0&21638.0\\
CN + H$_2$CO $\rightarrow$ HCO + HCN & 2.6e-10&-0.47&826.0\\
CN + HC$_4$H $\rightarrow$ HC$_5$N + H &2.72e-10&-0.52&19.0\\
CN + HCN $\rightarrow$ NCCN + H &2.5e-17&1.71&770.0\\
CN + HCO $\rightarrow$ CO + HCN & 1.e-10&0&0\\
CN + NO$_2$ $\rightarrow$ NO + OCN & 7.02e-11&-0.27&8.30 \\
CN + SiH$_4$ $\rightarrow$ HCN + SiH$_3$ &2.2e-10&0&0\\
CN + NH$_3$ $\rightarrow$ NH$_2$CN + H &1.38e-11&-1.14&0 \\
C$^-$ + N $\rightarrow$ CN + e$^-$ & 5.0e-10&0&0\\
N + C$_4$H$^-$ $\rightarrow$ CN + C$_3$H$^+$ + e$^-$ &3.0e-12&0&0 \\
C$^+$ + NCCN $\rightarrow$ C$_2$N$^+$ + CN & 2.0e-10&0&0\\
CN$^+$ + CO$_2$ $\rightarrow$ CO$_2^+$ + CN &3.0e-10&0&0 \\
CN$^+$ + H$_2$CO $\rightarrow$ H$_2$CO$^+$ + CN & 5.2e-10&-0.5&0\\
CN$^+$ + HCO $\rightarrow$ HCO$^+$ + CN &3.7e-10&-0.5&0 \\
C$_2$N$^+$ + e$^-$ $\rightarrow$ CN$^+$ + C &1.5e-7&-0.5&0 \\
C$_2$N$_2$+ + e$^-$ $\rightarrow$ CN + CN &1.5e-7&-0.5&0 \\
C$_3$N$^+$ + e$^-$ $\rightarrow$ CN + C$_2$ & 3.e-7&-0.5&0\\
CH$_2$CN$^+$ + e$^-$ $\rightarrow$ CN + CH$_2$ &1.5e-7&-0.5&0\\
CH$_4$N$^+$ + e$^-$ $\rightarrow$ CN + H$_2$ + H$_2$ & 3.e-8&-0.5&0\\
CNC$^+$ + e$^-$ $\rightarrow$ CN + C & 3.8e-7&-0.6&0\\
H$_2$NC$^+$ + e$^-$ $\rightarrow$ CN + H$_2$ & 1.8e-8&-0.5&0\\
H$_2$OCN$^+$ + e$^-$ $\rightarrow$ H$_2$O + CN & 1.5e-7&-0.5&0\\
HC$_3$N$^+$ + e$^-$ $\rightarrow$ CN + C$_2$H & 3.6e-7&-0.6&0\\
HC$_3$NN$^+$ + e$^-$ $\rightarrow$ C$_2$H$_2$ + CN & 7.2e-7&-0.58&0\\
HC$_7$N$^+$ + e$^-$ $\rightarrow$ C$_6$H + CN & 1.e-6&-0.3&0\\
HC$_9$N$^+$ + e$^-$ $\rightarrow$ C$_8$H + CN & 1.e-6&-0.3&0\\
HOCN$^+$ + e$^-$ $\rightarrow$ OH + CN & 1.5e-7&-0.5&0\\
HONC$^+$ + e$^-$ $\rightarrow$ OH + CN & 1.5e-7&-0.5&0\\
NCCNH$^+$ + e$^-$ $\rightarrow$ HNC + CN &3.e-7&-0.5&0 \\
SiNC$^+$ + e$^-$ $\rightarrow$ Si + CN &3.e-7&-0.5&0 \\
C$^+$ + CH$_2$CHCN $\rightarrow$ C$_3$H$_3^+$ + CN &1.8e-9&-0.5&0\\
C$^+$ + CH$_3$C$_3$N $\rightarrow$ C$_4$H$_3^+$ + CN & 5.e-9&0.&0\\
C$^+$ + CH$_3$C$_5$N $\rightarrow$ C$_6$H$_3^+$ + CN & 5.e-9&-0.5&0\\
C$^+$ + CH$_3$C$_7$N $\rightarrow$ C$_8$H$_3^+$ + CN & 5.e-9&-0.5&0\\
C$^+$ + CH$_3$CN $\rightarrow$ C$_2$H$_3^+$ + CN &5.6e-9&-0.5&0\\
C$^+$ + CNO $\rightarrow$ CO$^+$ + CN & 8.98e-9&-0.5&0\\
C$^+$ + HC$_5$N $\rightarrow$ C$_5$H$^+$ + CN & 6.e-9&-0.5&0\\
C$^+$ + HC$_7$N $\rightarrow$ C$_7$H$^+$ + CN & 6.e-9&-0.5&0\\
C$^+$ + HC$_9$N $\rightarrow$ C$_9$H$^+$ + CN & 6.e-9&-0.5&0\\
C$_2$H$_2^+$ + CH$_3$CN $\rightarrow$ C$_3$H$_5^+$ + CN &1.06e-9&-0.5&0\\
CH$_4$ + HC$_3$N$^+$ $\rightarrow$  C$_3$H$_5^+$ + CN &4.57e-10&0.&0\\
H$^+$ + HOCN $\rightarrow$ H$_2$O$^+$ + CN & 6.25e-9&-0.5&0\\
H$^+$ + HONC $\rightarrow$ H$_2$O$^+$ + CN & 1.45e-8&-0.5&0\\
H + C$_2$N$_2^+$ $\rightarrow$ HCN$^+$ + CN & 6.2e-10&0&0\\
HCN$^+$ + CO$_2$ $\rightarrow$ HCO$_2^+$ + CN & 2.1e-10&0&0\\
HCN$^+$ + H$_2$CO $\rightarrow$ H$_3$CO$^+$ + CN &1.e-9&-0.5&0\\
HCN$^+$ + HCO $\rightarrow$ H$_2$CO$^+$ + CN &3.7e-10&-0.5&0\\
HCN + C$_2$N$_2^+$ $\rightarrow$ NCCNH$^+$ + CN &2.3e-9&-0.5&0\\
HCN + C$_4$H$^+$ $\rightarrow$ C$_4$H$_2^+$ + CN & 9.45e-11&-0.5&0\\
HCN + C5$^+$ $\rightarrow$ C$_5$H$^+$ + CN & 3.3e-10&-0.5&0\\
He$^+$ + C$_2$N $\rightarrow$ CN + C$^+$ + He & 8.e-9&0&0\\
He$^+$ + C$_3$N $\rightarrow$ CN + C$_2$+ + He & 8.e-9&-0.5&0\\
He$^+$ + C$_5$N $\rightarrow$ C$_4^+$ + CN + He & 3.e-9&-0.5&0\\
He$^+$ + C$_7$N $\rightarrow$ C$_6^+$ + CN + He & 3.e-9&-0.5&0\\
He$^+$ + C$_9$N $\rightarrow$ C$_8^+$ + CN + He & 4.e-9&-0.5&0\\
He$^+$ + CH$_2$CHCN $\rightarrow$ C$_2$H$_3^+$ + CN + He & 5.25e-10&-0.5&0\\
He$^+$ + CH$_2$CN $\rightarrow$ CN + CH$_2+$ + He & 3e-9&-0.5&0\\
He$^+$ + CH$_3$CN $\rightarrow$ CN + CH$_3^+$ + He & 1.2e-9&-0.5&0\\
He$^+$ + CNO $\rightarrow$ O$^+$ + CN + He & 1.99e-8&-0.5&0\\
He$^+$ + HC$_5$N $\rightarrow$ C$_4$H$^+$ + CN + He &1.e-9&-0.5&0\\
He$^+$ + HC$_7$N $\rightarrow$ C$_6$H$^+$ + CN + He & 7.e-9&-0.5&0\\
He$^+$ + HC$_9$N $\rightarrow$ C$_8$H$^+$ + CN + He & 7.e-9&-0.5&0\\
He$^+$ + NCCN $\rightarrow$ CN$^+$ + CN + He & 2.7e-9&0&0\\
He$^+$ + NH$_2$CN $\rightarrow$ NH$_2^+$ + CN + He & 1.e-9&0&0\\
N + C$_10^-$ $\rightarrow$ CN + C$_9^-$ & 1.e-10&0&0\\
N + C$_3^-$ $\rightarrow$ CN + C$_2^-$ &5.e-11&0&0 \\
N + C$_3$H$^-$ $\rightarrow$ CN + C$_2$H$^-$ & 2.5e-12&0&0\\
N + C$_3$H$_2^+$ $\rightarrow$ C$_2$H$_2^+$ + CN & 3.74e-11&0&0\\
N + C$_4^-$ $\rightarrow$ CN + C$_3^-$ & 5.e-11&0&0\\
N + C5$^-$ $\rightarrow$ CN + C$_4^-$ & 2.7e-11&0&0\\
N + C$_5$H$^-$ $\rightarrow$ CN + C4H- & 2.5e-12&0&0\\
N + C6$^-$ $\rightarrow$ CN + C5$^-$ & 1.5e-10&0&0\\
N + C$_7^-$ $\rightarrow$ CN + C$_6^-$ & 5.06e-11&0&0\\
N + C$_8^-$ $\rightarrow$ CN + C$_7^-$ &1.e-10&0&0\\
N + C$_9^-$ $\rightarrow$ CN + C$_8^-$ & 5.e-11&0&0\\
N + SiC$^+$ $\rightarrow$ Si$^+$ + CN & 7.7e-10&0&0\\
Si$^+$ + NCCN $\rightarrow$ SiNC$^+$ + CN & 1.5e-10&0&0\\
CN$^-$ + C$^+$ $\rightarrow$ C + CN &7.51e-8&-0.5&0\\
CN$^-$ + C$_2$H$_2^+$ $\rightarrow$ CN + C$_2$H$_2$ & 7.51e-8&-0.5&0\\
CN$^-$ + C$_2$H$_3^+$ $\rightarrow$ CN + C$_2$H$_3$ & 7.51e-8&-0.5&0\\
CN$^-$ + C$_4$H$_2^+$ $\rightarrow$ CN + HC$_4$H & 7.51e-8&-0.5&0\\
CN$^-$ + C$_4$H$_3^+$ $\rightarrow$ CN + C$_4$H$_3$ & 7.51e-8&-0.5&0\\
CN$^-$ + C$_4$S$^+$ $\rightarrow$ CN + C$_4$S & 7.51e-8&-0.5&0\\
CN$^-$ + CH$_2$CCH$^+$ $\rightarrow$ CN + CH$_2$CCH & 7.51e-8&-0.5&0\\
CN$^-$ + CH$_3^+$ $\rightarrow$ CN + CH$_3$ & 7.51e-8&-0.5&0\\
CN$^-$ + CNC$^+$ $\rightarrow$ CN + C2N & 7.51e-8&-0.5&0\\
CN$^-$ + Fe$^+$ $\rightarrow$ CN + Fe & 7.51e-8&-0.5&0\\
CN$^-$ + H$^+$ $\rightarrow$ CN + H & 7.51e-8&-0.5&0\\
CN$^-$ + H2CO$^+$ $\rightarrow$ CN + H$_2$CO & 7.51e-8&-0.5&0\\
CN$^-$ + H$_2$S$^+$ $\rightarrow$ CN + H$_2$S &7.51e-8&-0.5&0\\
CN$^-$ + H$_3^+$ $\rightarrow$ CN + H$_2$ + H & 7.51e-8&-0.5&0\\
CN$^-$ + H$_3$O$^+$ $\rightarrow$ CN + H + H$_2$O &7.51e-8&-0.5&0\\
CN$^-$ + HC$_2$S$^+$ $\rightarrow$ CN + C$_2$S + H &7.51e-8&-0.5&0\\
CN$^-$ + HCNH$^+$ $\rightarrow$ CN + HCN + H &3.76e-8&-0.5&0\\
CN$^-$ + HCNH$^+$ $\rightarrow$ CN + HNC + H &3.76e-8&-0.5&0\\
CN$^-$ + HCO$^+$ $\rightarrow$ CN + H + CO &3.76e-8&-0.5&0\\
CN$^-$ + HCO$^+$ $\rightarrow$ CN + HCO & 7.51e-8&-0.5&0\\
CN$^-$ + Mg$^+$ $\rightarrow$ CN + Mg &7.51e-8&-0.5&0\\
CN$^-$ + N$^+$ $\rightarrow$ CN + N &7.51e-8&-0.5&0\\
CN$^-$ + N$_2$H$^+$ $\rightarrow$ CN + N$_2$ + H &7.51e-8&-0.5&0 \\
CN$^-$ + NH$_3^+$ $\rightarrow$ CN + NH$_3$ & 7.51e-8&-0.5&0\\
CN$^-$ + NH$_4^+$ $\rightarrow$ CN + NH$_3$ + H & 7.51e-8&-0.5&0\\
CN$^-$ + NO$^+$ $\rightarrow$ CN + NO & 7.51e-8&-0.5&0\\
CN$^-$ + Na$^+$ $\rightarrow$ CN + Na & 7.51e-8&-0.5&0\\
CN$^-$ + O$^+$ $\rightarrow$ CN + O &7.51e-8&-0.5&0\\
CN$^-$ + S$^+$ $\rightarrow$ CN + S & 7.51e-8&-0.5&0\\
CN$^-$ + SO$^+$ $\rightarrow$ CN + SO & 7.51e-8&-0.5&0\\
CN$^-$ + Si$^+$ $\rightarrow$ CN + Si & 7.51e-8&-0.5&0\\
CN$^-$ + SiO$^+$ $\rightarrow$ CN + SiO &7.51e-8&-0.5&0\\
CN$^-$ + SiOH$^+$ $\rightarrow$ CN + SiO + H &7.51e-8&-0.5&0\\
CN$^-$ + SiS$^+$ $\rightarrow$ CN + SiS &7.51e-8&-0.5&0\\
C$_4$H + NCCN $\rightarrow$ HC$_5$N + CN & 2.0e-13&0&0\\
C + C$_2$N $\rightarrow$ C$_2$ + CN &1.e-10&0&0\\
C + C$_3$N $\rightarrow$ C$_3$ + CN &1.e-13&0&0\\
C + C$_5$N $\rightarrow$ C$_5$ + CN & 1.e-13&0&0\\
C + C$_7$N $\rightarrow$  C$_7$ + CN &1.e-13&0&0\\
C + C$_9$N $\rightarrow$ C$_9$ + CN & 1.e-13&0&0\\
C + NCCN $\rightarrow$ CN + C$_2$N &3.e-11&0&0\\
H + CNO $\rightarrow$ OH + CN &1.e-10&0&0\\
H + HCN $\rightarrow$ CN + H$_2$ &6.2e-10&0&12500\\
H + NCCN $\rightarrow$ HCN + CN & 1.48e-10&0&3588\\
H + OCN $\rightarrow$ OH + CN &1.e-10&0&0\\
N + C$_2$N $\rightarrow$ CN + CN &1.e-10&0&0\\
N + C$_2$O $\rightarrow$ CN + CO &5.5e-10&0&0\\
N + C$_3$ $\rightarrow$ CN + C$_2$ &1.e-13&0&0\\
N + C$_3$N $\rightarrow$ CN + C$_2$N &1.e-10&0&0\\
N + C$_4$ $\rightarrow$ C$_3$ + CN &1.e-10&0&0\\
N + C$_4$N $\rightarrow$ CN + C$_3$N &1.e-10&0&0\\
N + C$_5$ $\rightarrow$ CN + N$_4$ &1.e-13&0&0\\
N + C$_5$N $\rightarrow$ CN + C$_4$N &1.e-10&0&0\\
N + C$_6$ $\rightarrow$ CN + C$_5$ &1.e-10&0&0\\
N + C$_6$H $\rightarrow$ CN + C$_5$H &1.e-13&0&0\\
N + C$_7$ $\rightarrow$ CN + C$_6$ &1.e-13&0&0\\
N + C$_8$ $\rightarrow$ CN + C$_7$ &1.e-10&0&0\\
N + C$_8$H $\rightarrow$ CN + C$_7$H& 1.e-10&0&0\\
N + C$_9$ $\rightarrow$ CN + C$_8$ & 1.e-13&0&0\\
N + SiC  $\rightarrow$ Si + CN &5.e-1&0&0\\
O + C$_2$N $\rightarrow$ CO + CN &6.e-12&0&0\\
C$_2$H$_5$CN + PHOTON $\rightarrow$ CN + C$_2$H$_5$ & 3.4e-9&0&2.\\
C$_2$N + PHOTON $\rightarrow$ CN + C &5.e-10&0&1.7\\
C$_3$N + PHOTON $\rightarrow$ CN + C$_2$ & 5.e-10&0&1.8\\
C$_4$N + PHOTON $\rightarrow$ C$_3$ + CN &5.e-10&0&1.7\\
C$_5$N + PHOTON $\rightarrow$ C$_4$ + CN &5.e-10&0&1.7\\
C$_7$N + PHOTON $\rightarrow$ C$_6$ + CN &5.e-10&0&1.7\\
C$_9$N + PHOTON $\rightarrow$ C$_8$ + CN &1.e-9&0&1.7\\
CH$_2$CHCN + PHOTON $\rightarrow$ C$_2$H$_3$ + CN &1.e-10&0.&1.7\\
CH$_2$CN + PHOTON $\rightarrow$ CH$_2$ + CN & 1.56e-9&0&1.9\\
CH$_3$CN + PHOTON $\rightarrow$ CN + CH$_3$ &2.5e-9&0&2.6\\
CN$^-$ + PHOTON $\rightarrow$ CN + e$^-$ &2.22e-9&0&2.\\
CNO + PHOTON $\rightarrow$ CN + O &1.e-11&0&2.\\
HC$_5$N + PHOTON $\rightarrow$ C$_4$H + CN &5.e-10&.&1.8\\
HC$_7$N + PHOTON $\rightarrow$ C$_6$H + CN &1.e-9&0&1.7\\
HC$_9$N + PHOTON $\rightarrow$ C$_8$H + CN &1.e-9&0&1.7\\
HOCN + PHOTON $\rightarrow$ OH + CN &1.e-9&0&1.7\\
HONC + PHOTON $\rightarrow$ OH + CN &1.e-9&0&1.7\\
NCCN + PHOTON $\rightarrow$ CN + CN &4.7e-11&0&2.6\\
NH$_2$CN + PHOTON $\rightarrow$ NH$_2$ + CN & 1.38e-9&0&1.7\\
SiNC + PHOTON $\rightarrow$ CN + Si &1.e-9&0&1.7\\
\enddata
\tablecomments{Rate coefficients are from RATE12  }
\end{deluxetable*}
\startlongtable
\begin{deluxetable*}{lccc}
\tablecaption{List of updated chemical reactions involving CN\label{tab:18}}
\tablehead{
\colhead{Reactions}&\colhead{$\alpha$}&\colhead{$\beta$}&\colhead{$\gamma$}
}
\startdata
CN + NO $\rightarrow$ N$_2$ + CO &1.6e-13&0&0 \\
CN + O$_2$ $\rightarrow$ NO + CO &5.12e-12&-.49&-5.2\\
H$_2$+ + CN $\rightarrow$ HCN$^+$ + H &1.2e-9&-0.5&0\\
H$_3+$ + CN  $\rightarrow$ HCN$^+$ + H$_2$ & 2.0e-9&-0.5&0\\  
He$^+$ + CN $\rightarrow$ N + C$^+$ + He &8.8e-10&-0.5&0\\
He$^+$ + CN $\rightarrow$ N$^+$ + C + He &8.8e-10&0&0\\ 
H$_2^+$ + CN $\rightarrow$ CN$^+$ + H$_2$ & 1.2e-9&-0.5&0\\
OH$^+$ + CN $\rightarrow$ HCN$^+$ + O &1.2e-9&-0.5&0\\
CN + HNO$^+$ $\rightarrow$ NO + HCN$^+$ &8.7e-10&-0.5&0\\ 
CN + N$_2^+$ $\rightarrow$ N$_2$ + CN$^+$ & 1.0e-10&-0.5&0\\ 
O + CN $\rightarrow$ NO + C &5.37e-11&0&13800.0\\
NH$_3$ + CN $\rightarrow$ HCN + NH$_2$ &2.75e-11&-1.14&0\\
CN + HNO $\rightarrow$ NO + HCN &3.0e-11&0&0\\ 
NH$^+$ + CN $\rightarrow$ HCN$^+$ + N &1.6e-9&-0.5&0\\
O$^+$ + CN $\rightarrow$ NO$^+$ + C &1.0e-9&0&0\\
NH + HCN$^+$ $\rightarrow$ CN + NH$_2$+ & 6.5e-10&-0.5&0\\
CH + HCN$^+$ $\rightarrow$ CN + CH$_2$+ &6.3e-10&-0.5&0\\
N$^+$ + CN $\rightarrow$ CN$^+$ + N &1.1e-9&-0.5&0\\
C + NH $\rightarrow$ CN + H &1.2e-10&0&0\\
C + NO $\rightarrow$ CN + O &6.0e-11&-0.16&0\\
C + NS $\rightarrow$ S + CN &1.5e-10&-0.16&0\\
HCN$^+$ + HNC $\rightarrow$ HCNH$^+$ + CN &1.0e-9&-0.5&0\\
HCN$^+$ + HCN $\rightarrow$ HCNH$^+$ + CN &1.6e-9&-0.5&0\\
C + N $\rightarrow$ CN + PHOTON &5.72e-19&0.37&51.0\\
CN$^+$ + HCN $\rightarrow$ HCN$^+$ + CN &1.79e-9&-0.5&0\\
OH + CN$^+$ $\rightarrow$ CN + OH$^+$ &6.4e-10&-0.5&0\\
NH$_2$ + CN$^+$ $\rightarrow$ CN + NH$_2^+$ &9.1e-10&-0.5&0\\
NH + CN$^+$ $\rightarrow$ CN + NH$^+$ &6.5e-10&-0.5&0\\
CH + CN$^+$ $\rightarrow$ CN + CH$^+$ &6.4e-10&-0.5&0\\
H$_2$O + HCN$^+$ $\rightarrow$ CN + H$_3$O$^+$ &1.8e-9&0&0\\
OH + HCN$^+$ $\rightarrow$ CN + H$_2$O$^+$ &6.3e-10&-0.5&0\\
NH$_2$ + HCN$^+$ $\rightarrow$ CN + NH$_3^+$ &9.0e-10&-0.5&0\\
H + CN$^+$ $\rightarrow$ CN + H$^+$ &6.4e-10&0&0 \\
N + C$_2$ $\rightarrow$ CN + C & 5.e-11&0.&0\\
C + OCN $\rightarrow$ CO + CN &1.0e-10&0.&0\\
CH + N$^+$ $\rightarrow$ CN$^+$ + H&3.6e-10&-0.5&0\\
C$_2$H$^+$ + HCN $\rightarrow$ C$_2$H$_2^+$ + CN &1.4e-9&-0.5&0\\
\enddata
\tablecomments{Rate coefficients are from RATE12  }
\end{deluxetable*}
\bibliography{update}{}

\begin{thebibliography}{}
\expandafter\ifx\csname natexlab\endcsname\relax\def\natexlab#1{#1}\fi

\bibitem[{{Abel} {et~al.}(2005){Abel}, {Ferland}, {Shaw}, \& {van
  Hoof}}]{2005ApJS..161...65A}
{Abel}, N.~P., {Ferland}, G.~J., {Shaw}, G., \& {van Hoof}, P.~A.~M. 2005,
  \apjs, 161, 65

\bibitem[{{Anicich}(1993)}]{1993JPCRD..22.1469A}
{Anicich}, V.~G. 1993, Journal of Physical and Chemical Reference Data, 22,
  1469

\bibitem[{{Baldwin} {et~al.}(1991){Baldwin}, {Ferland}, {Martin}, {Corbin},
  {Cota}, {Peterson}, \& {Slettebak}}]{1991ApJ...374..580B}
{Baldwin}, J.~A., {Ferland}, G.~J., {Martin}, P.~G., {et~al.} 1991, \apj, 374,
  580

\bibitem[{{Barlow} {et~al.}(2013){Barlow}, {Swinyard}, {Owen}, {Cernicharo},
  {Gomez}, {Ivison}, {Krause}, {Lim}, {Matsuura}, {Miller}, {Olofsson}, \&
  {Polehampton}}]{2013Sci...342.1343B}
{Barlow}, M.~J., {Swinyard}, B.~M., {Owen}, P.~J., {et~al.} 2013, Science, 342,
  1343

\bibitem[{{Bialy} {et~al.}(2019){Bialy}, {Neufeld}, {Wolfire}, {Sternberg}, \&
  {Burkhart}}]{2019ApJ...885..109B}
{Bialy}, S., {Neufeld}, D., {Wolfire}, M., {Sternberg}, A., \& {Burkhart}, B.
  2019, \apj, 885, 109

\bibitem[{{Chapillon} {et~al.}(2012){Chapillon}, {Dutrey}, {Guilloteau},
  {Pi{\'e}tu}, {Wakelam}, {Hersant}, {Gueth}, {Henning}, {Launhardt},
  {Schreyer}, \& {Semenov}}]{2012ApJ...756...58C}
{Chapillon}, E., {Dutrey}, A., {Guilloteau}, S., {et~al.} 2012, \apj, 756, 58

\bibitem[{{Cowie} \& {Songaila}(1986)}]{1986ARA&A..24..499C}
{Cowie}, L.~L., \& {Songaila}, A. 1986, \araa, 24, 499

\bibitem[{{Dagdigian}(2018)}]{2018MNRAS.475.5480D}
{Dagdigian}, P.~J. 2018, \mnras, 475, 5480

\bibitem[{{Dagdigian} {et~al.}(2019){Dagdigian}, {K{\l}os}, {Wolfire}, \&
  {Neufeld}}]{2019ApJ...872..203D}
{Dagdigian}, P.~J., {K{\l}os}, J., {Wolfire}, M., \& {Neufeld}, D.~A. 2019,
  \apj, 872, 203

\bibitem[{{Dalgarno}(2006)}]{2006PNAS..10312269D}
{Dalgarno}, A. 2006, Proceedings of the National Academy of Science, 103, 12269

\bibitem[{{Dalgarno} {et~al.}(1974){Dalgarno}, {de Jong}, {Oppenheimer}, \&
  {Black}}]{1974ApJ...192L..37D}
{Dalgarno}, A., {de Jong}, T., {Oppenheimer}, M., \& {Black}, J.~H. 1974,
  \apjl, 192, L37

\bibitem[{{Ferland} {et~al.}(2013){Ferland}, {Porter}, {van Hoof}, {Williams},
  {Abel}, {Lykins}, {Shaw}, {Henney}, \& {Stancil}}]{2013RMxAA..49..137F}
{Ferland}, G.~J., {Porter}, R.~L., {van Hoof}, P.~A.~M., {et~al.} 2013, \rmxaa,
  49, 137

\bibitem[{{Ferland} {et~al.}(2017){Ferland}, {Chatzikos}, {Guzm{\'a}n},
  {Lykins}, {van Hoof}, {Williams}, {Abel}, {Badnell}, {Keenan}, {Porter}, \&
  {Stancil}}]{2017RMxAA..53..385F}
{Ferland}, G.~J., {Chatzikos}, M., {Guzm{\'a}n}, F., {et~al.} 2017, \rmxaa, 53,
  385

\bibitem[{{Fish} {et~al.}(2003){Fish}, {Reid}, {Wilner}, \&
  {Churchwell}}]{2003ApJ...587..701F}
{Fish}, V.~L., {Reid}, M.~J., {Wilner}, D.~J., \& {Churchwell}, E. 2003, \apj,
  587, 701

\bibitem[{{Flower} \& {Pineau des Forets}(1998)}]{1998MNRAS.297.1182F}
{Flower}, D.~R., \& {Pineau des Forets}, G. 1998, \mnras, 297, 1182

\bibitem[{{Gay} {et~al.}(2012){Gay}, {Abel}, {Porter}, {Stancil}, {Ferland},
  {Shaw}, {van Hoof}, \& {Williams}}]{2012ApJ...746...78G}
{Gay}, C.~D., {Abel}, N.~P., {Porter}, R.~L., {et~al.} 2012, \apj, 746, 78

\bibitem[{{Gerin} {et~al.}(2010){Gerin}, {de Luca}, {Goicoechea}, {Herbst},
  {Falgarone}, {Godard}, {Bell}, {Coutens}, {Ka{\'z}mierczak}, {Sonnentrucker},
  {Black}, {Neufeld}, {Phillips}, {Pearson}, {Rimmer}, {Hassel}, {Lis},
  {Vastel}, {Boulanger}, {Cernicharo}, {Dartois}, {Encrenaz}, {Giesen},
  {Goldsmith}, {Gupta}, {Gry}, {Hennebelle}, {Hily-Blant}, {Joblin},
  {Ko{\l}os}, {Kre{\l}owski}, {Mart{\'\i}n-Pintado}, {Monje}, {Mookerjea},
  {Perault}, {Persson}, {Plume}, {Salez}, {Schmidt}, {Stutzki}, {Teyssier},
  {Yu}, {Contursi}, {Menten}, {Geballe}, {Schlemmer}, {Morris}, {Hatch},
  {Imram}, {Ward}, {Caux}, {G{\"u}sten}, {Klein}, {Roelfsema}, {Dieleman},
  {Schieder}, {Honingh}, \& {Zmuidzinas}}]{2010A&A...521L..16G}
{Gerin}, M., {de Luca}, M., {Goicoechea}, J.~R., {et~al.} 2010, \aap, 521, L16

\bibitem[{{Godard} {et~al.}(2010){Godard}, {Falgarone}, {Gerin}, {Hily-Blant},
  \& {de Luca}}]{2010A&A...520A..20G}
{Godard}, B., {Falgarone}, E., {Gerin}, M., {Hily-Blant}, P., \& {de Luca}, M.
  2010, \aap, 520, A20

\bibitem[{{Goddi} {et~al.}(2011){Goddi}, {Greenhill}, {Humphreys}, {Chandler},
  \& {Matthews}}]{2011ApJ...739L..13G}
{Goddi}, C., {Greenhill}, L.~J., {Humphreys}, E.~M.~L., {Chandler}, C.~J., \&
  {Matthews}, L.~D. 2011, \apjl, 739, L13

\bibitem[{{Goldsmith} \& {Langer}(1999)}]{1999ApJ...517..209G}
{Goldsmith}, P.~F., \& {Langer}, W.~D. 1999, \apj, 517, 209

\bibitem[{{Hacar} {et~al.}(2020){Hacar}, {Bosman}, \& {van
  Dishoeck}}]{2020A&A...635A...4H}
{Hacar}, A., {Bosman}, A.~D., \& {van Dishoeck}, E.~F. 2020, \aap, 635, A4

\bibitem[{{Harada} {et~al.}(2021){Harada}, {Mart{\'\i}n}, {Mangum}, {Sakamoto},
  {Muller}, {Tanaka}, {Nakanishi}, {Herrero-Illana}, {Yoshimura}, {M{\"u}hle},
  {Aladro}, {Colzi}, {Rivilla}, {Aalto}, {Behrens}, {Henkel}, {Holdship},
  {Humire}, {Meier}, {Nishimura}, {van der Werf}, \&
  {Viti}}]{2021ApJ...923...24H}
{Harada}, N., {Mart{\'\i}n}, S., {Mangum}, J.~G., {et~al.} 2021, \apj, 923, 24

\bibitem[{{H{\'e}brard} {et~al.}(2012){H{\'e}brard}, {Dobrijevic}, {Loison},
  {Bergeat}, \& {Hickson}}]{2012A&A...541A..21H}
{H{\'e}brard}, E., {Dobrijevic}, M., {Loison}, J.~C., {Bergeat}, A., \&
  {Hickson}, K.~M. 2012, \aap, 541, A21

\bibitem[{{Herbst} \& {van Dishoeck}(2009)}]{2009ARA&A..47..427H}
{Herbst}, E., \& {van Dishoeck}, E.~F. 2009, \araa, 47, 427

\bibitem[{{Hern{\'a}ndez Vera} {et~al.}(2017){Hern{\'a}ndez Vera}, {Lique},
  {Dumouchel}, {Hily-Blant}, \& {Faure}}]{2017MNRAS.468.1084H}
{Hern{\'a}ndez Vera}, M., {Lique}, F., {Dumouchel}, F., {Hily-Blant}, P., \&
  {Faure}, A. 2017, \mnras, 468, 1084

\bibitem[{{Indriolo} {et~al.}(2007){Indriolo}, {Geballe}, {Oka}, \&
  {McCall}}]{2007ApJ...671.1736I}
{Indriolo}, N., {Geballe}, T.~R., {Oka}, T., \& {McCall}, B.~J. 2007, \apj,
  671, 1736

\bibitem[{{Jacob} {et~al.}(2020){Jacob}, {Menten}, {Wyrowski}, {Winkel}, \&
  {Neufeld}}]{2020A&A...643A..91J}
{Jacob}, A.~M., {Menten}, K.~M., {Wyrowski}, F., {Winkel}, B., \& {Neufeld},
  D.~A. 2020, \aap, 643, A91

\bibitem[{{Jacob} {et~al.}(2021){Jacob}, {Menten}, {Wyrowski}, {Winkel},
  {Neufeld}, \& {Koribalski}}]{2021arXiv211215546J}
{Jacob}, A.~M., {Menten}, K.~M., {Wyrowski}, F., {et~al.} 2021, arXiv e-prints,
  arXiv:2112.15546

\bibitem[{{Karachentsev} {et~al.}(2003){Karachentsev}, {Grebel}, {Sharina},
  {Dolphin}, {Geisler}, {Guhathakurta}, {Hodge}, {Karachentseva}, {Sarajedini},
  \& {Seitzer}}]{2003A&A...404...93K}
{Karachentsev}, I.~D., {Grebel}, E.~K., {Sharina}, M.~E., {et~al.} 2003, \aap,
  404, 93

\bibitem[{{Landi} {et~al.}(2012){Landi}, {Del Zanna}, {Young}, {Dere}, \&
  {Mason}}]{2012ApJ...744...99L}
{Landi}, E., {Del Zanna}, G., {Young}, P.~R., {Dere}, K.~P., \& {Mason}, H.~E.
  2012, \apj, 744, 99

\bibitem[{{Le Gal} {et~al.}(2017){Le Gal}, {Herbst}, {Dufour}, {Gratier},
  {Ruaud}, {Vidal}, \& {Wakelam}}]{2017A&A...605A..88L}
{Le Gal}, R., {Herbst}, E., {Dufour}, G., {et~al.} 2017, \aap, 605, A88

\bibitem[{{Lee} {et~al.}(2008){Lee}, {Balakrishnan}, {Forrey}, {Stancil},
  {Shaw}, {Schultz}, \& {Ferland}}]{2008ApJ...689.1105L}
{Lee}, T.~G., {Balakrishnan}, N., {Forrey}, R.~C., {et~al.} 2008, \apj, 689,
  1105

\bibitem[{{Lindberg} {et~al.}(2011){Lindberg}, {Aalto}, {Costagliola},
  {P{\'e}rez-Beaupuits}, {Monje}, \& {Muller}}]{2011A&A...527A.150L}
{Lindberg}, J.~E., {Aalto}, S., {Costagliola}, F., {et~al.} 2011, \aap, 527,
  A150

\bibitem[{{Liszt} \& {Gerin}(2016)}]{2016A&A...585A..80L}
{Liszt}, H.~S., \& {Gerin}, M. 2016, \aap, 585, A80

\bibitem[{{Liszt} {et~al.}(2015){Liszt}, {Guzm{\'a}n}, {Pety}, {Gerin},
  {Neufeld}, \& {Gratier}}]{2015A&A...579A..12L}
{Liszt}, H.~S., {Guzm{\'a}n}, V.~V., {Pety}, J., {et~al.} 2015, \aap, 579, A12

\bibitem[{{Lykins} {et~al.}(2015){Lykins}, {Ferland}, {Kisielius}, {Chatzikos},
  {Porter}, {van Hoof}, {Williams}, {Keenan}, \&
  {Stancil}}]{2015ApJ...807..118L}
{Lykins}, M.~L., {Ferland}, G.~J., {Kisielius}, R., {et~al.} 2015, \apj, 807,
  118

\bibitem[{{Marinakis} {et~al.}(2015){Marinakis}, {Dean}, {K{\l}os}, \&
  {Lique}}]{2015PCCP...1721583M}
{Marinakis}, S., {Dean}, I.~L., {K{\l}os}, J., \& {Lique}, F. 2015, Physical
  Chemistry Chemical Physics (Incorporating Faraday Transactions), 17, 21583

\bibitem[{{Mathis} {et~al.}(1977){Mathis}, {Rumpl}, \&
  {Nordsieck}}]{1977ApJ...217..425M}
{Mathis}, J.~S., {Rumpl}, W., \& {Nordsieck}, K.~H. 1977, \apj, 217, 425

\bibitem[{{McElroy} {et~al.}(2013){McElroy}, {Walsh}, {Markwick}, {Cordiner},
  {Smith}, \& {Millar}}]{2013A&A...550A..36M}
{McElroy}, D., {Walsh}, C., {Markwick}, A.~J., {et~al.} 2013, \aap, 550, A36

\bibitem[{{Meijerink} \& {Spaans}(2005)}]{2005A&A...436..397M}
{Meijerink}, R., \& {Spaans}, M. 2005, \aap, 436, 397

\bibitem[{{Millar} {et~al.}(1997){Millar}, {Farquhar}, \&
  {Willacy}}]{1997A&AS..121..139M}
{Millar}, T.~J., {Farquhar}, P.~R.~A., \& {Willacy}, K. 1997, \aaps, 121, 139

\bibitem[{{Monje} {et~al.}(2013){Monje}, {Lis}, {Roueff}, {Gerin}, {De Luca},
  {Neufeld}, {Godard}, \& {Phillips}}]{2013ApJ...767...81M}
{Monje}, R.~R., {Lis}, D.~C., {Roueff}, E., {et~al.} 2013, \apj, 767, 81

\bibitem[{{Muller} {et~al.}(2016){Muller}, {Kawaguchi}, {Black}, \&
  {Amano}}]{2016A&A...589L...5M}
{Muller}, S., {Kawaguchi}, K., {Black}, J.~H., \& {Amano}, T. 2016, \aap, 589,
  L5

\bibitem[{{Neufeld} \& {Wolfire}(2016)}]{2016ApJ...826..183N}
{Neufeld}, D.~A., \& {Wolfire}, M.~G. 2016, \apj, 826, 183

\bibitem[{{Neufeld} \& {Wolfire}(2017)}]{2017ApJ...845..163N}
---. 2017, \apj, 845, 163

\bibitem[{{Neufeld} {et~al.}(2006){Neufeld}, {Schilke}, {Menten}, {Wolfire},
  {Black}, {Schuller}, {M{\"u}ller}, {Thorwirth}, {G{\"u}sten}, \&
  {Philipp}}]{2006A&A...454L..37N}
{Neufeld}, D.~A., {Schilke}, P., {Menten}, K.~M., {et~al.} 2006, \aap, 454, L37

\bibitem[{{Neufeld} {et~al.}(2010){Neufeld}, {Sonnentrucker}, {Phillips},
  {Lis}, {de Luca}, {Goicoechea}, {Black}, {Gerin}, {Bell}, {Boulanger},
  {Cernicharo}, {Coutens}, {Dartois}, {Kazmierczak}, {Encrenaz}, {Falgarone},
  {Geballe}, {Giesen}, {Godard}, {Goldsmith}, {Gry}, {Gupta}, {Hennebelle},
  {Herbst}, {Hily-Blant}, {Joblin}, {Ko{\l}os}, {Kre{\l}owski},
  {Mart{\'\i}n-Pintado}, {Menten}, {Monje}, {Mookerjea}, {Pearson}, {Perault},
  {Persson}, {Plume}, {Salez}, {Schlemmer}, {Schmidt}, {Stutzki}, {Teyssier},
  {Vastel}, {Yu}, {Cais}, {Caux}, {Liseau}, {Morris}, \&
  {Planesas}}]{2010A&A...518L.108N}
{Neufeld}, D.~A., {Sonnentrucker}, P., {Phillips}, T.~G., {et~al.} 2010, \aap,
  518, L108

\bibitem[{{Osterbrock} \& {Ferland}(2006)}]{2006agna.book.....O}
{Osterbrock}, D.~E., \& {Ferland}, G.~J. 2006, {Astrophysics of gaseous nebulae
  and active galactic nuclei}

\bibitem[{{Pellegrini} {et~al.}(2009){Pellegrini}, {Baldwin}, {Ferland},
  {Shaw}, \& {Heathcote}}]{2009ApJ...693..285P}
{Pellegrini}, E.~W., {Baldwin}, J.~A., {Ferland}, G.~J., {Shaw}, G., \&
  {Heathcote}, S. 2009, \apj, 693, 285

\bibitem[{{Peng} {et~al.}(2012){Peng}, {Zapata}, {Wyrowski}, {G{\"u}sten}, \&
  {Menten}}]{2012A&A...544L..19P}
{Peng}, T.~C., {Zapata}, L.~A., {Wyrowski}, F., {G{\"u}sten}, R., \& {Menten},
  K.~M. 2012, \aap, 544, L19

\bibitem[{{Plume} {et~al.}(2012){Plume}, {Bergin}, {Phillips}, {Lis}, {Wang},
  {Crockett}, {Caux}, {Comito}, {Goldsmith}, \&
  {Schilke}}]{2012ApJ...744...28P}
{Plume}, R., {Bergin}, E.~A., {Phillips}, T.~G., {et~al.} 2012, \apj, 744, 28

\bibitem[{{Porter} {et~al.}(2012){Porter}, {Ferland}, {Storey}, \&
  {Detisch}}]{2012MNRAS.425L..28P}
{Porter}, R.~L., {Ferland}, G.~J., {Storey}, P.~J., \& {Detisch}, M.~J. 2012,
  \mnras, 425, L28

\bibitem[{{Priestley} {et~al.}(2017){Priestley}, {Barlow}, \&
  {Viti}}]{2017MNRAS.472.4444P}
{Priestley}, F.~D., {Barlow}, M.~J., \& {Viti}, S. 2017, \mnras, 472, 4444

\bibitem[{{Richardson} {et~al.}(2013){Richardson}, {Baldwin}, {Ferland}, {Loh},
  {Kuehn}, {Fabian}, \& {Salom{\'e}}}]{2013MNRAS.430.1257R}
{Richardson}, C.~T., {Baldwin}, J.~A., {Ferland}, G.~J., {et~al.} 2013, \mnras,
  430, 1257

\bibitem[{{R{\"o}llig} {et~al.}(2007){R{\"o}llig}, {Abel}, {Bell}, {Bensch},
  {Black}, {Ferland}, {Jonkheid}, {Kamp}, {Kaufman}, {Le Bourlot}, {Le Petit},
  {Meijerink}, {Morata}, {Ossenkopf}, {Roueff}, {Shaw}, {Spaans}, {Sternberg},
  {Stutzki}, {Thi}, {van Dishoeck}, {van Hoof}, {Viti}, \&
  {Wolfire}}]{2007A&A...467..187R}
{R{\"o}llig}, M., {Abel}, N.~P., {Bell}, T., {et~al.} 2007, \aap, 467, 187

\bibitem[{{Roueff} {et~al.}(2014){Roueff}, {Alekseyev}, \& {Le
  Bourlot}}]{2014A&A...566A..30R}
{Roueff}, E., {Alekseyev}, A.~B., \& {Le Bourlot}, J. 2014, \aap, 566, A30

\bibitem[{{Sarrasin} {et~al.}(2010){Sarrasin}, {Abdallah}, {Wernli}, {Faure},
  {Cernicharo}, \& {Lique}}]{2010MNRAS.404..518S}
{Sarrasin}, E., {Abdallah}, D.~B., {Wernli}, M., {et~al.} 2010, \mnras, 404,
  518

\bibitem[{{Savage} \& {Sembach}(1996)}]{1996ARA&A..34..279S}
{Savage}, B.~D., \& {Sembach}, K.~R. 1996, \araa, 34, 279

\bibitem[{{Schilke} {et~al.}(2014){Schilke}, {Neufeld}, {M{\"u}ller}, {Comito},
  {Bergin}, {Lis}, {Gerin}, {Black}, {Wolfire}, {Indriolo}, {Pearson},
  {Menten}, {Winkel}, {S{\'a}nchez-Monge}, {M{\"o}ller}, {Godard}, \&
  {Falgarone}}]{2014A&A...566A..29S}
{Schilke}, P., {Neufeld}, D.~A., {M{\"u}ller}, H.~S.~P., {et~al.} 2014, \aap,
  566, A29

\bibitem[{{Sch{\"o}ier} {et~al.}(2005){Sch{\"o}ier}, {van der Tak}, {van
  Dishoeck}, \& {Black}}]{2005A&A...432..369S}
{Sch{\"o}ier}, F.~L., {van der Tak}, F.~F.~S., {van Dishoeck}, E.~F., \&
  {Black}, J.~H. 2005, \aap, 432, 369

\bibitem[{{Shaw} \& {Ferland}(2021)}]{2021ApJ...908..138S}
{Shaw}, G., \& {Ferland}, G.~J. 2021, \apj, 908, 138

\bibitem[{{Shaw} {et~al.}(2005){Shaw}, {Ferland}, {Abel}, {Stancil}, \& {van
  Hoof}}]{2005ApJ...624..794S}
{Shaw}, G., {Ferland}, G.~J., {Abel}, N.~P., {Stancil}, P.~C., \& {van Hoof},
  P.~A.~M. 2005, \apj, 624, 794

\bibitem[{{Shaw} {et~al.}(2009){Shaw}, {Ferland}, {Henney}, {Stancil}, {Abel},
  {Pellegrini}, {Baldwin}, \& {van Hoof}}]{2009ApJ...701..677S}
{Shaw}, G., {Ferland}, G.~J., {Henney}, W.~J., {et~al.} 2009, \apj, 701, 677

\bibitem[{{Shaw} {et~al.}(2020){Shaw}, {Ferland}, \&
  {Ploeckinger}}]{2020RNAAS...4...78S}
{Shaw}, G., {Ferland}, G.~J., \& {Ploeckinger}, S. 2020, Research Notes of the
  American Astronomical Society, 4, 78

\bibitem[{{Shaw} {et~al.}(2006){Shaw}, {Ferland}, {Srianand}, \&
  {Abel}}]{2006ApJ...639..941S}
{Shaw}, G., {Ferland}, G.~J., {Srianand}, R., \& {Abel}, N.~P. 2006, \apj, 639,
  941

\bibitem[{{Smith}(2011)}]{2011IAUS..280..361S}
{Smith}, I. W.~M. 2011, in The Molecular Universe, ed. J.~{Cernicharo} \&
  R.~{Bachiller}, Vol. 280, 361--371

\bibitem[{{Smith} {et~al.}(2004){Smith}, {Herbst}, \&
  {Chang}}]{2004MNRAS.350..323S}
{Smith}, I. W.~M., {Herbst}, E., \& {Chang}, Q. 2004, \mnras, 350, 323

\bibitem[{{Snow} {et~al.}(2007){Snow}, {Destree}, \&
  {Jensen}}]{2007ApJ...655..285S}
{Snow}, T.~P., {Destree}, J.~D., \& {Jensen}, A.~G. 2007, \apj, 655, 285

\bibitem[{{Sonnentrucker} {et~al.}(2003){Sonnentrucker}, {Friedman}, {Welty},
  {York}, \& {Snow}}]{2003ApJ...596..350S}
{Sonnentrucker}, P., {Friedman}, S.~D., {Welty}, D.~E., {York}, D.~G., \&
  {Snow}, T.~P. 2003, \apj, 596, 350

\bibitem[{{Taniguchi} {et~al.}(2016){Taniguchi}, {Saito}, \&
  {Ozeki}}]{2016ApJ...830..106T}
{Taniguchi}, K., {Saito}, M., \& {Ozeki}, H. 2016, \apj, 830, 106

\bibitem[{{van der Tak} {et~al.}(2007){van der Tak}, {Black}, {Sch{\"o}ier},
  {Jansen}, \& {van Dishoeck}}]{2007A&A...468..627V}
{van der Tak}, F.~F.~S., {Black}, J.~H., {Sch{\"o}ier}, F.~L., {Jansen}, D.~J.,
  \& {van Dishoeck}, E.~F. 2007, \aap, 468, 627

\bibitem[{{van der Tak} {et~al.}(2020){van der Tak}, {Lique}, {Faure}, {Black},
  \& {van Dishoeck}}]{2020Atoms...8...15V}
{van der Tak}, F. F.~S., {Lique}, F., {Faure}, A., {Black}, J.~H., \& {van
  Dishoeck}, E.~F. 2020, Atoms, 8, 15

\bibitem[{{van Dishoeck} {et~al.}(1996){van Dishoeck}, {Beaerda}, \& {van
  Hemert}}]{1996A&A...307..645V}
{van Dishoeck}, E.~F., {Beaerda}, R.~A., \& {van Hemert}, M.~C. 1996, \aap,
  307, 645

\bibitem[{{van Dishoeck} \& {Black}(1989)}]{1989ApJ...340..273V}
{van Dishoeck}, E.~F., \& {Black}, J.~H. 1989, \apj, 340, 273

\bibitem[{{van Regemorter}(1962)}]{1962ApJ...136..906V}
{van Regemorter}, H. 1962, \apj, 136, 906

\bibitem[{{Vasyunin} {et~al.}(2007){Vasyunin}, {Semenov}, {Henning}, {Wakelam},
  {Herbst}, \& {Sobolev}}]{2007msl..confE.112V}
{Vasyunin}, A.~I., {Semenov}, D.~A., {Henning}, T., {et~al.} 2007, in Molecules
  in Space and Laboratory, ed. J.~L. {Lemaire} \& F.~{Combes}, 112

\bibitem[{{Vasyunin} {et~al.}(2004){Vasyunin}, {Sobolev}, {Wiebe}, \&
  {Semenov}}]{2004AstL...30..566V}
{Vasyunin}, A.~I., {Sobolev}, A.~M., {Wiebe}, D.~S., \& {Semenov}, D.~A. 2004,
  Astronomy Letters, 30, 566

\bibitem[{{Wakelam} {et~al.}(2010){Wakelam}, {Smith}, {Herbst}, {Troe},
  {Geppert}, {Linnartz}, {{\"O}berg}, {Roueff}, {Ag{\'u}ndez}, {Pernot},
  {Cuppen}, {Loison}, \& {Talbi}}]{2010SSRv..156...13W}
{Wakelam}, V., {Smith}, I.~W.~M., {Herbst}, E., {et~al.} 2010, \ssr, 156, 13

\bibitem[{{Wan} {et~al.}(2018){Wan}, {Yang}, {Stancil}, {Balakrishnan},
  {Parekh}, \& {Forrey}}]{2018ApJ...862..132W}
{Wan}, Y., {Yang}, B.~H., {Stancil}, P.~C., {et~al.} 2018, \apj, 862, 132

\bibitem[{{Woon} \& {Herbst}(2009)}]{2009ApJS..185..273W}
{Woon}, D.~E., \& {Herbst}, E. 2009, \apjs, 185, 273

\bibitem[{{Wootten} {et~al.}(2021){Wootten}, {Bentley}, {Baldwin}, {Combes},
  {Fabian}, {Ferland}, {Loh}, {Salome}, {Shingledecker}, \&
  {Castro-Carrizo}}]{2021arXiv211106033W}
{Wootten}, A., {Bentley}, R.~O., {Baldwin}, J., {et~al.} 2021, arXiv e-prints,
  arXiv:2111.06033

\bibitem[{{Zhang} {et~al.}(2021){Zhang}, {Cummings}, {Wan}, {Yang}, \&
  {Stancil}}]{2021ApJ...912..116Z}
{Zhang}, Z.~E., {Cummings}, S.~J., {Wan}, Y., {Yang}, B., \& {Stancil}, P.~C.
  2021, \apj, 912, 116

\end{thebibliography}
\bibliographystyle{aasjournal}
\end{document}